\documentclass[12pt,twoside, openright]{report}
\usepackage[centertags]{amsmath}
\usepackage{amsfonts}
\usepackage{amssymb}
\usepackage{amsthm}
\usepackage{newlfont}
\usepackage{xspace}
\usepackage{graphicx}
\usepackage{psfig}
\usepackage{subfigure}
\usepackage{amsbsy}
\usepackage{amstext}
\usepackage{psfrag}
\usepackage{multirow}
\usepackage{rotating}
\usepackage{multibox}
\usepackage{curves}
\usepackage{threeparttable}
\usepackage{supertabular}
\usepackage{afterpage}
\usepackage{hyperref}
\usepackage{writer}

\usepackage{layout}
\usepackage{calc}

\begin{document}
\bibliographystyle{plain}

\newcommand{\piz}{$\pi^0$}
\newcommand{\trig}{$2\times 2$}

\newcommand{\mytitle}{Measurement of the Transverse Single-Spin Asymmetry for Mid-rapidity Production
of Neutral Pions in Polarized \emph{p+p} Collisions at 200 GeV
Center-of-Mass Energy}

\newcommand{\myname}{Christine Angela Aidala}

%
%
\pagestyle{empty}
%
%
%
%
%
%

\begin{center}
\vspace*{0.25\textheight}
{\Large \mytitle} \\
\vspace*{1in}
\myname \\
\vspace*{1in}
Submitted in partial fulfillment of the \\
requirements for the degree \\
of Doctor of Philosophy \\
in the Graduate School of Arts and Sciences \\
\vspace*{1.0in}
COLUMBIA UNIVERSITY \\
2006
\end{center}

%
%

%
%

\vspace*{0.55\textheight}

\setlength{\baselineskip}{2.0\baselineskip}

\begin{center}
\copyright \ 2006 \\
\myname \\
All Rights Reserved
\end{center}

\setlength{\baselineskip}{0.5\baselineskip}

%
%

\setlength{\baselineskip}{1.4\baselineskip}
%
%

\begin{center}
{\Large ABSTRACT} \\
\vspace{0.25in}
{\large \mytitle} \\
\vspace{0.25in}
\myname \\
\vspace*{0.5in}
\end{center}

The spin structure of the proton has revealed itself to be extremely
complex and is an area of ongoing research.  The Relativistic Heavy
Ion Collider (RHIC) at Brookhaven National Laboratory (BNL)
inaugurated its operation as the first polarized-proton collider
during the 2001-2002 run, marking the beginning of a new era in the
study of proton spin structure.

From the data collected in this run, the PHENIX experiment measured
the transverse single-spin asymmetry ($A_N$) for neutral pion
production at $x_F\approx$0.0 over a transverse momentum range of 1
to 5~GeV/$c$ from polarized proton-proton interactions at a
center-of-mass energy ($\sqrt{s}$) of 200~GeV and found it to be
zero within a few percent.  Interest in these measurements arises
from the observation of large ($\sim$30\%) transverse single-spin
asymmetries in $p+p^{\uparrow}\!\!\rightarrow\!\!\pi + X$ at forward
angles by the E704 collaboration at Fermilab ($\sqrt{s} =
19.4$~GeV), found by the STAR and BRAHMS experiments to persist at
RHIC energies, as well as single-spin, azimuthal asymmetries
observed recently in semi-inclusive deep-inelastic scattering
experiments.  Such large asymmetries were initially surprising
because at leading order, perturbative quantum chromodynamics (pQCD)
predicted only small effects.

Several possible origins of these large asymmetries have been
proposed.  Despite great theoretical progress in recent years, no
single, clear formalism has emerged in which to interpret the
available data.  Further theoretical work and a variety of
additional experimental measurements will be necessary to understand
current results and elucidate the transverse spin structure of the
proton.



%
%
 \pagestyle{plain} \pagenumbering{roman}
\newpage

\thispagestyle{empty}

\vfill \vspace*{18cm}

\begin{center}For technical reasons, there have been minor textual
modifications to the Table of Contents, List of Figures, and List of
Tables in the present online version with respect to the final
version of this thesis on record with Columbia University.\\
\end{center}

\vspace{2cm} \tableofcontents \listoffigures \listoftables
\begin{center}
{\Large \bf Acknowledgements}
\end{center}

First and foremost, I have to express my gratitude to my husband,
Gabriele Carcassi, for all the sacrifices he has made in order to
make it possible for me to pursue a Ph.D. in physics here in the
U.S., my home country.  Without his dedicated support and infinite
patience, in particular through rough times in the past, I would not
be where I am today.  He inspires me to achieve more, to think big,
to dream what I never otherwise would have imagined for myself. I
only hope that his own immense talents and creativity can find a
worthy outlet, his own great ambitions fulfilled, through his life
with me. I could hardly acknowledge Gabriele without recognizing how
rewarding it has been to share the experience of watching our son,
Matteo, grow and develop since he came into our world in November
2004. Becoming parents has rounded out our lives, bringing new
adventures and experiences, leading us to grow in turn as adults. I
look forward to many more years drawing energy and strength from my
family.

I would like to thank my own parents, Paul and Susan Aidala, who
instilled in me an appreciation of nature, science, and mathematics.
Through them I learned integrity, the value of hard work, and the
importance of attention to detail. They taught me to be active in
the pursuit of knowledge.  Countless questions of mine were answered
by trips to our shelves of reference books; thus were my humble
beginnings as an elementary-school researcher. The influence of my
parents is made even more evident by the fact that my sister, Kathy,
shares a similar interest in science.  She is currently a Ph.D.
candidate in the Applied Physics Department at Harvard, and it is a
pleasure to be able to continue to share so much with her
intellectually.

I am grateful to Brian Cole as my thesis sponsor.  He has always
been ready with a wealth of ideas and suggestions whenever I have
discussed my work with him.  During the course of my three years at
Columbia, he has provided detailed input and feedback on my efforts
involving PHENIX software triggers, data analysis, and this
dissertation. He has been extremely flexible regarding the details
of where, how, and when I have worked.  He has listened to
understand my needs and goals and helped me to further my own
interests.  Such an accommodating and supportive supervisor is not
something a graduate student can take for granted.

Being a part of the PHENIX group at Columbia has been a rewarding
experience overall.  Brian, along with Bill Zajc, graciously allowed
me to pursue a physics topic outside the group's area of focus; I
feel that the intellectual independence granted to me has served me
well.  My professional development has also been enriched by
extensive travel.  I have been able to take advantage of numerous
conferences, workshops, and summer schools, thanks largely to
financial support provided by the group.  Moreover, Brian and Bill
have both provided invaluable career advice and have consistently
encouraged me to pursue my professional interests.  The other junior
members of the group with whom I have overlapped, especially Justin
Frantz, Sotiria Batsouli, Mickey Chiu, and David d'Enterria, have
been a friendly support network and resource for me to draw upon.
Furthermore, I was surprised and touched by the overwhelming support
I received from the group when I informed them of my decision to
have Matteo during the last year of my Ph.D.  The success I have had
in balancing my studies with my family life is due in part to their
unhesitating confidence in me.

Bill Zajc, as spokesperson of PHENIX, has provided a role model of
great responsibility, amazing time management, and politics without
pettiness.  His ability to keep a handle on the experiment and
collaboration on a multitude of focus levels, from the place of
PHENIX in the context of the broader field of nuclear physics to the
details of many individual careers, is impressive.  He has made me
feel that even in a large collaboration, individuals matter.

I would like to thank Matthias Grosse Perdekamp for the role he has
played as my "spin advisor" on PHENIX.  His in-depth knowledge and
understanding of a wide spectrum of physics has set a standard for
me to aspire towards as a young experimentalist.  He has been
extremely supportive of my career and has provided a number of
useful career discussions.

I would like to express my sincere appreciation to Werner Vogelsang
for a wealth of ready physics information and advice over the past
several years.  My quests for knowledge, big and small, were always
received with respect and interest, and frequently with enthusiasm.
He has been a valuable asset throughout my time as a student, and I
expect to continue to learn from him in the years to come.

I would like to acknowledge Sam Aronson for the role he played in
bringing me back to physics in 2001 after I spent nearly two years
out of the field.  I found myself with limited options at that time,
and if he had not hired me to do research at BNL, I feel that life,
for better or worse, would have most likely taken me down a
completely different path. He opened the door to all the positive
experiences I have had as a part of PHENIX for the past four years.
I appreciate the gamble he took, hiring me based on such little
information, and I have done my best to prove myself a worthwhile
investment.

Recalling my early days in PHENIX, warm memories of all the support
I received from Saskia Mioduszewski come immediately to mind.  I was
indeed lucky to find myself sharing an office with her when I came
to BNL.  She made me feel welcome when I first arrived, and over
time she became a friend and mentor, offering ready and reliable
advice on technical problems, analysis issues, career paths, or life
in general.

After Saskia was promoted and moved down the hallway, my good
fortune in officemates at BNL continued.  Henner B\"{u}sching has
become a valued friend and colleague.  He has provided an example of
dedication to his work and of caring about doing it well.  Through
him, I have witnessed the contribution a single young scientist can
make toward the successful functioning of a large collaboration. I
greatly appreciate the technical and moral support he has offered
for nearly three years now, as well as the positive and pleasant
work environment he has created.

Also at BNL, I would like to thank Craig Woody and Gabor David for
all that I learned through them during my first year on PHENIX. They
helped lay the foundation for my ability to work relatively
independently in subsequent years.

Other PHENIXians who supported my efforts, in particular on my
thesis analysis, were Frank Bauer and Hisayuki Torii. Frank provided
numerous helpful conversations regarding transverse spin physics and
analysis as well as moral support.  I was pleased to bring his own
work to eventual publication after he moved on to pursue a career
outside of physics. Hisa furnished lots of help and technical
support in the early stages of my thesis analysis.

Turning back to Columbia, I would like to thank Lalla Grimes, the
Administrative Coordinator in the Physics Department, for being so
warm, welcoming, competent, efficient, and simply a delight to deal
with. She has done a great deal to make me feel at home in the
Department right from the beginning.

I am also grateful to many of my fellow classmates for making my
experience at Columbia such a positive one, especially Mike Cheng,
Oleg Loktik, Sasha Lyulko, Alexis Aguilar, Chad Johnson, Azfar Adil,
Bahar Moezzi, and Christina Tosti.  They created a friendly and
cooperative atmosphere during the time we shared in classes and, of
course, working on problem sets.  It was great to know there were so
many people I could reliably turn to when I wanted to discuss course
work or physics in general.  I would specifically like to thank
Sasha for hosting me periodically in the city and Alexis for
inspiring me to study so hard for the quals.

During the three semesters in which I was taking classes, I was
fortunate to have the local hospitality of Angela and Genevieve
Aidala and Jean Schmidt available.  Their kindness in opening their
Upper West Side home to me gave some respite from the challenge of
commuting into campus several days a week from eastern Long Island.

Last but not least, I would like to express my deep appreciation to
Stefan Bathe for his staunch encouragement and support over the past
several years we have spent together on PHENIX.  The close
friendship we have developed has uplifted my spirits time and again.
I am grateful for the countless, valuable conversations we have had
regarding careers and life paths in general. I am also indebted to
him for his careful reading of and extensive feedback on earlier
drafts of this dissertation; the present version is greatly improved
thanks to his thorough comments.

%
%

\begin{center}
\vspace*{0.4\textheight} To Yale University Prof. Emer. Frank W. K.
Firk, who, ten years ago, inspired the beginnings of a career.
\end{center}

%
%

%
%
%
%
\makeatletter
\renewcommand{\ps@plain}{%
\renewcommand{\@oddhead}{\hfil\textit{\thepage}}%
\renewcommand{\@evenhead}{\@oddhead}%
\renewcommand{\@oddfoot}{}%
\renewcommand{\@evenfoot}{}}
\makeatother
\pagestyle{headings} \pagenumbering{arabic}

\chapter{Introduction}
\section{Proton structure}

The proton, together with the neutron and electron, is one of the
basic building blocks of everyday atomic matter. Far from the point
particle it was once believed to be, it has proven to be an
extremely complex entity, and more than 80 years after it was
discovered in the first decades of the twentieth century, the
composition of the proton is still not completely understood. A very
rich structure has gradually been uncovered over the past 40 years
of research, with the appropriate description depending on the
energy scale at which the proton is probed.  The composition of the
proton is now described by \emph{partons}, including \emph{quarks}
and the particles carrying the force that binds them, \emph{gluons}.
More specifically, quarks can be categorized as either
\emph{valence} or \emph{sea}.  In the simplest composite model of
the proton, it can be viewed as three valence quarks, each carrying
1/3 of the proton's linear momentum. In reality, measurements have
demonstrated that there is a multitude of gluons and sea
quark-antiquark pairs present as well, each carrying generally a
small fraction of the proton's momentum but with a large summed
momentum contribution overall.

\section{The "proton spin crisis"}

In the naive model of the proton as simply three valence quarks of
spin $\frac{1}{2}\hbar$ each, one might expect the proton's spin to
be the straightforward sum of two parallel quark spins and one
antiparallel.  However, in the late 1980's it was discovered that in
fact only a small fraction of the proton's spin, less than 30\%, was
carried by quarks.  This revelation, surprising at the time, came to
be known as the "proton spin crisis."  In retrospect, considering
the complex linear-momentum structure of the proton, it is
reasonable to expect a complex angular-momentum structure as well.
Not only the spin of the partons is involved, but also their orbital
angular momentum.

Due to the fact that spatial rotations and Lorentz boosts do not
commute, polarized proton structure must be considered separately
for a proton with spin vector parallel to or perpendicular to its
(linear) momentum.  This difference between the longitudinal and
transverse spin structure adds further complexity to the problem. As
will be discussed, significant progress has been made in
understanding the longitudinal spin structure of the proton, while
transverse structure remains a largely open field.  Despite
impressive advances in just the past few years, numerous additional
experimental measurements will need to go hand-in-hand with further
theoretical investigation in order to elucidate the transverse spin
structure of the proton.

\section{Studying proton spin structure at the Relativistic Heavy Ion Collider}

Resolution of the proton spin crisis, in particular determination of
contributions from sea quarks and gluons, remains the goal of
extensive ongoing study. The PHENIX experiment at the Relativistic
Heavy Ion Collider (RHIC) at Brookhaven National Laboratory is in a
unique position to make significant contributions to improve our
understanding of the origin of the proton's spin.

RHIC is the most versatile hadron collider in the world.  It is
capable of colliding heavy ions with energies as high as $\sqrt{s} =
200$~GeV per colliding nucleon pair and polarized protons anywhere
from 50 to 500~GeV, as well as different species in the two beams.
In the first five years of running, RHIC has provided gold
collisions at four different energies, copper collisions at three
energies, deuteron-gold collisions, and polarized-proton collisions.
The flexibility of RHIC allows for a very diverse physics program.
The heavy ion physics program investigates strongly-interacting
matter at extreme temperatures and energy densities, seeking to
create and study the properties of a state of matter known as the
quark-gluon plasma (QGP).

The polarized proton program seeks a better understanding of the
proton's spin structure, in particular contributions to its
longitudinal spin structure from the gluons and sea quarks.  By
studying the proton using hadronic collisions rather than
electromagnetic probes, which do not couple to the
electromagnetically neutral gluon, RHIC experiments may directly
observe gluon-scattering processes. As a collider, RHIC can provide
collisions at much higher energy than can be achieved in
fixed-target measurements.  As a result hard processes, describable
by perturbative quantum chromodynamics (pQCD), can be studied, and
new probes such as $W$ bosons will eventually become available.

\section{Aims and outline of this thesis}

This thesis aims to motivate the study of the structure of the
proton, in particular the polarized structure, as a fundamental
question in QCD. It also seeks to describe how spin-dependent
observables can be and have already been measured at RHIC, with
focus on a transverse single-spin asymmetry (SSA) measurement,
providing information on the transverse spin structure of the
proton.

This thesis will present a review of proton structure, both
unpolarized and polarized, in terms of history and the current
status. An overview of pQCD as applicable to proton-proton
collisions at RHIC will be given. The RHIC polarized-proton
accelerator complex and the PHENIX experiment and detector will be
described. A measurement of the transverse single-spin asymmetry of
neutral pions will be presented and discussed.  Finally, the
prospects for future measurements to shed further light on the
transverse spin structure of the proton will be explored.

\chapter{Nucleon structure}

\section{Unpolarized nucleon structure}

\subsection{Elastic structure}

\subsubsection{Magnetic moments}

The first evidence of proton substructure came from a measurement of
its magnetic moment in 1933 by Esterman, Frisch, and Stern. It was
found to be anomalously large and is now known to be approximately
2.79 times the Dirac magnetic moment, given by
$\overrightarrow{\mu}_p = \frac{e}{Mc}\overrightarrow{S}$, for a
point-like spin-$\frac{1}{2}$ particle of the same mass.  The
anomalous magnetic moment of the proton is now understood in terms
of its valence quark structure (see Section~\ref{section:QCD} below)
and can be given by $\mu_p = \frac{1}{3}(4\mu_u - \mu_d)$, where
$\mu_u$ and $\mu_d$ are the magnetic moments of the up and down
valence quarks, respectively.

Similarly, the magnetic moment of the neutron was also found to be
anomalous by Esterman and Stern in 1934 and is now understood in
terms of its own valence quark structure.

\subsubsection{Form factors}
Charge and current distributions within the nucleon can be described
by electromagnetic form factors, measurable via elastic
electron-proton scattering.  Viewed in a particular frame known as
the Breit frame ($\overrightarrow{p}_{\textrm{final}} =
-\overrightarrow{p}_{\textrm{initial}}$ for the proton), the form
factors $G_E$ and $G_M$ are proportional to the Fourier transforms
of the charge and magnetization distributions, respectively.

The cross section for elastic electron-proton scattering can be
expressed in terms of the form factors, as given in Eq.
\ref{eq:elasticCrossSection},
\begin{equation}\label{eq:elasticCrossSection}
    \frac{d\sigma}{d\Omega}\vert_{lab} =
    \frac{\alpha^2}{4E^2\sin^4\frac{\theta}{2}}
    \frac{E'}{E}\left(\frac{G_E^2+\tau
    G_M^2}{1+\tau}\cos^2\frac{\theta}{2} + 2\tau
    G_M^2\sin^2\frac{\theta}{2}\right)
\end{equation}
where $\tau \equiv -q^2/4M^2$, $q$ is the four-momentum transfer in
the scattering, $M$ is the proton mass, $\alpha$ is the fine
structure constant, $\theta$ is the electron scattering angle in the
laboratory frame, and $E$ and $E'$ are the incident and scattered
electron energies. The electric and magnetic form factors can be
determined either via differential cross section measurements of
unpolarized $e+p$ scattering or via measurement of the recoil proton
polarization in the reaction $\overrightarrow{e}+p\rightarrow
e+\overrightarrow{p}$, where the arrows indicate polarization. For a
relatively recent summary of proton electromagnetic form factor
measurements performed at Jefferson Lab, see
\cite{Perdrisat:2003en}.

\subsection{Inelastic structure:  QCD and the quark-parton model}
\label{section:QCD}

In the 1960's deep-inelastic lepton-nucleon scattering (DIS)
experiments at SLAC, analogous to the famous Rutherford scattering
experiment that led to the discovery of the atom's hard core, found
that protons also had "hard" subcomponents \cite{Bloom:1969kc,
Breidenbach:1969kd}. These hard subcomponents came to be known as
partons.  It took some time before the experimentally observed
partons inside the proton were identified as the so-called "quarks,"
which had been theoretically hypothesized based on hadron
spectroscopy data as part of the "Eightfold Way" by Gell-Mann and
Ne'eman independently in the early 1960's \cite{Gell-Mann:1964xx}.
But eventually the quark-parton model of the proton came into being.
As experimental work progressed and higher-energy lepton beams were
used as probes, the proton came to reveal a much more intricate
structure than that of the three so-called "valence" quarks.  These
other subcomponents are now known to be sea quarks and gluons.

The experimental and theoretical work in the 1960's and 1970's
regarding hadronic interactions and structure led to the development
of the theory of quantum chromodynamics (QCD), describing the
behavior of the strong force.  A central concept of QCD is that of
\emph{asymptotic freedom}.  While quarks are strongly bound at
distance scales larger than a typical hadron radius ($r \approx
10^{-15}$~m), at shorter distances they behave as nearly free.

\subsubsection{Structure functions}
\label{section:structureFunctions}

The nucleon structure functions describe the inelastic structure of
the proton and neutron, probed principally via DIS.  The double
differential cross section for inelastic cross section for
electron-proton scattering can be expressed as in
Eq.~\ref{eq:DISCrossSection},

\begin{equation} \label{eq:DISCrossSection}
\frac{d^2\sigma}{dE'd\Omega}|_\textrm{lab} = \frac{4\alpha^2
E'^2}{q^4}\left[W_2(\nu,q^2)\cos^2 \frac{\theta}{2} +
2W_1(\nu,q^2)\sin^2 \frac{\theta}{2}\right]
\end{equation}
in which $q$, $\alpha$, $\theta$, and $E'$ are as in
Eq.~\ref{eq:elasticCrossSection}, $\nu = (p \cdot q)/M$ with $p$
being the initial nucleon four-momentum, and $W_1$ and $W_2$ are the
proton structure functions. Note the similarity of
Eq.~\ref{eq:DISCrossSection} to Eq.~\ref{eq:elasticCrossSection},
the cross section for elastic electron-proton scattering, with the
structure functions playing the role of the form factors.

It is common to express the proton structure functions slightly
differently, as given in Eq.~\ref{eq:structureFunctions}.
\begin{eqnarray}\label{eq:structureFunctions}
\nonumber
F_1 &=& MW_1 \\
F_2 &=& \nu W_2
\end{eqnarray}
$F_1$ and $F_2$ can be written as functions of $Q^2 = -q^2 > 0$ and
the dimensionless variable $x = Q^2/2p\cdot q$.  They can be related
to the cross sections for scattering of transversely and
longitudinally polarized virtual photons off of the proton,
$\sigma_T$ and $\sigma_L$, as given in
Eq.~\ref{eq:longAndTransverseDISCrossSections}.
\begin{eqnarray}\label{eq:longAndTransverseDISCrossSections}
\nonumber
\sigma_T &\propto& F_1 \\
\sigma_L &\propto& \left( \frac{F_2}{2x} - F_1 \right)
\end{eqnarray}
The total virtual photon-proton cross section is the sum of these
components and is proportional only to $F_2$.

In 1969 Bjorken predicted that at large $Q^2$, scattering off of
"point-like" subcomponents that were approximately free in the
proton would lead to proton structure functions with no $Q^2$
dependence for a given value of $x$, i.e.~that scale with just this
single, dimensionless variable \cite{Bjorken:1968dy}. That is,
assuming point-like constituents of the proton, at large $Q^2$,
\emph{inelastic} electron-\emph{proton} scattering could be viewed
as \emph{elastic} scattering of an electron off of a hard,
point-like particle \emph{within} the proton.

The experiments performed at SLAC mentioned above
\cite{Bloom:1969kc, Breidenbach:1969kd} discovered the scaling
behavior predicted by Bjorken. The structure functions they measured
had very little explicit dependence on $Q^2$ and could in fact be
written simply as functions of $x$.

\begin{figure}
\centering
\includegraphics[height=0.8\textheight]{%
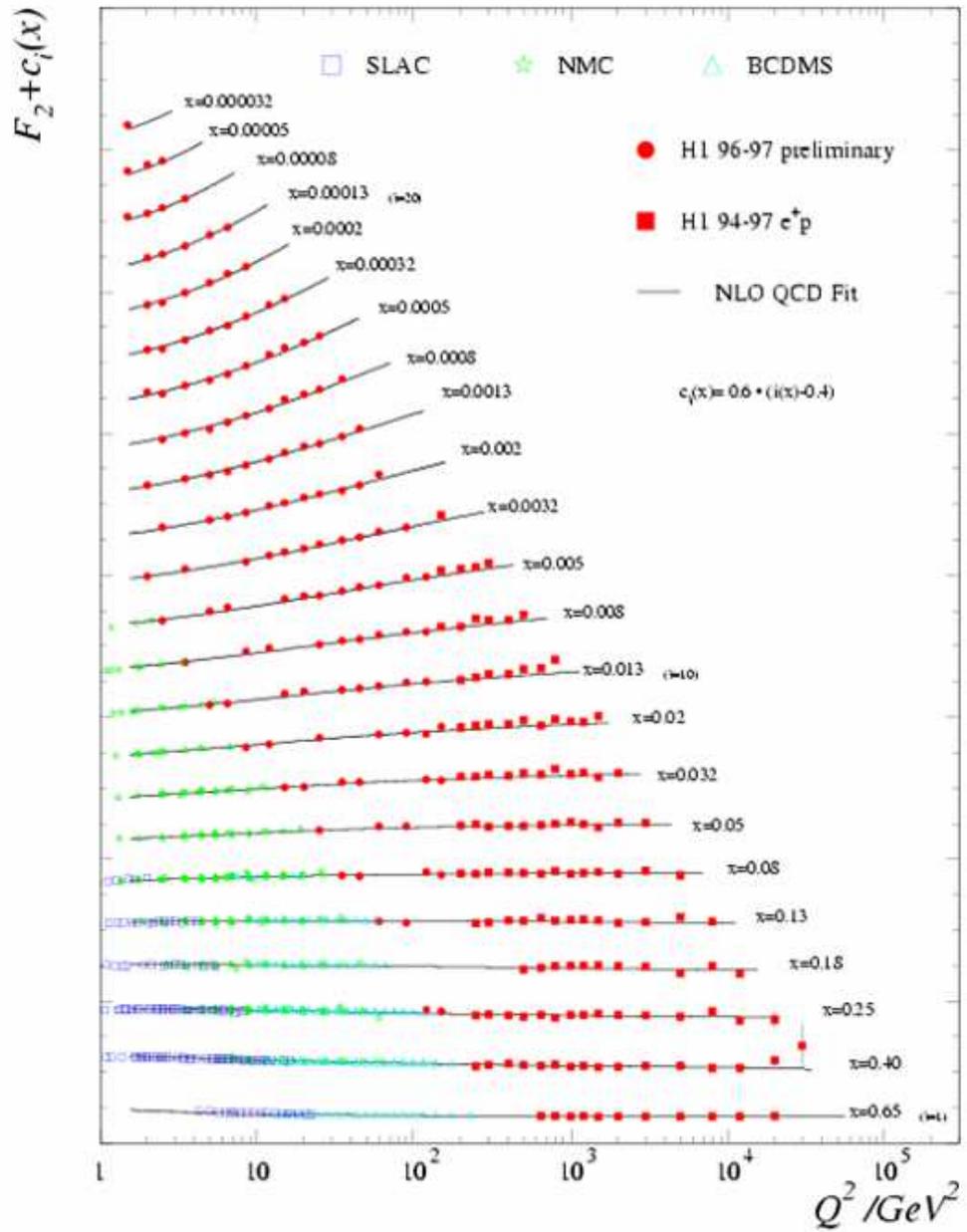} \caption[World DIS data on the unpol. structure
function of the proton, $F_2$.]{World DIS data on the unpolarized
structure function of the proton, shown compared to next-to-leading
order (NLO) pQCD fits. } \label{figure:unpolStructure}
\end{figure}

In Figure~\ref{figure:unpolStructure} showing world DIS data for
$F_2$ of the proton, one sees that for $x \gtrsim 0.02$, $F_2$ is
nearly flat in $Q^2$. This indicates that the hard subcomponents
being probed are approximately free. The early measurements included
here provided evidence leading to development of the concept of
asymptotic freedom in QCD.  Note, however, the scaling violations
observed at low $x$ and low $Q^2$. These are now understood in terms
of gluon radiation emitted by the parton prior to the hard
scattering. See Section \ref{section:pdfs} for discussion of the
relation of scaling violations to the gluon distribution function.

In 1969 Callan and Gross predicted that for spin-$\frac{1}{2}$
charged components within the nucleon, the scaling structure
functions would be related as given in Eq.~\ref{eq:callanGross},
known as the Callan-Gross relation \cite{Callan:1969uq}.
\begin{equation}\label{eq:callanGross}
F_2(x) = 2xF_1(x)
\end{equation}
Experimental confirmation of the Callan-Gross relation came from
SLAC in the late 1970's \cite{Bodek:1979rx} and thus provided strong
evidence for the spin-$\frac{1}{2}$ nature of what are now known to
be quarks.

\subsubsection{Parton distribution functions}
\label{section:pdfs}

Feynman's introduction of the quark-parton model (QPM) in 1969
\cite{Feynman:1969ej} offered a relatively intuitive explanation of
Bjorken scaling.  The virtual photon in DIS could be viewed as
scattering elastically off of a collection of hard partons within
the proton; the DIS cross section is the incoherent sum of the
individual cross sections.  As the proton momentum approaches
infinity, $x$ can be seen as the fraction of the proton's linear
momentum carried by the parton. The size of the cross section for
scattering off of a particular parton is proportional to the
probability, $q(x)$, of hitting a quark of flavor $q$ carrying
momentum fraction $x$ of the proton. $q(x)$ is known as a parton
distribution function (pdf). The scaling structure functions,
$F_1(x)$ and $F_2(x)$, can then be viewed as representing the
probability of scattering off of a parton within the proton carrying
momentum fraction $x$.  They can be expressed in terms of the pdf's
for different quark flavors as follows, where $e_i$ indicates the
electromagnetic charge of the quark of flavor $i$.

\begin{eqnarray}
\nonumber
  F_1(x) &=& \frac{1}{2}\sum_i e_i^2q_i(x) \\
  F_2(x) &=& \sum_i e_i^2xq_i(x)
\end{eqnarray}

Including all partons in the proton and not only the charged
particles, which couple electromagnetically and can be probed
directly by DIS, a momentum sum rule is obtained
(Eq.~\ref{eq:momSumRule}).

\begin{equation}\label{eq:momSumRule}
\sum_i\int dx xf_i(x) = 1
\end{equation}
It has been found experimentally that approximately half of the
proton's total momentum is carried by (electromagnetically neutral)
gluons, which are present as mediators of the strong interactions
among the quarks in the nucleon and dominate at low $x$ values.  In
the Bjorken-scaling regime, at large $x$, the interactions among the
quarks, i.e. the gluons, do not play a significant role.  While DIS
experiments cannot directly probe the gluon distribution function,
$g$, the scaling \emph{violations} and the evolution of the
structure functions in $Q^2$ provide information on the gluon. The
gluon distribution can be obtained from the logarithmic scaling
violations of the structure function $F_2$, as given in
Eq.~\ref{eq:gluonDist}.

\begin{equation}\label{eq:gluonDist}
g \propto dF_2/d(\ln Q^2)
\end{equation}
See Section \ref{section:unpolStatus} for a discussion of the
current status of pdf measurements.

\subsection{pQCD, factorization, and universality}

Performing calculations in QCD presents a number of challenges that
quantum \emph{electro}dynamic (QED) calculations do not. In QCD the
force carriers themselves are charged; gluons carry color charge,
whereas photons, the force transmitters in QED, are electrically
neutral. Contributions from higher-order Feynman scattering diagrams
in QED, i.e.~higher powers in the electromagnetic coupling constant,
$\alpha \approx 1/137$, representing additional lepton-photon
vertices, quickly become negligibly small, due to the fact that
$\alpha$ is much less than one. An analogous expansion in QCD is
only possible in the regime where the coupling, $\alpha_s(Q^2)$, is
small, which is generally the case for processes involving a large
momentum transfer.  Perturbative QCD (pQCD) is the calculation
technique used in this kinematic regime.

A hadron can be viewed as a collection of free, massless partons
with parallel momenta.  The collinear factorization theorem in pQCD
starts from this assumption of \emph{collinearity} of the partons
and hadrons, i.e.~no transverse momentum of the partons in the
proton with respect to the initial proton momentum, and no
transverse momentum of the final-state hadron with respect to the
scattered parton momentum. The collinear factorization theorem
separates cross sections for hard-scattering processes into parts
that are soft, or non-perturbative, and hard, or perturbative, in a
self-consistent way. The soft components, pdf's and fragmentation
functions (FF's), must be obtained from experimental measurements.
The hard components, partonic hard-scattering cross sections, are
directly calculable in pQCD. Parton distribution functions,
discussed above in Section~\ref{section:pdfs}, correspond to the
probability of striking a particular parton carrying momentum
fraction $x$ of the proton; FF's represent the probability of the
scattered parton fragmenting into a particular final-state hadron,
as a function of the fraction $z$ of the scattered parton's momentum
passed along to the final-state hadron.

The factorization theorem was developed and proven over the course
of the late 1970's to the mid-1980's. Early work can be found in
\cite{Libby:1978qf,Ellis:1978sf,Ellis:1978ty,Amati:1978wx,Amati:1978by,Curci:1980uw};
complete proofs are available in
\cite{Collins:1985ue,Collins:1988ig,Mueller:1989,Bodwin:1984hc}. The
factorized cross section for hard scattering in hadron-hadron
collisions ($A + B \rightarrow C$) is given by
Eq.~\ref{eq:factCrossSect},

\begin{equation}\label{eq:factCrossSect}
d\sigma = \sum_{abc} f_a(x_a,\mu_f) \otimes f_b(x_b,\mu_f) \otimes d
\hat{\sigma}_{ab}(x_a, x_b, z_c, \mu_f, \mu_{f'}) \otimes D_c^C(z_c,
\mu_{f'})
\end{equation}
in which $f_a$ ($f_b$) is the density of parton $a$ ($b$) in hadron
$A$ ($B$),  $D_c^C$ is the fragmentation function of parton $c$ into
hadron $C$, and $\mu_f$ and $\mu_{f'}$ are arbitrary scales known as
factorization scales, which can be thought of as the amount of
parton radiation incorporated into the pdf's and FF, or as the
separation scale chosen to distinguish between the hard and soft
components of the cross section.  While the scales chosen are
arbitrary, they must be chosen consistently between the soft
components and the partonic hard-scattering cross section,
$d\hat{\sigma}_{ab}$.  The partonic cross section depends on an
additional arbitrary scale, the renormalization scale, which
controls the running of the strong coupling, $\alpha_s$. It is
common practice to set the factorization and renormalization scales
to be equal; a typical value chosen is one close to the momentum
transfer of the process.

The principle of \emph{universality} in conjunction with the
factorization theorem makes the formalism of pQCD extremely
powerful. The principle of universality states that pdf's and FF's
are the same regardless of the scattering processes involved.
Universality implies the dominance at high momentum transfer of
leading-twist (twist-two) contributions, with interactions only
between the two hard-scattering partons. A higher-twist calculation
takes into account the exchange of additional gluons between the
hard-scattering partons and the nucleon remnants.  The twist
expansion is in successive powers of $1/Q^2$; therefore,
higher-twist contributions are suppressed for processes with large
momentum transfer.  Because of universality, pdf's and FF's can be
measured in the environment which allows the most accurate
determination and then utilized as input for pQCD calculations in
other processes. For example quark distribution functions can be
measured in DIS experiments, in which the kinematics and thus the
probed $x$ values are straightforward to understand, then utilized
in calculations for the more complicated environment of
hadron-hadron collisions. FF's are most easily measured in $e^+ +
e^-$ collisions because the four-momenta of the outgoing quarks are
well known. Decades of comparison between experimental cross section
measurements and pQCD have provided a testing ground for the
assumption of universality, and by now it is a well established and
accepted principle.

While pdf's are not calculable in pQCD and are typically obtained
from experiment, they are in principle calculable using other
theoretical techniques such as lattice QCD. For recent calculations
of structure and distribution functions on the lattice, see
\cite{Orginos:2005uy} and references therein.

Over the course of the 1970's, a formalism emerged in which it was
possible to take a measurement of a pdf at a particular value of $x$
and $Q^2$ and predict the pdf at the same $x$ but different $Q^2$.
This formalism is known as DGLAP, acknowledging important
contributions from Dokshitzer, Gribov, Lipatov, Altarelli, and
Parisi
\cite{Gribov:1972ri,Lipatov:1974qm,Dokshitzer:1977sg,Altarelli:1977zs}.
DGLAP has been essential to the relevant application of factorized
pQCD, which requires as input experimentally measured pdf's, which
necessarily are available at only a finite set of $x$ and $Q^2$
values. Using DGLAP, calculations can be done for any $Q^2$ value
desired.

There is additionally a prescription for evolution of measured pdf's
to different values of $x$ for fixed $Q^2$, formulated by Balitsky,
Fadin, Kuraev, and Lipatov, also in the 1970's
\cite{Balitsky:1978ic,Kuraev:1976ge,Kuraev:1977fs}.  The BFKL
technique has demonstrated itself to be similarly useful in
performing calculations at desired values of $x$ and $Q^2$.

Factorized pQCD has proven to be a valuable and successful
theoretical technique for many years now.  Its applicability to
measurements in $p+p$ collisions at RHIC will be discussed in
Chapter~\ref{section:QCDatRHIC}.

\subsection{Current status of the unpolarized structure of the
proton} \label{section:unpolStatus}

A long history of experiments, in particular DIS experiments at
SLAC, CERN, and DESY, has measured the unpolarized structure of the
proton well.  Gluons have been found to play an important role,
carrying approximately 50\% of the proton's momentum. A
comprehensive review of the contributions to unpolarized proton
structure made by the experiments at the HERA electron-proton
collider at DESY is given in \cite{Abramowicz:1998ii}.  A thorough
review of nucleon structure functions and pdf's is available in
\cite{Cooper-Sarkar:1997jk}.

The wealth of accumulated data regarding the unpolarized structure
of the proton is evident in Figure~\ref{figure:unpolStructure}.
Measured $x$ values range from deep into the sea at $x = 3.2 \times
10^{-5}$ to well into the valence region at $x = 0.65$, with $Q^2$
values as high as $10^4~\textrm{GeV}^2$. Periodic efforts have been
made to examine all data available and perform a global analysis in
order to obtain the best-fit pdf's; see for example
\cite{Pumplin:2002vw}.

In the 1990's it was discovered by the NMC experiment at CERN that
there is a flavor asymmetry in the unpolarized sea of light quarks
in the proton \cite{Amaudruz:1991at,Arneodo:1994sh}.  There is a
significant excess of $\bar{d}$ with respect to $\bar{u}$. Although
no known symmetry requires $\bar{d} / \bar{u} = 1$, the experimental
result was unexpected.  It has since been confirmed by other
experiments at CERN \cite{Baldit:1994jk}, Fermilab
\cite{Hawker:1998ty,Towell:2001nh}, and DESY
\cite{Ackerstaff:1998sr} but is still not well understood.

\section{Polarized proton structure}
\label{section:polStructure}

\subsection{Historical overview}

For many years it was assumed that the proton's spin of
$\frac{1}{2}\hbar $ was due to the spins of the three
spin-$\frac{1}{2}$ valence quarks, with two oriented in one
direction and one in the other.  In the late 1980's, however, the
EMC experiment at CERN \cite{Ashman:1987hv,Ashman:1989ig} discovered
that only approximately $13\pm 16\%$ of the proton's spin was due to
the spin of the quarks. This surprising result became known as the
"proton spin crisis." With so little of the proton's spin coming
from the total quark spin ($\Delta \Sigma$), the remainder is
expected to come from gluon spin contributions ($\Delta g$) and the
orbital angular momentum (OAM) of both quarks and gluons
($L_{g+q}$), as indicated in the spin sum rule given in
Eq.~\ref{eq:longitudinalSumRule}, which is valid in the infinite
momentum frame.

\begin{equation}\label{eq:longitudinalSumRule}
\frac{1}{2} = \frac{1}{2}\Delta \Sigma + \Delta g + L_{g+q}
\end{equation}

Experimental work following the EMC discovery, mostly exploiting
DIS, has continued to explore this problem for more than 15 years,
yet there remains much to be understood.  In particular, the
magnitude and even sign of the gluon spin contribution to the spin
of the proton remains to be determined, the flavor breakdown of the
sea quark spin contributions is largely unknown, and no definitive
experimental technique with which to access OAM directly has yet
been proposed.

\subsection{Polarized structure functions}

\begin{figure}
\centering
\includegraphics[height=0.8\textheight]{%
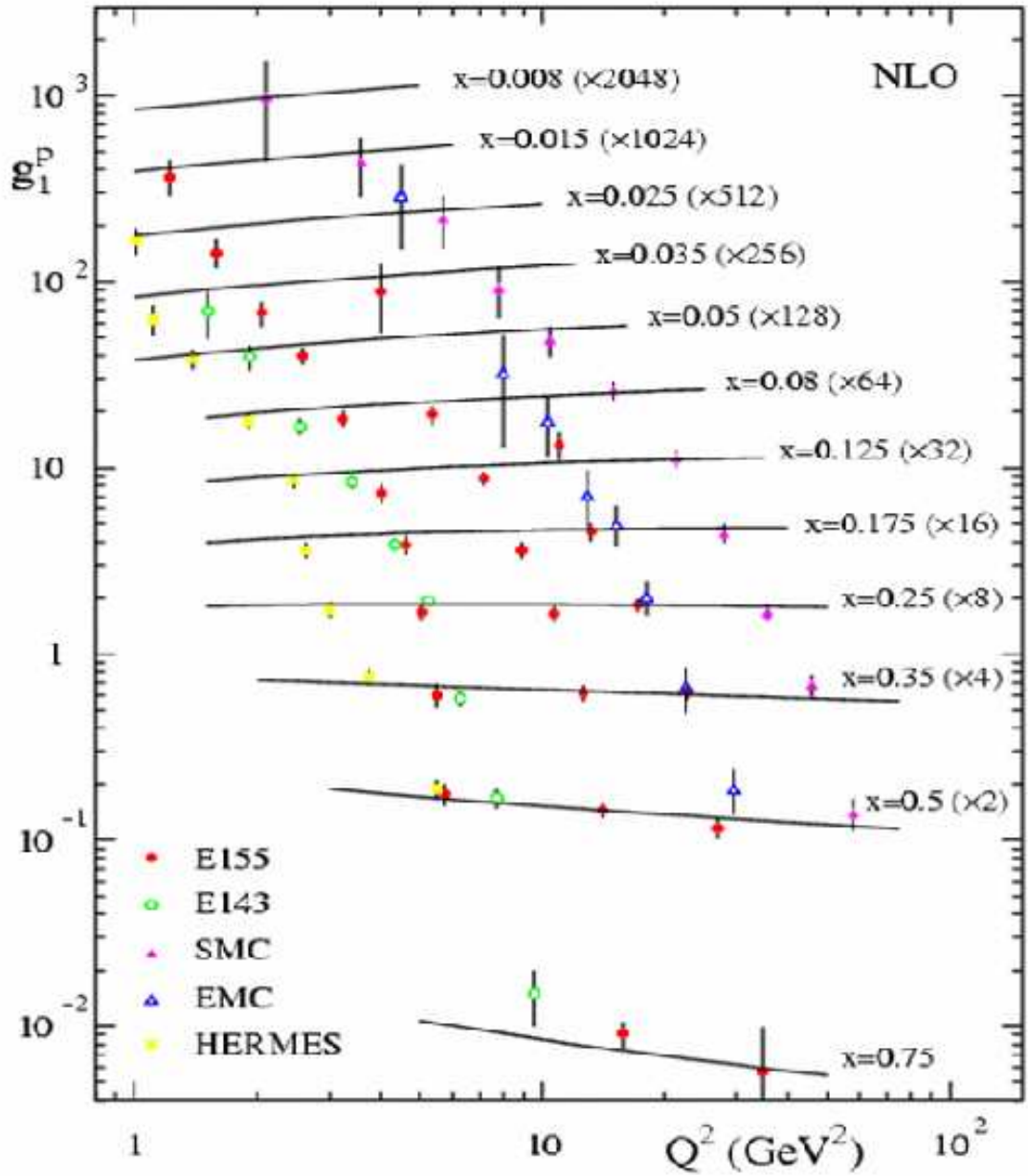} \caption[World DIS data on the polarized structure
function of the proton, $g_1$.]{World DIS data on the polarized
structure function of the proton. } \label{figure:polStructure}
\end{figure}

Similar to the unpolarized structure functions discussed in Section
\ref{section:structureFunctions}, spin-dependent structure functions
can also be defined. The difference in cross sections for
deep-inelastic scattering of leptons polarized antiparallel and
parallel to the spin of the target proton can be written as in
Eq.~\ref{eq:polCrossSectionDifference},

\begin{equation}\label{eq:polCrossSectionDifference}
\frac{d^2\sigma^{+-}}{dQ^2d\nu} - \frac{d^2\sigma^{++}}{dQ^2d\nu} =
\frac{4\pi \alpha^2}{E^2Q^2}[M(E + E'\cos \theta) G_1(\nu,Q^2) - Q^2
G_2(\nu,Q^2)]
\end{equation}
in which the kinematic variables are defined as in Equations
\ref{eq:elasticCrossSection} and \ref{eq:DISCrossSection} and $G_1$
and $G_2$ represent polarized structure functions of the proton.  In
the Bjorken scaling limit of large $Q^2$ and $\nu$, these structure
functions depend only on $x$ and can be given as in
Eq.~\ref{eq:polStructureFunctions}.

\begin{eqnarray}\label{eq:polStructureFunctions}
\nonumber
g_1(x) &=& M^2\nu G_1(\nu, Q^2) \\
g_2(x) &=& M\nu^2 G_2(\nu, Q^2)
\end{eqnarray}
$g_1(x)$ can be viewed as the difference in probability of
scattering off of a parton carrying momentum fraction $x$ of the
proton with parton helicity antiparallel versus parallel to the
proton spin. Figure~\ref{figure:polStructure} shows the world DIS
data for $g_1$ as a function of $Q^2$; the scaling behavior can be
seen over most of the kinematic range that has been measured, which
is limited compared to the kinematic range over which $F_2$ has been
measured (refer back to Figure~\ref{figure:unpolStructure}).

\subsection{Polarized parton distribution functions}
\label{section:polpdf}

Polarized pdf's, or helicity distribution functions, denoted $\Delta
f$, represent the \emph{difference} in probability of scattering off
of a parton $f$ with its spin vector parallel versus antiparallel to
the proton's spin, in the case of longitudinal polarization,
i.e.~polarization along the direction of proton motion. The
transverse spin structure of the proton is discussed separately in
Section~\ref{section:transverseStructure}. Helicity distributions
can be obtained for example from global fits to DIS measurements of
$g_1(x)$ for the proton and neutron.  It is well accepted that
polarized pdf's can be used as input to calculations in pQCD in a
similar fashion to unpolarized pdf's. Arguments for collinear
factorization involving \emph{spin-dependent} processes are given in
\cite{Collins:1992xw}.

\subsection{Current status of the longitudinally pol. structure
of the proton}

There are multiple ongoing experiments making measurements to study
the polarized structure of the nucleon.  The HERMES experiment at
DESY makes use of the longitudinally polarized electron (or
positron) beam at the HERA collider and performs DIS measurements on
nucleon targets which can be longitudinally or transversely
polarized.  The COMPASS experiment at CERN also performs spin
measurements through DIS, using a polarized muon beam from pion
decays on a polarized fixed target.  The PHENIX and STAR experiments
at RHIC study polarized proton-proton collisions and have the
ability to choose either longitudinal or transverse beam
polarization, while the BRAHMS experiment at RHIC can only study
transverse spin physics (see Section~\ref{section:collider}).

In 2004 HERMES published its final longitudinal results, the first
five-flavor fit for quark polarizations measured in DIS
\cite{Airapetian:2004zf}. Their results yield $\Delta u/u > 0$ and
increasing with $x$, $\Delta d/d < 0$, and polarizations consistent
with zero for $\bar{u}$, $\bar{d}$, and $s$ quarks.  In contrast to
the unpolarized case, the polarized light-quark sea does not appear
to have a significant flavor asymmetry.

\begin{figure}
\centering
\includegraphics[height=0.6\textheight]{%
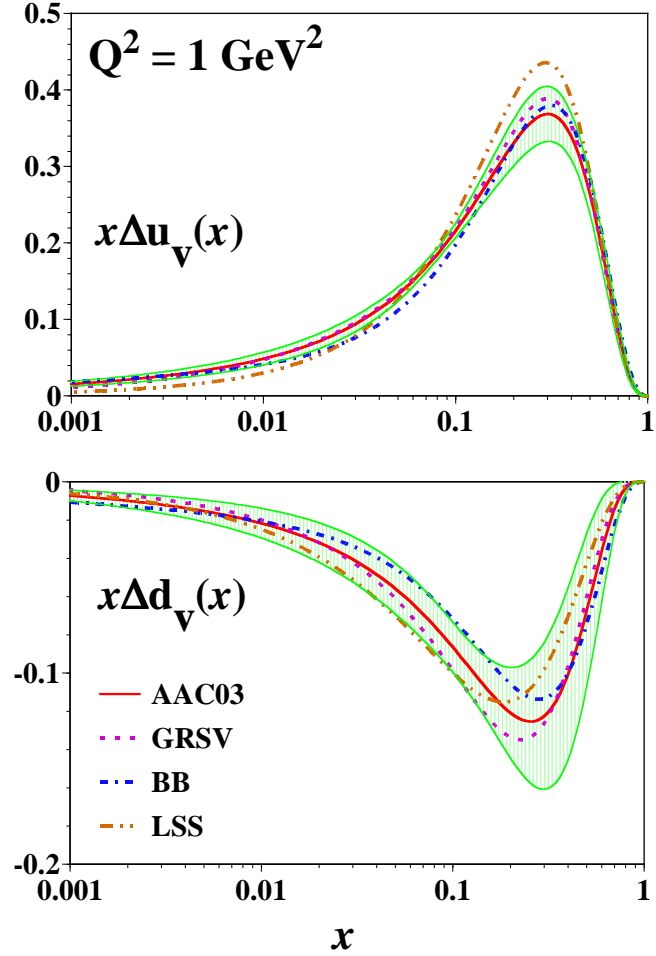} \caption[Global fits to world data for the
polarized pdf's of valence quarks.]{Global fits to world data for
the polarized pdf's of valence quarks, taken from
\cite{Hirai:2003pm}.} \label{figure:polGlobalFitsValence}
\end{figure}

\begin{figure}
\centering
\includegraphics[height=0.6\textheight]{%
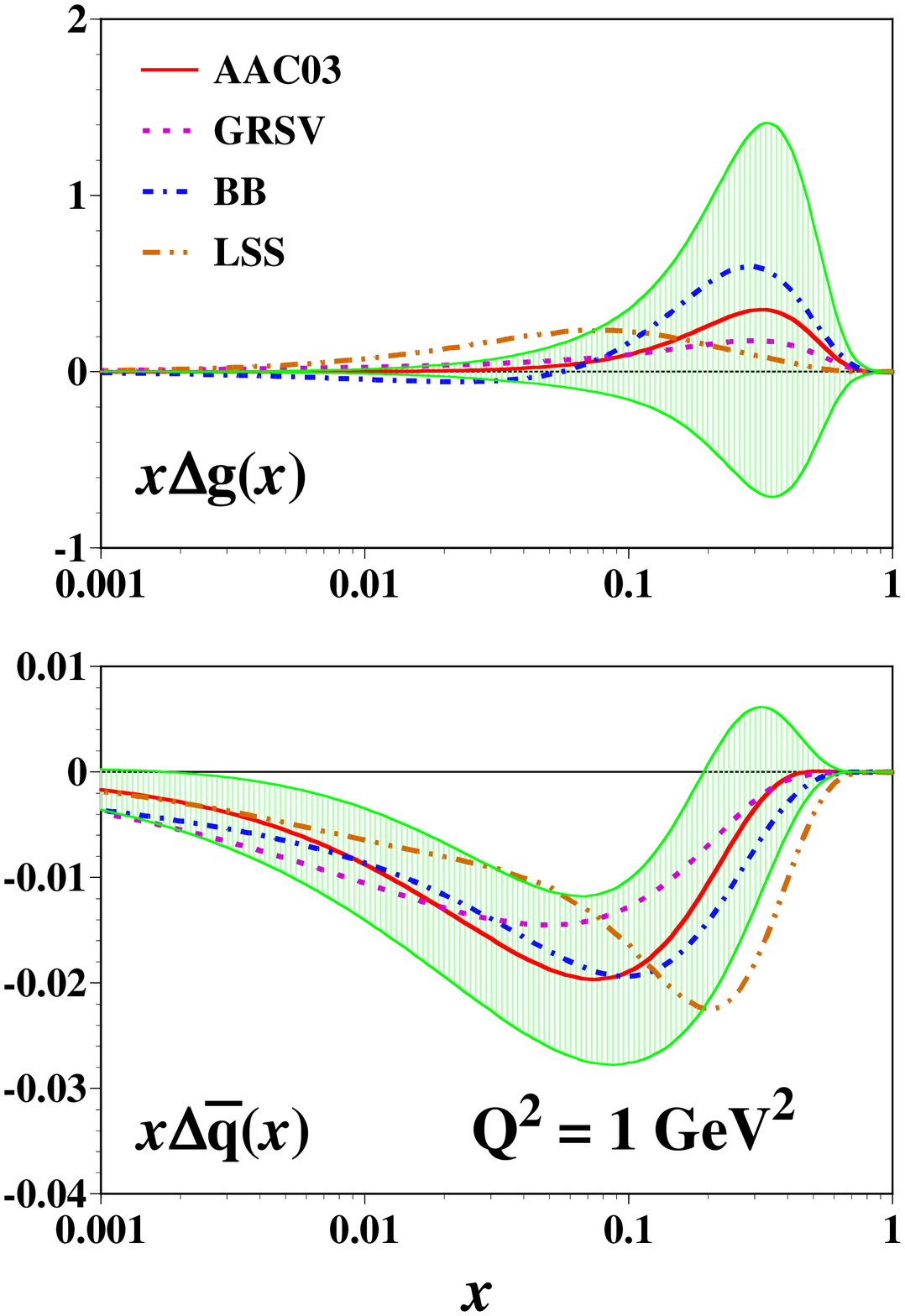} \caption[Global fits to world data for
the pol. pdf's of gluons and sea quarks.]{Global fits to world data
for the polarized pdf's of gluons and sea quarks, taken from
\cite{Hirai:2003pm}.} \label{figure:polGlobalFitsSea}
\end{figure}

Various different groups have performed global fits to the world
data available from DIS experiments on the helicity structure of the
proton \cite{Blumlein:2002be,Gluck:2000dy,Hirai:2003pm}.
Figure~\ref{figure:polGlobalFitsValence} depicts the results of
several such fits for valence quarks;
Figure~\ref{figure:polGlobalFitsSea} for gluons and sea quarks. Both
figures are taken from \cite{Hirai:2003pm}; see the reference for
details regarding the different curves. The best constraints are
naturally available for valence quarks. As can be seen, the
polarization for valence up quarks is significant and positive,
while for valence down quarks it is significant and negative,
consistent with the HERMES results mentioned above, which were not
entirely available at the time of these fits. A global analysis of
the sea quarks yields a negative polarization of smaller magnitude
than for valence down quarks and with a larger relative uncertainty.

As mentioned above for the unpolarized case, DIS experiments cannot
access the gluon directly because it does not couple
electromagnetically.  Consequently, $\Delta g$ can only be inferred
through scaling violations of the $g_1$ polarized structure
function, as given in Eq.~\ref{eq:polGluonDist},

\begin{equation}\label{eq:polGluonDist}
\Delta g \propto dg_1/d(\ln Q^2)
\end{equation}
or via di-hadron or heavy flavor production, both of which provide
some sensitivity to the gluon. The magnitude of $\Delta g$ remains
almost completely unknown, and its sign is not yet clear, as can be
seen in Figure~\ref{figure:polGlobalFitsSea}.  A recent
comprehensive review of the longitudinal spin structure of the
proton can be found in \cite{Bass:2004xa}.

\subsubsection{Recent results from RHIC}

\begin{figure}
\centering
\includegraphics[height=0.6\textheight]{%
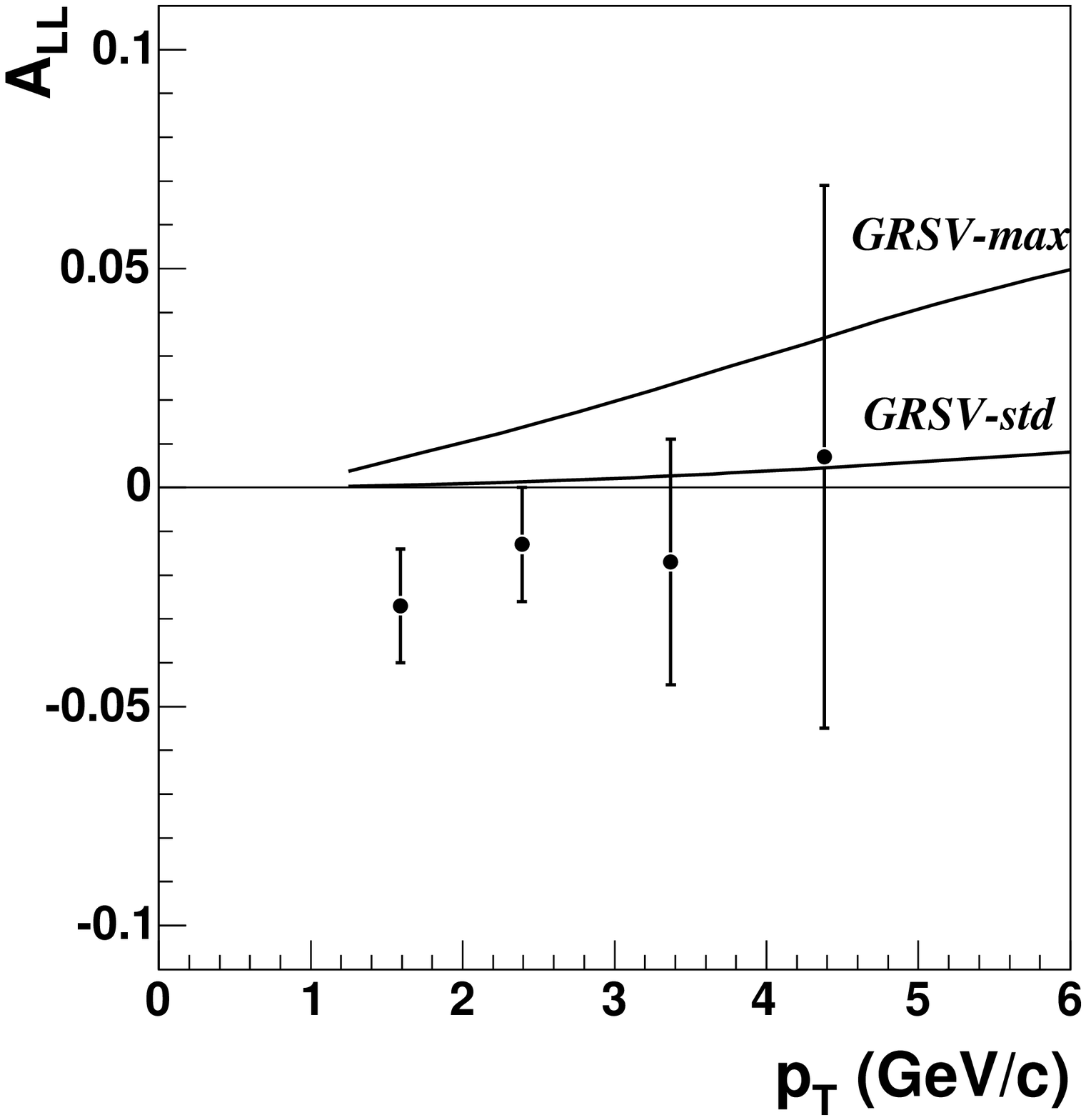} \caption[$A_{LL}$ for neutral pions]{Longitudinal
double-spin asymmetry for neutral pion production at PHENIX,
compared to predicted asymmetries assuming various values of $\Delta
g$ (see text), taken from \cite{Adler:2004ps}. }
\label{figure:pubPi0DoubleAsym}
\end{figure}

From the 2003 polarized proton run at RHIC, a measurement of the
longitudinal double-spin asymmetry ($A_{LL}$) of neutral pions at
mid-rapidity has been made by PHENIX \cite{Adler:2004ps} for a
transverse-momentum ($p_T$) range of 1 to 5~GeV/$c$. Pion production
in this kinematic region is due mainly to gluon-gluon and
gluon-quark scattering; thus, this measurement is sensitive to the
polarized gluon distribution.  (See
Figure~\ref{figure:partonicProcesses} in
Chapter~\ref{section:conclusions} indicating the relative
contributions of different partonic scattering processes to $\pi^0$
production as a function of $p_T$.) The asymmetry is shown in
Figure~\ref{figure:pubPi0DoubleAsym}.  The theoretical curves in the
figure represent next-to-leading-order (NLO) pQCD calculations using
two different assumptions for $\Delta g$
\cite{Gluck:2000dy,Jager:2002xm}. GRSV-std takes $\Delta g$ as the
value that best fits the world DIS data, and GRSV-max takes $\Delta
g$ to be equal to the unpolarized gluon distribution at a scale of
$Q^2 = 0.6~\textrm{GeV}^2$.

\begin{figure}
\centering
\includegraphics[height=0.3\textheight]{%
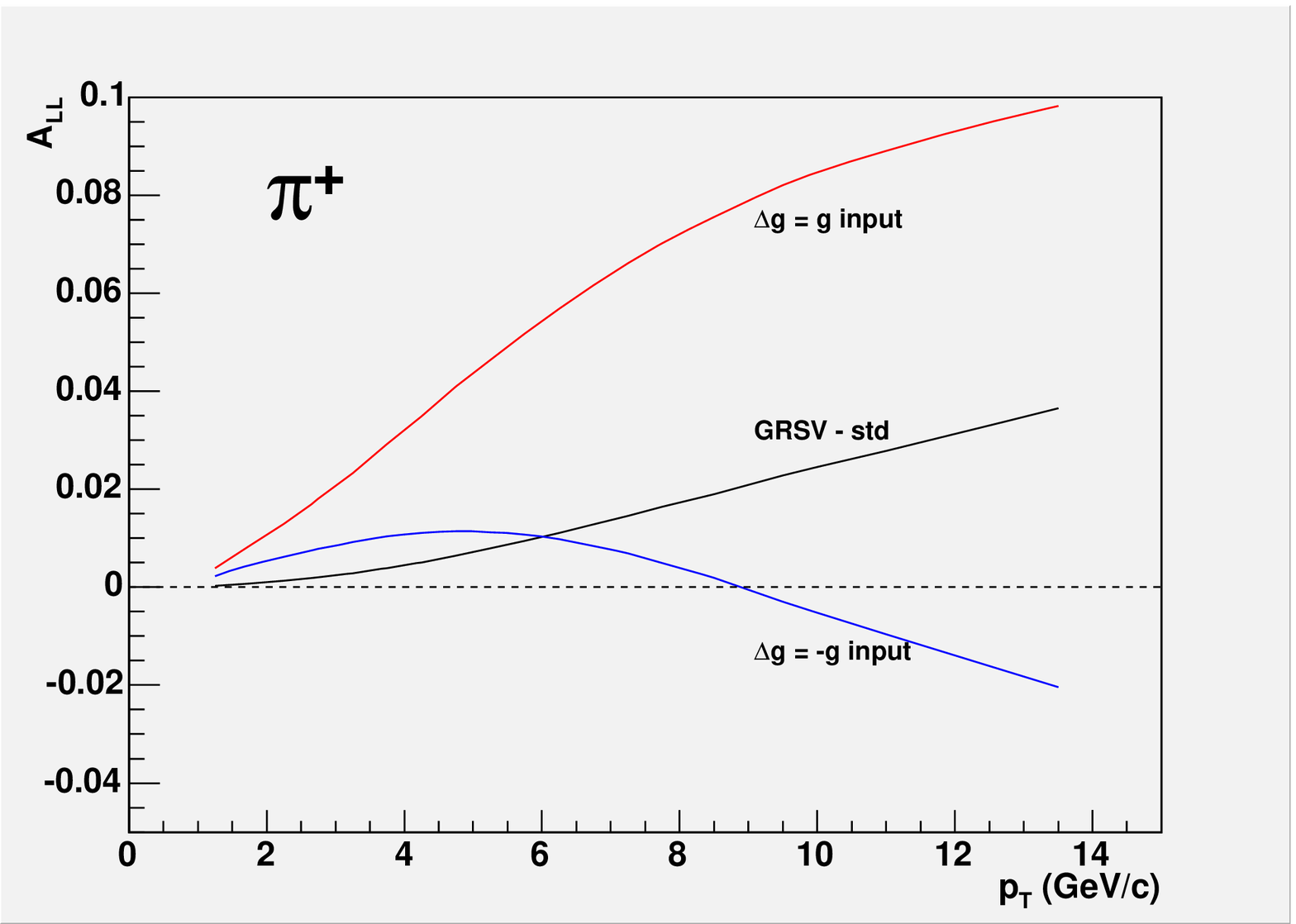} \caption[Predicted $A_{LL}$ of positive pions
for various $\Delta g$ assumptions]{ Predicted $A_{LL}$ of positive
pions for various $\Delta g$ assumptions, from M. Stratmann.}
\label{figure:piPlusTheory}
\end{figure}

\begin{figure}
\centering
\includegraphics[height=0.3\textheight]{%
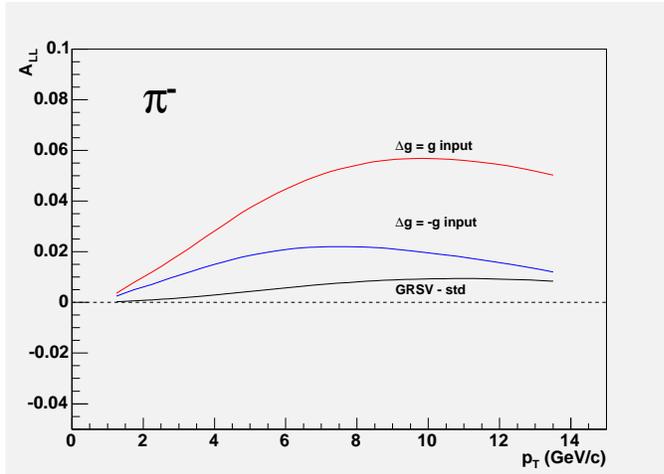} \caption[Predicted $A_{LL}$ of negative pions
for various $\Delta g$ assumptions]{ Predicted $A_{LL}$ of positive
pions for various $\Delta g$ assumptions, from M. Stratmann.}
\label{figure:piMinusTheory}
\end{figure}

Due to the significant contribution of gluon-gluon scattering and
the isospin symmetry of the neutral pion, its double-spin asymmetry
is largely insensitive to the sign of the polarized gluon
distribution.  Future results for $A_{LL}$ of positive and negative
pions for $p_T \gtrsim 5$~GeV/$c$ will provide independent
measurements of $\Delta g$ and allow determination of its sign. See
Figures \ref{figure:piPlusTheory} and \ref{figure:piMinusTheory} for
expected charged pion asymmetries as a function of $p_T$ for
different-sign polarized gluon distributions, calculated by M.
Stratmann. PHENIX expects to be able to make a significant
measurement of charged pions in the next long polarized proton run
at RHIC, expected to occur in 2006 or 2007.  $A_{LL}$ of direct
photon production, measurable on a slightly longer time scale, will
provide a clean measurement of both the magnitude and sign of
$\Delta g$, as discussed in Section~\ref{section:asymInFactQCD}.

\subsection{Understanding the transverse spin structure of the proton}
\label{section:transverseStructure}

\subsubsection{Longitudinal versus transverse spin structure of the proton}

The transverse spin structure of the proton cannot be determined
from its longitudinal spin structure.  A simple explanation for this
fact is the non-commutation of Lorentz boosts and spatial rotations.
Differences between the transverse and longitudinal polarization
structure of the nucleon provide insight into the relativistic
nature of partons bound within the nucleon.  A concise description
of transverse spin structure functions and their relation to the
longitudinal structure of the proton can be found in
\cite{Jaffe:1991kp}.

Inelastic and elastic scattering can be related by the optical
theorem.  Thus inelastic proton-proton or quark-proton scattering
can be considered in terms of elastic quark-proton scattering.  In
the elastic scattering of two spin-$\frac{1}{2}$ particles, there
are three different possibilities for the initial- and final-state
helicities.  The particles may start and end with the same helicity
($++ \rightarrow ++$), start and end with opposite helicities ($+-
\rightarrow +-$), or start with opposite helicities and change
helicities in the scattering ($+- \rightarrow -+$).  Linear
combinations of these three helicity configurations in the
scattering can be formed, corresponding to the momentum ($q$),
helicity ($\Delta q$), and transversity ($\delta q$) distribution
functions given in Eq.~\ref{eq:helicitiesAndPdfs}.

\begin{eqnarray}\label{eq:helicitiesAndPdfs}
\nonumber
  q &:& (++ \rightarrow ++) + (+- \rightarrow +-) \\
\nonumber
  \Delta q &:& (++ \rightarrow ++) - (+- \rightarrow +-) \\
  \delta q &:& (+- \rightarrow -+)
\end{eqnarray}

Transversity is therefore a chiral-odd, or "helicity-flip,"
distribution. In a transverse basis, it represents the difference in
probability of scattering off a quark with transverse spin parallel
versus antiparallel to a transversely polarized proton. This
interpretation is directly analogous to the meaning of helicity
distributions in a helicity basis.  The transversity distribution
was first discussed in \cite{Ralston:1979ys}. The Soffer bound,
given by Eq.~\ref{eq:sofferBound}, relates the transversity,
helicity, and momentum distributions of the nucleon
\cite{Soffer:1994ww}.

\begin{equation}\label{eq:sofferBound}
|2\delta q(x)| \leq q(x) + \Delta q(x)
\end{equation}

Analogous to the longitudinal spin sum rule
(Eq.~\ref{eq:longitudinalSumRule}), there is also a transverse spin
sum rule \cite{Bakker:2004ib}, given in
Eq.~\ref{eq:transverseSumRule}, in which $L_{S_T}$ is the transverse
component of the partonic orbital angular momentum.

\begin{equation}\label{eq:transverseSumRule}
\frac{1}{2} = \frac{1}{2}\sum_{a=q,\bar{q}}\int dx\delta q_a(x,Q^2)
+ \sum_{a=q,\bar{q},g} \langle L_{S_T} \rangle_a(Q^2)
\end{equation}
Note the absence of a gluon spin contribution; there is no
transversity distribution for gluons at leading twist because there
is no mechanism to flip the helicity of (spin-1) gluons in the
scattering.

While transversity is a non-perturbative object as any other pdf,
the nucleon tensor charge, related to transversity ($\int_0^1
dx(\delta q(x) - \delta \bar{q}(x))$), can be calculated in lattice
QCD. For a description of recent work, see \cite{Orginos:2005uy}.

In pQCD, chiral-odd functions must appear in pairs because hard
scattering processes conserve helicity.  One possibility is to look
for observables that represent the convolution of two transversity
distributions, i.e.~double transverse-spin asymmetries, $A_{TT}$.
Another possibility is to convolute transversity with a chiral-odd
fragmentation function, one example of which will be discussed
below.

\subsubsection{Observation of large transverse single-spin asymmetries}
The measurement of transverse single-spin asymmetries (SSA's), for
example in proton-proton collisions or DIS, represents one way of
probing the quark and gluon structure of transversely polarized
nucleons and is the approach exploited for the measurement in this
thesis. Interest in these measurements is heightened by the large
transverse SSA's observed in spin-dependent proton-proton scattering
experiments spanning a wide range of energies.  The experimental
observation of large asymmetries, with the first measurements being
in the late 1970's, was initially a surprise. The leading-twist pQCD
expectation was that transverse SSA's should be suppressed as
$\frac{\alpha_s m_q}{\sqrt{s}}$, where $m_q$ is the quark mass
\cite{Kane:1978nd}.

Striking asymmetries were seen at a number of spin-dependent $p+p$
scattering experiments at energies ranging from $\sqrt{s} = 5 -
10$~GeV. Asymmetries approaching 30\% were observed in inclusive
pion production at large Feynman-$x$ ($x_F = 2p_L/\sqrt{s}$, where
$p_L$ is the component of particle momentum in the beam direction)
and $p_T$ up to 2~GeV/$c$ \cite{Dragoset:1978gg,Allgower:2002qi}. In
a different kinematic region at mid-rapidity and large $x_T =
2p_T/\sqrt{s}$, asymmetries were also observed in inclusive $\pi^0$
and $\pi^+$ production but not in $\pi^-$ production
\cite{Antille:1980th,Saroff:1989gn,Apokin:1990ik}. At a higher
center-of-mass energy of 19.4~GeV where pQCD may be applicable,
asymmetries at large $x_F$ persisted
\cite{Adams:1991rw,Adams:1991cs}; however, the asymmetry in $\pi^0$
production at mid-rapidity at this energy was found to be zero up to
$p_T$ of 4~GeV/$c$ \cite{Adams:1994yu}. Non-zero transverse
single-spin asymmetries were also observed in semi-inclusive DIS
\cite{Airapetian:1999tv,Airapetian:2001iy,Bravar:2000ti}.

\begin{figure}
\centering
\includegraphics[height=0.6\textheight]{%
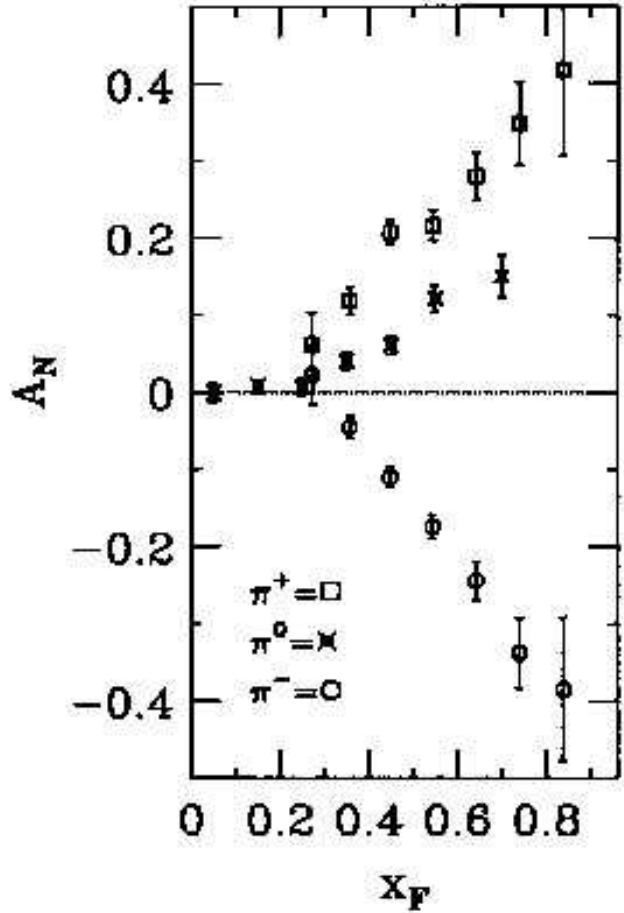} \caption[$A_N$ of high-$x_F$ neutral and charged pions
at $\sqrt{s} = 19.4$~GeV.]{Transverse SSA of high-$x_F$ neutral and
charged pions at $\sqrt{s} = 19.4$~GeV, taken from
\cite{Adams:1991cs}. } \label{figure:E704}
\end{figure}

The results for neutral and charged pions at high $x_F$ from the
E704 experiment at Fermilab \cite{Adams:1991cs} are shown in
Figure~\ref{figure:E704}.  The observed asymmetries are strikingly
large, reaching a magnitude of $\sim 40\%$ for charged pions at $x_F
\approx 0.8$.  There is a clear sign dependence of the asymmetry on
the pion charge, with $A_N^{\pi^+} > 0$ and $A_N^{\pi^-} < 0$ and
both of approximately equal magnitude and exhibiting the same
dependence on $x_F$.  The $\pi^0$ asymmetry is also positive but
with smaller magnitude.

Generally, transverse SSA's can be described by spin-momentum
correlations having the form $\overrightarrow{S} \cdot
(\overrightarrow{p_1} \times \overrightarrow{p_2})$, in which there
are different possibilities for the spin and momentum three-vectors.
Three different mechanisms, originating from different spin-momentum
correlations, have been studied extensively as the possible origin
of transverse SSA's in high-energy hadron collisions.
\begin{enumerate}
  \item Transversity distributions can give rise to SSA's in combination
with spin-dependent, chiral-odd FF's, which serve as analyzers for
the transverse spin of the struck quark.  The Collins function
\cite{Collins:1992kk} is an example of such a FF.
  \item Quark and gluon distributions that are asymmetric in the
intrinsic transverse momentum of the parton within the proton,
$k_T$, can lead to SSA's. This idea was first suggested by Sivers
\cite{Sivers:1989cc}.  The Sivers pdf can exist both for quarks and
gluons, and a relation to orbital angular momentum of partons in the
nucleon has been suggested \cite{Sivers:1989cc,Burkardt:2003je}.
  \item Interference between quark and gluon fields in the initial or final
state can generate SSA's \cite{Qiu:1998ia,Kanazawa:2000hz}.
\end{enumerate}

\subsubsection{Non-collinear pdf's and FF's}
The polarized pdf's and FF's relevant to collinear factorization
given in Section~\ref{section:polpdf} are integrated over all
possible values of $k_T$.  In a pdf, $k_T$ represents the transverse
momentum of the parton within the proton; in a FF, it indicates the
transverse momentum of the fragmenting hadron with respect to the
scattered quark, or jet axis. Transverse-momentum-dependent (TMD)
pdf's relate naturally to the orbital angular momentum of partons
within a proton; however, a precise understanding of this relation
remains unclear. If one does not assume collinearity but rather
takes $k_T$-dependent pdf's and FF's, a rich variety of new
possibilities emerges.

\begin{figure}
\centering
\includegraphics[height=0.4\textheight]{%
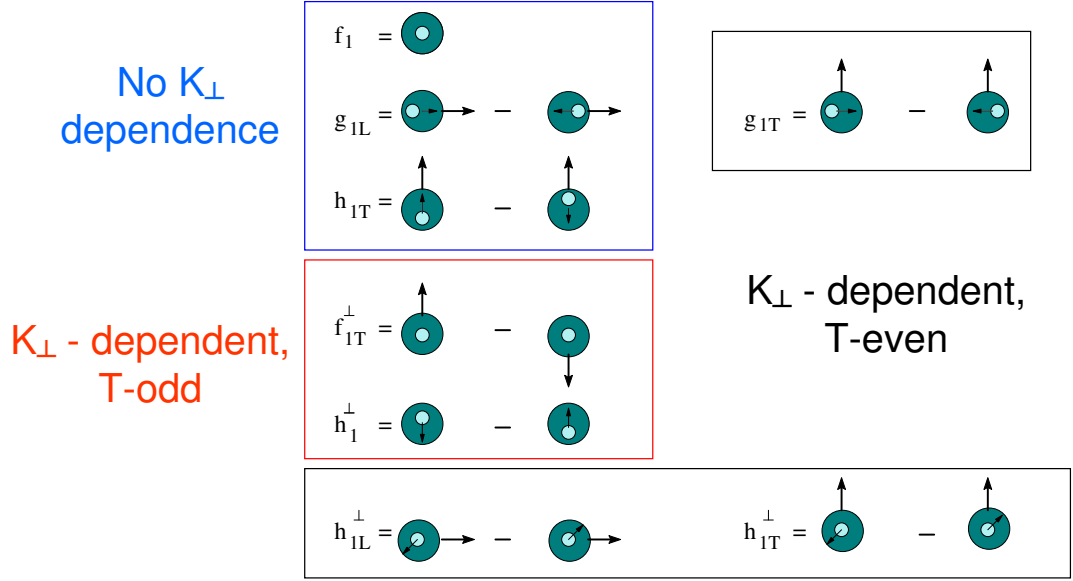} \caption[Collinear and $k_T$-dependent leading-twist
pdf's.]{The eight leading-twist quark distribution functions,
including collinear as well as $k_T$-dependent distributions. }
\label{figure:pdfs}
\end{figure}

There are a total of eight leading-twist (twist-two)
$k_T$-integrated and $k_T$-dependent quark distribution functions,
as shown in Figure~\ref{figure:pdfs}. Only three of these pdf's are
independent of $k_T$.  Parton distributions denoted by $f$ indicate
unpolarized quarks, $g$ longitudinally polarized quarks, and $h$
transversely polarized quarks.  The subscript 1 signifies leading
twist; the subscripts $L$ and $T$ denote longitudinal and transverse
proton polarization, respectively. The superscript $\bot$ indicates
explicit dependence on transverse momenta with a non-contracted
index, as described in \cite{Mulders:1995dh}.  Note that $h_{1T}$ is
the transversity distribution, an alternative notation for $\delta
q$. The field of transverse spin physics has been plagued for many
years by confusing and inconsistent notation in the literature for
relevant structure functions, pdf's, and FF's, as well as sign
conventions for azimuthal angles in semi-inclusive DIS.  One of the
outcomes of the Transversity 2004 workshop in Trento, Italy was an
examination and comparison of what existed in the literature and a
set of recommended notation and sign conventions
\cite{Bacchetta:2004jz}.

It should be pointed out that the factorization theorem has not been
proven generally for the case of non-collinear partons.  It has so
far only been proven for the Drell-Yan process ($q+\overline{q}
\rightarrow \ell^+ + \ell^-$) \cite{Collins:1984kg}, with notable
work also in $e^+ + e^- \rightarrow 2 h + X$ and semi-inclusive DIS
\cite{Collins:1981uk,Collins:1981va,Collins:1981uw}. $k_T$-dependent
factorization is therefore strictly speaking an assumption, albeit a
well-accepted one in the field.  Efforts are ongoing to establish
the theoretical basis more firmly. Recent work considering gauge
invariance in the cases of semi-inclusive DIS and Drell-Yan appears
in \cite{Ji:2004wu,Ji:2004xq,Collins:2004nx}.  Discussions of the
universality and the evolution of TMD pdf's can be found in
\cite{Collins:2002kn,Metz:2002iz,Boer:2003cm,Bomhof:2004aw} and
\cite{Boer:2001he,Henneman:2001ev,Kundu:2001pk,Idilbi:2004vb},
respectively.  In general, the role of $k_T$ in hard-scattering
processes is a vibrant area of research in pQCD.

\subsubsection{Sivers effect}

One of the pdf's that emerges if $k_T$ is left unintegrated is known
as the Sivers pdf, denoted as $f_{1T}^\bot$ and first proposed by
Sivers in 1989 \cite{Sivers:1989cc,Sivers:1990fh}. It represents the
spin-dependent asymmetry in the intrinsic $k_T$ of the (unpolarized)
partons in a transversely polarized proton.  The Sivers function was
for some time believed to be forbidden because it is time-reversal
odd (T-odd), but it was finally realized in 2002 that final-state
interactions via a soft gluon can create the necessary interference
of amplitudes for the Sivers function to exist
\cite{Brodsky:2002cx}.  The T-odd nature of the Sivers function is
now commonly referred to as "naive T-odd" to express the fact that
it is not in fact forbidden in QCD.

The Sivers function plays a central role in the phenomenological
Sivers effect, which has its origin in correlations of the form
$\overrightarrow{S} \cdot (\overrightarrow{P} \times
\overrightarrow{k_T})$, where $\overrightarrow{S}$ is the proton
spin, $\overrightarrow{P}$ the proton momentum, and
$\overrightarrow{k_T}$ the intrinsic transverse parton momentum in
the proton.  In a simplistic picture of the Sivers effect, the
transverse polarization of the proton can be viewed as originating
from the orbital angular momentum of the partons.  A spin-dependent
final-state azimuthal asymmetry is then generated by preferential
scattering off of the orbiting partons in the "front" or "back" of
the proton, with scattering off of the front of a proton with spin
up generating particle production preferentially to the left of the
polarized beam.  The mechanism responsible for scattering off of a
particular "side" of the proton is not entirely clear.  Additional
study of the Sivers effect has been performed by Burkardt as well as
Hwang \cite{Burkardt:2003uw,Burkardt:2003je}.

The absolute value of the Sivers function would provide a lower
bound on parton OAM.  The Sivers function was originally
investigated for quarks; that for gluons was first discussed only in
2003 \cite{Schmidt:2003wi}.  Asymmetry calculations based on the
Sivers effect to describe the E704 data and other results can be
found for example in \cite{Anselmino:1994tv,Anselmino:1998yz}.

\subsubsection{Collins effect}

As mentioned above, chiral-odd functions in pQCD must come in pairs
because helicity is conserved in hard scattering processes.
Therefore transversity, as a chiral-odd distribution, needs to be
convoluted with another chiral-odd function in order to be relevant
in physical processes.  For production of final-state hadrons, a
chiral-odd fragmentation function is one possibility.  Such a FF was
proposed in the early 1990's by Collins, Heppelmann, and Ladinsky
\cite{Collins:1992kk,Collins:1992xw,Collins:1993kq}.

The Collins FF, denoted $H_1^\bot$, represents the correlation
between the transverse polarization of the fragmenting quark and the
orientation of the hadron production plane, given by
$\overrightarrow{S} \cdot (\overrightarrow{k} \times
\overrightarrow{P}_h)$, in which $\overrightarrow{S}$ is the
transverse polarization of the scattered quark, $\overrightarrow{k}$
is its three-momentum, and $\overrightarrow{P}_h$ is the
three-momentum of the final-state hadron.

The Collins mechanism, a phenomenological mechanism incorporating
the Collins FF, describes a spin-dependent azimuthal asymmetry in
the distribution of hadrons within a jet. A relatively intuitive
model of the Collins mechanism for the production of pseudoscalar
mesons has been proposed by Artru \textit{et al.}
\cite{Artru:1995bh}. In this model, a transversely polarized quark
is scattered out of a transversely polarized proton, with the
probability of the direction of the scattered quark spin given by
the transversity distribution. In order to produce a pseudoscalar
(spin-0) meson such as a pion, the fragmenting quark must acquire an
oppositely polarized (anti-)quark from the vacuum.  If the
quark-antiquark pair from the vacuum is assumed to have a total spin
angular momentum of 1, conservation of angular momentum requires one
unit of orbital angular momentum in the opposite direction. This
orbital angular momentum of the (anti-)quark from the vacuum which
subsequently binds to the scattered quark then produces a preference
in azimuthal direction in the production of the final-state pion.

While it was originally believed that it was possible to explain the
large transverse SSA's observed entirely in terms of the Collins
effect (see for example \cite{Artru:1995bh}), recent work has
suggested that this might not be the case
\cite{Anselmino:2004ky,Ma:2004tr}.  As yet, no consensus has been
reached.

\subsubsection{Higher-twist effects}

It has been shown that higher-twist contributions and non-zero $k_T$
can produce the same effects in hard-scattering processes
\cite{Politzer:1974fr}.  As such, there have been studies of how
twist-three effects rather than TMD distributions can give rise to
the large SSA's observed.

Qiu and Sterman have examined higher-twist asymmetry contributions
in the initial state system, i.e.~interference between quark and
gluon fields in the polarized proton \cite{Qiu:1998ia}.  Similar
studies have been performed by Kanazawa and Koike for quark-gluon
interference in the final state, a parton fragmenting to a hadron
\cite{Kanazawa:2000hz}.  In the initial-state case, both chiral-even
and chiral-odd components are possible. It is believed that a
relation of the chiral-odd twist-three initial-state effect of
\cite{Qiu:1998ia} to the Sivers mechanism may exist, but this is not
completely understood.

\subsection{Recent experimental results in transverse spin physics}
\begin{figure}
\centering
\includegraphics[height=0.6\textheight]{%
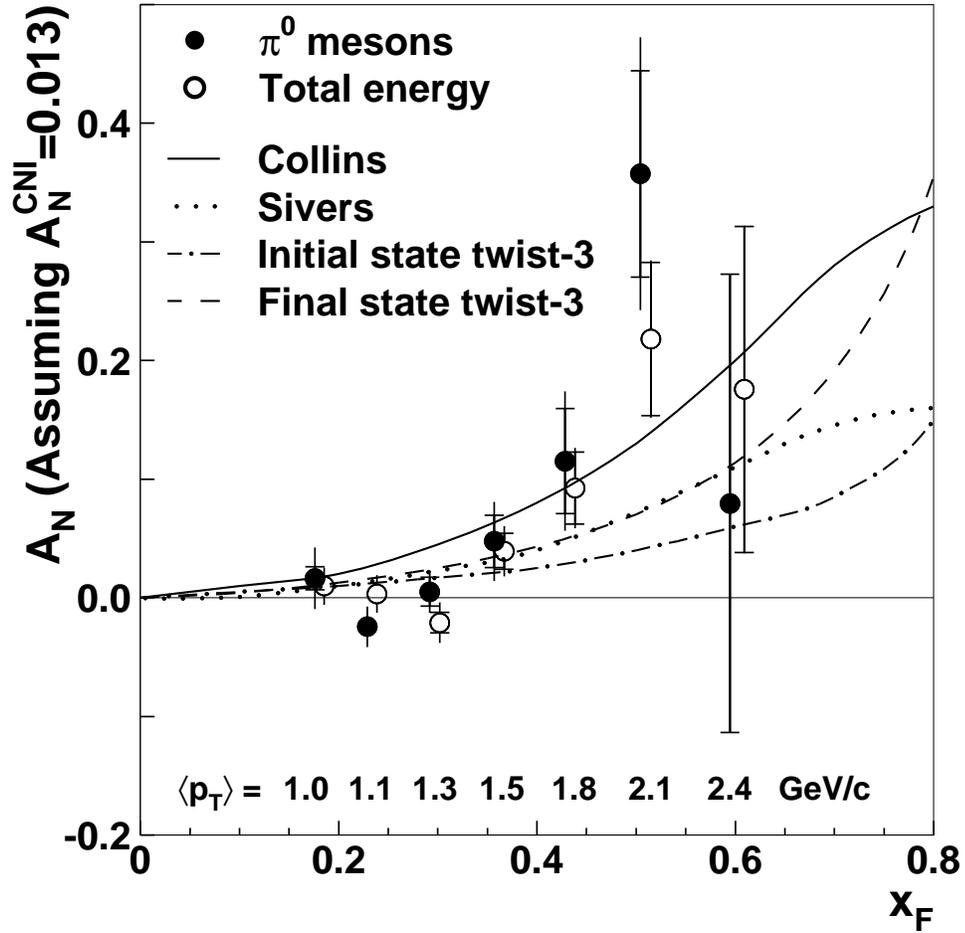} \caption[$A_N$ of high-$x_F$ neutral pions at
$\sqrt{s} = 200$~GeV.]{Transverse SSA of high-$x_F$ neutral pions at
$\sqrt{s} = 200$~GeV, taken from \cite{Adams:2003fx}. See text for
more details.} \label{figure:STARpi0Asym}
\end{figure}

RHIC data for polarized-proton collisions at $\sqrt{s} = 200$~GeV
first became available in late 2001.  In addition to the results
from the PHENIX experiment presented in this thesis, transverse spin
measurements have been made by the STAR and BRAHMS experiments at
RHIC.  At STAR it was discovered that large transverse SSA's persist
even at RHIC energies, an order of magnitude higher than the energy
for previous results \cite{Adams:2003fx}.  They found asymmetries of
up to $\sim 30\%$ in the forward production of neutral pions, as can
be seen in Figure~\ref{figure:STARpi0Asym}.  The theoretical curves
represent different fits to the E704 results discussed above
\cite{Adams:1991cs}, scaled in energy from 19.4 to 200~GeV.  The
general agreement of the scaled fits with the 200-GeV data suggests
that the asymmetries are generated by similar mechanisms at the two
energies; however, further study is needed before any definitive
conclusion can be made.  STAR also has preliminary asymmetry results
for the production of neutral pions in the backward direction with
respect to the polarized beam, for $-0.6 < x_F < -0.2$, which are
consistent with zero \cite{Morozov:2005tb}.

\begin{figure}
\centering
\includegraphics[height=0.4\textheight]{%
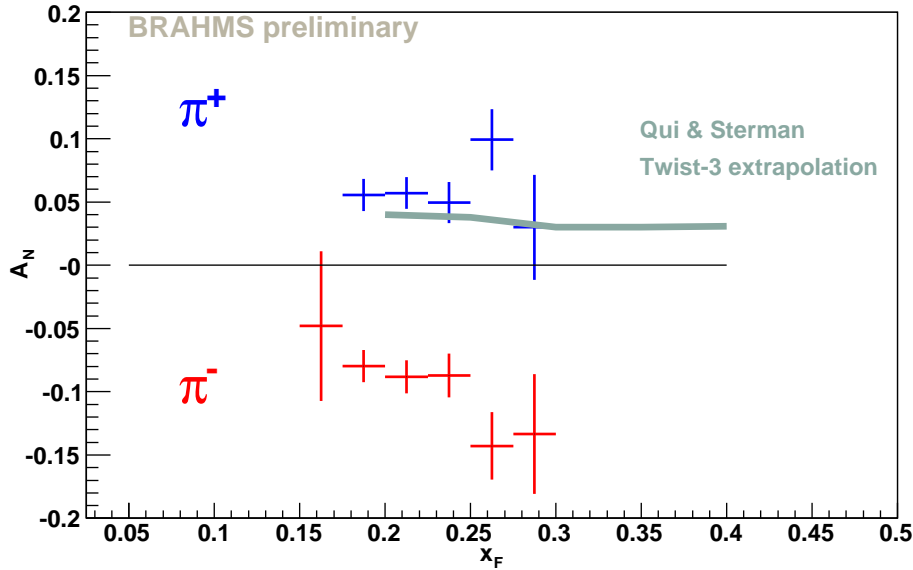} \caption[$A_N$ of moderate-$x_F$ charged pions
at $\sqrt{s} = 200$~GeV.]{Transverse SSA's of moderate-$x_F$ charged
pions at $\sqrt{s} = 200$~GeV, taken from \cite{Videbaek:2005fm}.
The theoretical curve is an extrapolation of initial-state
twist-three calculations by Qiu and Sterman.}
\label{figure:BRAHMSChargedPiAsym}
\end{figure}

BRAHMS has preliminary results for the transverse SSA's of charged
pions as well as protons \cite{Videbaek:2005fm}.  The charged pion
asymmetries are shown in Figure~\ref{figure:BRAHMSChargedPiAsym}.
The charge dependence of the sign of the asymmetry clearly follows
that observed by E704 \cite{Adams:1991cs}; however, the results are
for low transverse-momentum values ($p_T < 3$~GeV/$c$), so
pQCD-based interpretations such as the Sivers and Collins effects
may not be applicable. The results are in reasonable agreement with
extrapolations of initial-state twist-three calculations by Qiu and
Sterman.

The HERMES experiment at DESY has measured non-zero Collins and
Sivers moments via semi-inclusive deep-inelastic scattering of
positrons off of a hydrogen target \cite{Airapetian:2004tw}.  The
COMPASS experiment at CERN made a similar measurement with a muon
beam on a deuteron target and found results consistent with zero
\cite{Alexakhin:2005iw}.  The COMPASS observation is now understood
to be because of asymmetry cancelations due to the isospin symmetry
of the deuteron target at COMPASS.  For an interpretation of the
recent HERMES and COMPASS results and related predictions for
transverse SSA's at RHIC, see \cite{Vogelsang:2005cs}.

\begin{figure}
\centering
\includegraphics[height=0.5\textheight]{%
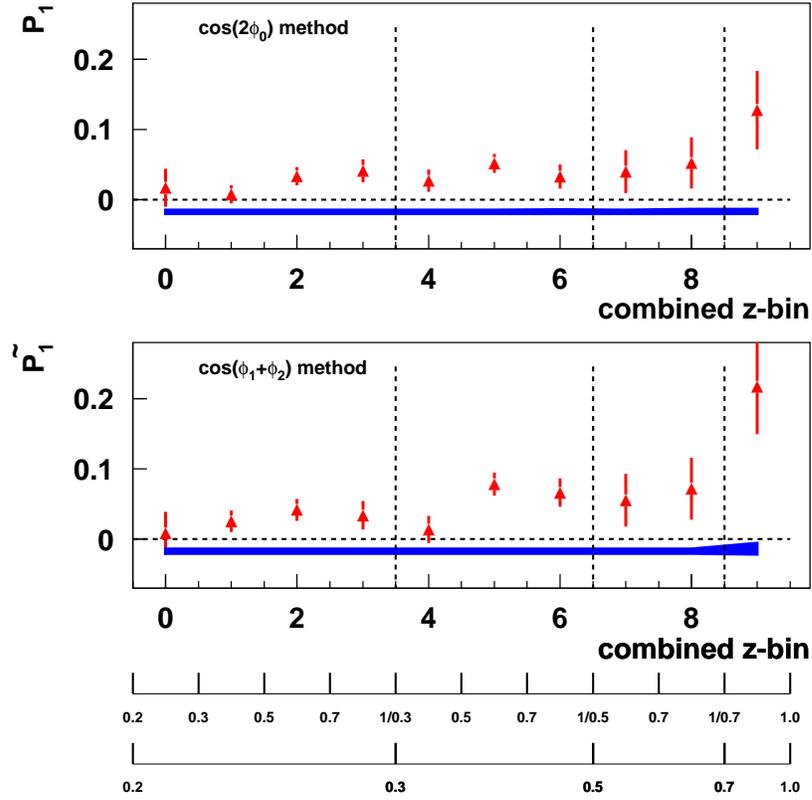} \caption[Measured analyzing power of the Collins
FF for pions.]{Analyzing power of the Collins FF for pions as a
function of $z$, measured by the BELLE experiment via two different
methods. See \cite{Abe:2005zx} for details.}
\label{figure:BELLECollinsFF}
\end{figure}

The Collins FF for pions was recently measured by the BELLE $e^+ +
e^-$ annihilation experiment at KEK \cite{Abe:2005zx}.  The
analyzing power of the Collins FF was determined to be significantly
non-zero, as can be seen in Figure~\ref{figure:BELLECollinsFF}. This
important measurement will provide vital input for factorized
calculations, allowing pion production processes sensitive to the
Collins mechanism to put constraints on the transversity
distribution.

Despite much progress within both the experimental and theoretical
arenas in recent years, no single, clear formalism has emerged in
which to interpret the notable transverse spin effects that have
been observed.  A number of mechanisms remain as possible origins
for the large transverse single-spin asymmetries, and a variety of
further experimental measurements in different kinematic regions and
sensitive to different partonic processes will be necessary in order
to disentangle them.

\chapter{QCD at RHIC}
\label{section:QCDatRHIC}

RHIC was designed to study numerous aspects of QCD. The most
flexible hadron collider in the world, it has so far produced
gold-gold, copper-copper, deuteron-gold, and polarized proton-proton
collisions at a variety of energies. Such a machine provides a rich
environment for the study of QCD. The goal of colliding heavy ions
at high energies is to create nuclear matter at extreme temperatures
and densities, high enough that the quarks and gluons are
(momentarily) not bound as hadrons but may co-exist rather as a
quark-gluon plasma (QGP). Variations in collision species size and
collision energy provide information on how the properties of the
created matter are related to the initial conditions. Studies of
small systems colliding with large nuclei (e.g.~deuteron-gold)
permit distinction between cold and hot nuclear effects.  For a
summary and review of what has been learned in the first few years
of the RHIC heavy-ion program, see the evaluations from the four
RHIC experiments published after the fourth year of running
\cite{Arsene:2004fa,Back:2004je,Adams:2005dq,Adcox:2004mh}.

The nucleon structure program at RHIC, with unique access to
high-energy polarized-proton collisions, seeks to measure the
helicity distributions of the partons within the proton, in
particular gluon and sea-quark distributions, and to improve
knowledge of the transverse spin structure of the proton.  Through
$W$ boson production in eventual 500-GeV running, it will not only
be possible and of interest to measure the flavor-separated helicity
distributions of the sea quarks ($\Delta \bar{u}(x)$, $\Delta
\bar{d}(x)$), but also the flavor-separated unpolarized pdf's
($\bar{u}(x)$, $\bar{d}(x)$). As stated in
Section~\ref{section:unpolStatus}, a large difference in the
$\bar{u}$ and $\bar{d}$ content of the (unpolarized) proton has been
observed and is still not well understood.

\section{Cross section measurements and NLO pQCD}
The goal of this thesis is to demonstrate how proton structure can
be investigated via proton-proton collisions at RHIC. It is
essential to understand how well factorized pQCD can be used to
describe and interpret the RHIC data.

\begin{figure}
\centering
\includegraphics[height=0.6\textheight]{%
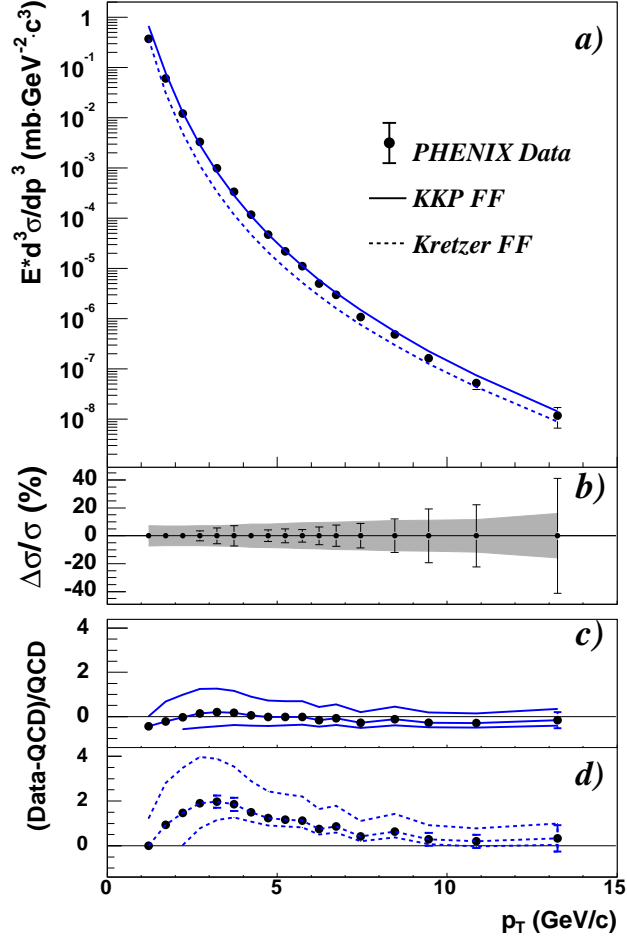} \caption[Mid-rapidity $\pi^0$ cross section
compared to NLO pQCD]{Invariant cross section versus transverse
momentum for mid-rapidity neutral pion production at PHENIX, taken
from \cite{Adler:2003pb}.  The data are compared to NLO pQCD
calculations using two different gluon-to-pion fragmentation
functions.  See text for further details. }
\label{figure:pi0CrossSection}
\end{figure}

\begin{figure}
\centering
\includegraphics[height=0.6\textheight]{%
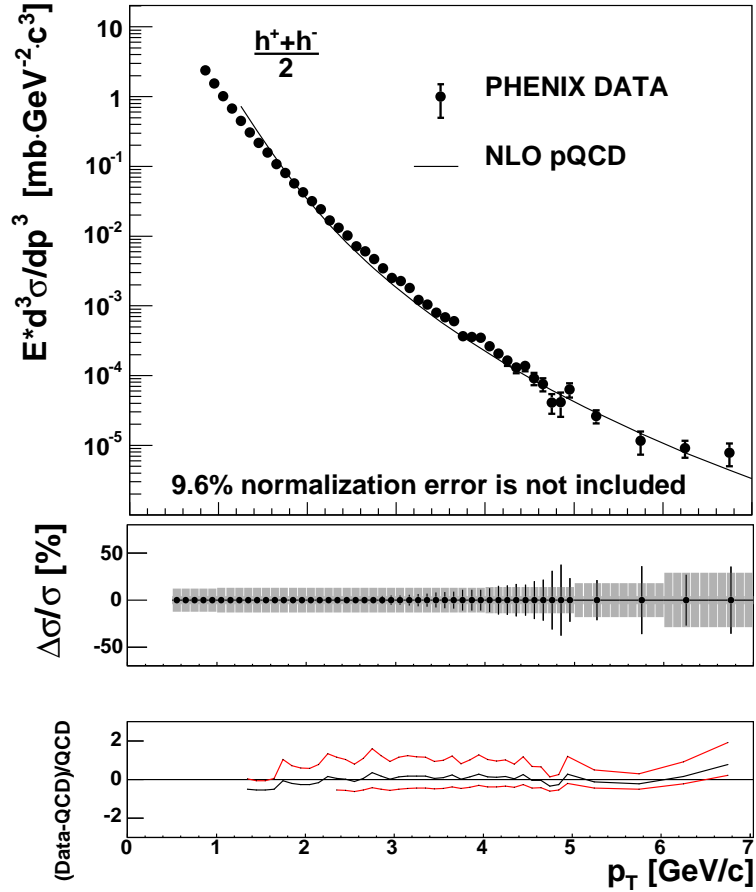} \caption[Mid-rapidity charged hadron cross
section compared to NLO pQCD]{Invariant cross section versus
transverse momentum for mid-rapidity charged hadron production at
PHENIX, compared to NLO pQCD, taken from \cite{Adler:2005in}.  See
text for further details. } \label{figure:chargedCrossSection}
\end{figure}

\begin{figure}
\centering
\includegraphics[height=0.6\textheight]{%
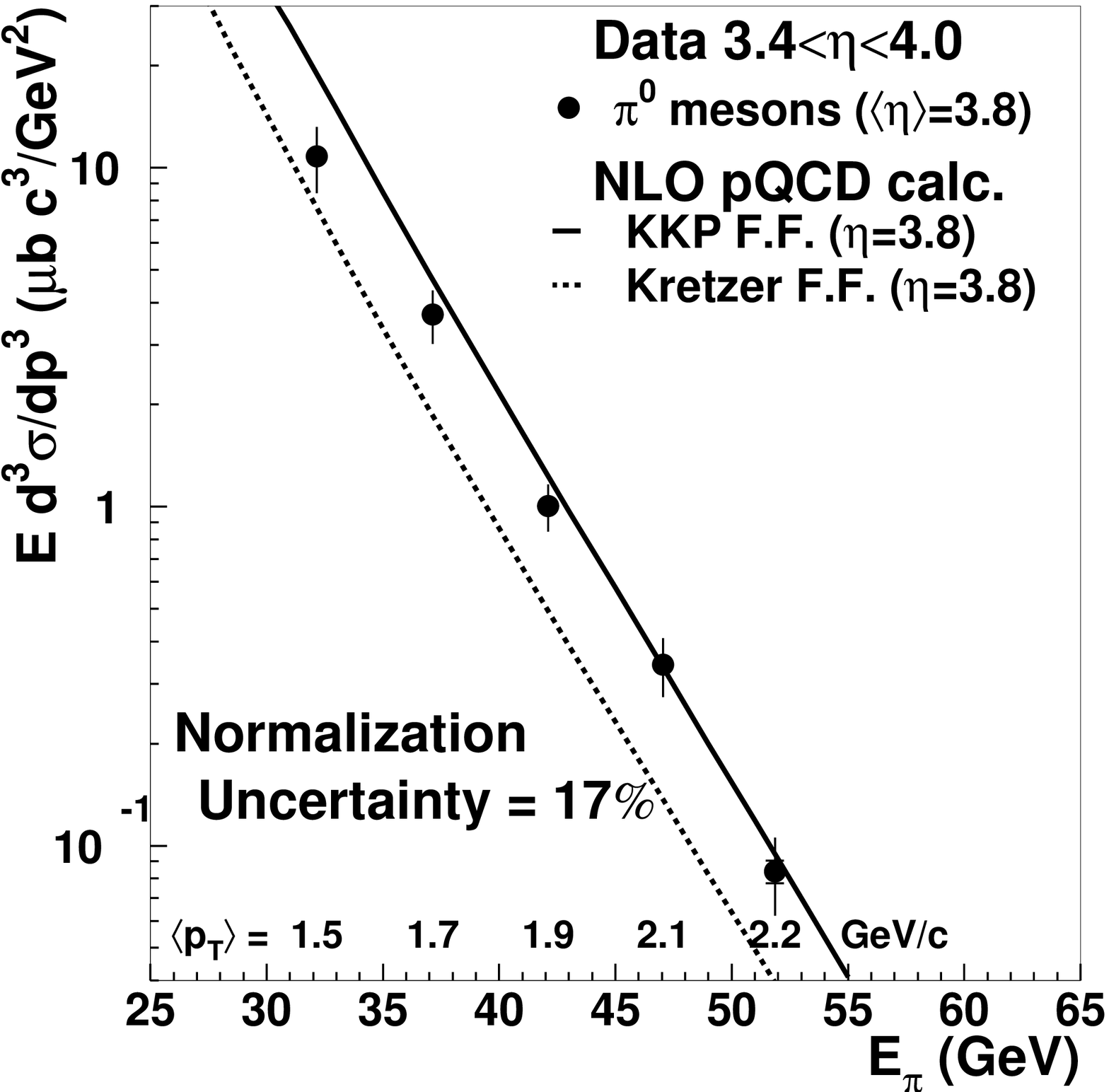} \caption[Forward $\pi^0$ cross section
compared to NLO pQCD]{Invariant cross section versus transverse
momentum for forward neutral pion production at STAR, compared to
NLO pQCD calculations using two different fragmentation functions,
taken from \cite{Adams:2003fx}. See text for further details.}
\label{figure:STARpi0CrossSection}
\end{figure}

A number of polarization-averaged cross section measurements have
been made at RHIC and compared to NLO pQCD calculations
\cite{Adler:2003pb,Adams:2003fx,Adler:2005in,Adler:2005qk}.
Comparisons of data to NLO pQCD calculations are shown in Figures
\ref{figure:pi0CrossSection}, \ref{figure:chargedCrossSection}, and
\ref{figure:STARpi0CrossSection}. The most spectacular example of
agreement between theory and data is seen in the mid-rapidity
neutral pion measurement (Fig.~\ref{figure:pi0CrossSection})
\cite{Adler:2003pb}, which spans eight orders of magnitude and
covers $1 < p_T < 15$ GeV/$c$. The data are compared to NLO pQCD
calculations utilizing the CTEQ6M pdf's \cite{Pumplin:2002vw} and
two different FF sets, differing principally in the gluon-to-pion
FF.  While both calculations describe the data well down to what are
perhaps surprisingly low values of $p_T$ ($\sim 1.5$~GeV/$c$), the
calculation incorporating the FF set of Kniel, Kramer, and
P\"{o}tter (KKP) \cite{Kniehl:2000hk} is in better agreement with
the data than that of Kretzer \cite{Kretzer:2000yf}. This is
consistent with a larger gluon-to-pion FF. The bottom two panels in
Fig.~\ref{figure:pi0CrossSection} indicate the sensitivity of the
calculations to the choice of factorization and renormalization
scales.  The calculations have been performed using equal
factorization and renormalization scales of $p_T$, $2p_T$, and
$p_T/2$.  The KKP FF set in particular exhibits relatively little
scale dependence.  For comparison, the renormalization and
factorization scale dependence of current DIS fixed-target
experiments performing spin physics measurements is several times
larger.  An alternative theoretical calculation is compared to these
neutral pion results in \cite{Bourrely:2003bw}.

There is similar agreement and similarly little scale dependence
when studying the mid-rapidity production of inclusive charged
hadrons, shown in Fig.~\ref{figure:chargedCrossSection}
\cite{Adler:2005in}. Here the calculations use the KKP FF set.  As
in the case of the neutral pions, the bottom panel shows the
difference between data and theory for three different scales of
$p_T$, $2p_T$, and $p_T/2$.

Even the forward production of neutral pions at RHIC, potentially
susceptible to soft (non-perturbative) contributions, has been
described by pQCD with relative success, as shown in
Fig.~\ref{figure:STARpi0CrossSection} from the STAR collaboration
\cite{Adams:2003fx}. The two calculations given in this figure
utilize the KKP and Kretzer FF sets, and similar to the mid-rapidity
data, the forward data, at least for $p_T \gtrsim 1.7$~GeV/$c$, are
in better agreement with the KKP set.

The establishment of the ability of NLO pQCD to describe RHIC cross
section data well and with little dependence on the choice of
factorization and renormalization scales provides a solid foundation
for using NLO pQCD to interpret in turn the \emph{polarized} data at
RHIC.

\section{Spin asymmetries in factorized QCD}
\label{section:asymInFactQCD}

Generally, a spin asymmetry is the ratio of the difference to the
sum of the spin-dependent cross sections for a particular process,
given for example by Eq.~\ref{eq:doubleAsym} in the case of a
double-spin asymmetry, with the arrow combinations representing
same-spin and opposite-spin collisions which could be transverse or
longitudinal.

\begin{equation}\label{eq:doubleAsym}
\varepsilon = \frac{\sigma^{\uparrow \uparrow} - \sigma^{\uparrow
\downarrow}}{\sigma^{\uparrow \uparrow} + \sigma^{\uparrow
\downarrow}}
\end{equation}
The denominator is simply the total unpolarized cross section and as
such is calculable in factorized QCD as described above.  The
numerator is instead the difference of a convolution of
spin-dependent pdf's, a spin-dependent partonic hard scattering
cross section, and spin-independent FF's.  As in the
polarization-averaged case, the partonic hard scattering cross
section is calculable directly from perturbative theory, while the
pdf's and FF's must be obtained from experimental measurements.  The
goal of the spin program at RHIC is to extract spin-dependent, or
polarized, pdf's from asymmetry measurements. A particularly clean
example is the extraction of the polarized gluon distribution,
$\Delta g$, from the longitudinal double-spin asymmetry in direct
photon production.  At mid-rapidity and over a wide $p_T$ range at
RHIC energies, about 75\% of direct photon production comes from
quark-gluon Compton scattering, $q+g \rightarrow q+\gamma$.
Equation~\ref{eq:photonAsym} gives the asymmetry in direct photon
production from this process.

\begin{equation}\label{eq:photonAsym}
A_{LL}^{q+g \rightarrow q+\gamma}(p_T) =  \frac{\sum_q \Delta q(x_1)
\bigotimes \Delta g(x_2) \bigotimes \Delta \sigma^{q+g \rightarrow
q+\gamma}(\hat{s})}{\sum_q q(x_1) \bigotimes g(x_2) \bigotimes
\sigma^{q+g \rightarrow q+\gamma}(\hat{s})}
\end{equation}
Assuming the polarized quark distributions have already been well
measured, e.g.~in DIS experiments, it is relatively straightforward
to extract $\Delta g$ from Eq.~\ref{eq:photonAsym}.  In practice, it
is slightly more complicated, given that gluon Compton scattering is
not the only process that contributes to direct photon production
and the exact $x$ values of the scattered partons are not known (see
related discussion in Chapter~\ref{section:future}). But generally,
the longitudinal double-spin asymmetry in direct photon production
at RHIC is expected to give a relatively clean measurement of
$\Delta g$ once enough statistics are available.

The above example illustrates how factorized QCD can be used to
extract polarized pdf's from experimental spin asymmetries.  In the
future, once a variety of well-constrained measurements are
available from RHIC, a global analysis of both DIS and RHIC results
will be performed in order to obtain the spin-dependent pdf's that
best describe all world data.

\chapter{Experimental setup}
\section{The Relativistic Heavy Ion Collider}

The Relativistic Heavy Ion Collider (RHIC) is located at Brookhaven
National Laboratory on Long Island, New York.  The RHIC storage ring
is 3.83~km in circumference and is designed with six interaction
points (IP's), at which beam collisions are possible. Up to 112
particle bunches per ring can be injected, in which case the time
between bunch crossings at the IP's is 106~ns. The design luminosity
for Au+Au collisions is $2 \times
10^{26}$~$\textrm{cm}^{-2}$~$\textrm{s}^{-2}$; for $p+p$ collisions
it is $2 \times 10^{32}$~$\textrm{cm}^{-2}$~$\textrm{s}^{-2}$.  The
design polarization for proton beams is 70\%.

RHIC was built to collide heavy ions at a center-of-mass energy of
up to 200~GeV per colliding nucleon pair and polarized protons at
energies ranging from 50 to 500~GeV. Collision of asymmetric
species, i.e.~different species in the two beams, is also possible
due to independent rings with independent steering magnets. The
first physics run took place in 2000, with Au+Au collisions at
130~GeV per nucleon.  The following four running periods included
Au+Au collisions at 200, 62.4, and 19.6~GeV/nucleon, Cu+Cu
collisions at 200, 62.4, and 22.4~GeV/nucleon, $d$+Au collisions at
200~GeV/nucleon, and polarized $p+p$ collisions at 200~GeV.

There were four major experiments developed for RHIC, three of which
will continue to take data after 2005.  There are two large
experiments, STAR \cite{Ackermann:2002ad} and PHENIX
\cite{Adcox:2003zm}, each of which have more than 500 collaborators,
and two smaller experiments, BRAHMS \cite{Adamczyk:2003sq} and
PHOBOS \cite{Back:2003sr}, having fewer than 100 collaborators each.
PHENIX, STAR, and BRAHMS all have a spin-physics program as well as
a heavy-ion program, while PHOBOS only studies heavy-ion physics.
PHOBOS finished taking data in 2005. The four experiments were
designed with some overlap and some complementarity in the physics
processes they could measure. In this way it is frequently possible
for one experiment to cross-check the results of another, yet each
experiment has its own area of specialization. The PHENIX
experiment, through which the measurement for this thesis was made,
is described in detail in Section~\ref{section:PHENIX} below.

\section{RHIC as a polarized p+p collider} \label{section:collider}
\begin{figure}
\centering
\includegraphics[height=0.7\textheight,angle=270]{%
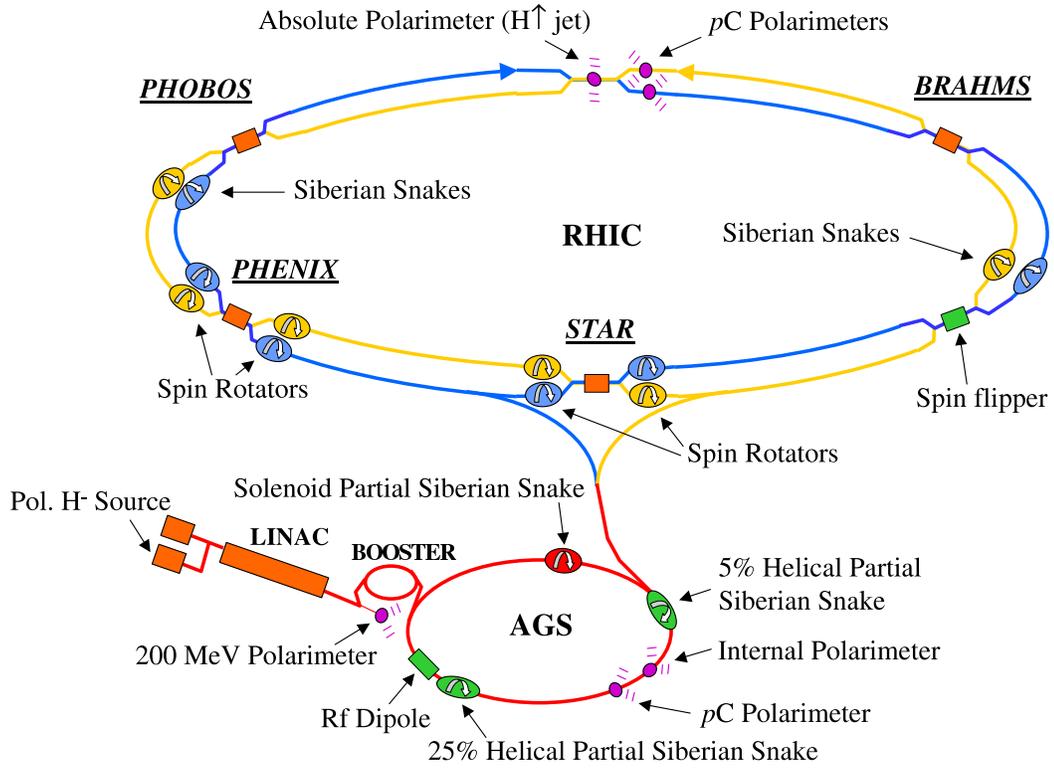} \caption[RHIC-AGS complex.]{The RHIC-AGS complex as
a polarized proton facility. } \label{figure:RHICComplex}
\end{figure}

RHIC is the first and only polarized proton-proton collider in the
world.  Figure~\ref{figure:RHICComplex} shows a diagram of the RHIC
complex including all equipment relevant for polarized proton beams.
A number of technological developments and advances have made it
possible to create a high-current polarized source, maintain the
beam polarization throughout acceleration and storage, and obtain
accurate measurements of the degree of beam polarization at various
stages from the source to full-energy beams in RHIC.  For an
overview of RHIC as a polarized-proton collider, refer to
\cite{Alekseev:2003sk}.

\subsection{RHIC-AGS complex}

In the case of polarized-proton running at RHIC, a pulse of
polarized $H^-$ ions from the source (see
Section~\ref{section:source}) is accelerated to 200~MeV in the
Linac, then stripped of its electrons as it is injected and captured
as a single bunch of polarized protons in the Booster, which
accelerates the protons to 1.5~GeV. The bunch of polarized protons
is then transferred to the Alternating Gradient Synchrotron (AGS)
and accelerated to 24 GeV before injection into RHIC.  Each bunch is
accelerated in the AGS and injected into RHIC independently, with
the two RHIC rings being filled one bunch at a time. The direction
of the spin vector is selected for each bunch separately.

\subsection{Polarized source}
\label{section:source}

Polarized proton injection uses an optically-pumped polarized $H^-$
ion source (OPPIS) constructed at TRIUMF in Canada from an OPPIS
source previously used at KEK in Japan \cite{Zelenski:2002gb}.
Polarization of 85\% has been reached.  In order to achieve a
polarized proton beam, there are several steps.  Unpolarized protons
first pick up polarized electrons from an optically pumped rubidium
vapor in a high magnetic field, forming hydrogen atoms with
polarized electrons.  The electron polarization is then transferred
to the nucleus via static magnetic fields, creating
nuclear-spin-polarized atoms. These atoms pick up a second
unpolarized electron in a sodium vapor negative-ionizer cell.  A
pulse of polarized $H^-$ ions is then accelerated to 200~MeV in the
Linac and injected as a single bunch of polarized protons in the
Booster.

\subsection{Siberian snakes}

The precession of the spin vector of a proton in a planar circular
accelerator ring is given by the Thomas-BMT equation
\cite{Thomas:1927,Bargmann:1959gz} (Eq.~\ref{eq:spinEvolution}),

\begin{equation}\label{eq:spinEvolution}
\frac{d\overrightarrow{P}}{dt} = -(\frac{e}{\gamma M})(G\gamma
\overrightarrow{B}_\bot + (1 + G)\overrightarrow{B}_\|)\times
\overrightarrow{P}
\end{equation}
in which $e$ is the proton charge, $M$ is the proton mass, $\gamma =
E/M$ is the relativistic boost, $G = 1.7928$ is the anomalous
magnetic moment of the proton, $B_\bot$ indicates the magnetic field
perpendicular to the plane of proton motion, typically the vertical
guide field, and $B_\|$ is the longitudinal field.
$\overrightarrow{P}$ here is in the frame of the proton.  Note that
for highly energetic protons ($\gamma$ large), the $B_\bot$ term
dominates.

Comparing the spin evolution equation to the Lorentz force equation
of motion for a particle orbit in a magnetic field
(Eq.~\ref{eq:particleInField}),

\begin{equation}\label{eq:particleInField}
\frac{d\overrightarrow{v}}{dt} = -(\frac{e}{\gamma
M})\overrightarrow{B}_\bot \times \overrightarrow{v}
\end{equation}
it becomes evident that for highly energetic protons or in a purely
vertical field, the spin precesses a factor of $G \gamma$ faster
than the orbital motion, meaning that the spin precesses $G \gamma$
times in a single revolution around the RHIC ring.  This number is
referred to as the spin tune, $\nu_{sp}$.

There are two principal types of depolarizing resonances that may be
encountered during acceleration, imperfection and intrinsic
resonances.  Imperfection resonances refer to those due to errors in
magnet currents or alignments; intrinsic resonances are due to the
(desired) beam focusing fields.  A depolarizing resonance is
encountered whenever the spin precession frequency is equal to the
frequency with which a depolarizing field is crossed, leading to
additive depolarization effects.  Imperfect fields exist with a more
or less random distribution around each ring.  As such, resonances
occur when $\nu_{sp} = G \gamma = n$, an integer, and the spin
vector is at the same phase in its precession every time it
encounters the same imperfect field.  Intrinsic resonances occur
when $\nu_{sp} = G \gamma = kP \pm \nu_y$, where $k$ is an integer,
$P$ is the superperiodicity, or degree of regularity in the
focusing-defocusing lattice, and $\nu_y$ is the vertical betatron
tune, the number of oscillations around the stable beam orbit per
beam revolution, in the vertical plane.  (The $z$-axis is taken to
be in direction of proton motion, in the frame of the proton.)

Siberian snakes \cite{Derbenev:1978hv}, a series of spin-rotating
dipoles, so named because of the beam trajectory through the magnets
and the fact that they were developed at Novosibirsk, are used to
overcome both imperfection and intrinsic resonances in RHIC. There
are two snakes installed in each RHIC ring at diametrically opposite
points along the rings. The two snakes in each ring rotate the spin
vector $180^\circ$ about perpendicular horizontal axes, without
perturbing the stable spin direction and with only local distortion
of the beam orbit. In this way, all additive depolarizing effects
from resonances are avoided.

In the AGS there is not enough space to permit a full snake.  Only a
partial snake \cite{Roser:1988bk}, which rotates the spin vector by
less than $180^\circ$, is possible, and complete reversal of the
spin direction only occurs over the course of multiple revolutions.
A partial snake of less than 10\% is sufficient to overcome
imperfection resonances but not intrinsic resonances in the AGS; a
partial snake of 20\% or more is expected to overcome both types of
resonances. In the 2001-02 run, a solenoidal "warm"
(non-superconducting) snake was available in the AGS.  This magnet
was a 5\% snake; therefore, it rotated the spin direction by
$9^\circ$ each revolution.  In order to handle intrinsic resonances,
the technique used was to artificially enhance the resonances such
that they were tuned to produce a complete spin flip each time one
is encountered, rather than depolarization.  An AC dipole magnet was
used to achieve this. In 2004 the solenoidal snake was upgraded to a
warm helical snake, which aided in decoupling the snake's effects on
the $x$ and $y$ motion of the beam.

In 2005 a superconducting "cold" snake was installed in the AGS and
partially commissioned; this final piece of equipment should allow
achievement of the RHIC design polarization of 70\%. Superconducting
magnets allow a larger magnetic induction, thus a larger $\int
B\cdot dl$ and greater spin rotation per revolution. The cold snake
is a 25\% snake, rotating the spin vector $45^\circ$ each time.  In
2006 the plan is to use both the warm and cold snakes in
conjunction, rotating the spin by 5\% and 15\% respectively,
achieving the minimum rotation of 20\% required to handle all
resonances in the AGS.  A description of the technique of using
multiple partial snakes in conjunction can be found in
\cite{Roser:2005}.

The first polarized-proton run in 2001-2002 achieved an average beam
polarization in RHIC of $\sim 15$\%.  Improvements in both available
hardware and machine understanding led to measured polarization
values in excess of 50\% in 2005.  In addition, significant
polarization ($\sim 30$\%) at a center-of-mass energy of 410~GeV was
achieved in 2005. Further high-energy commissioning in preparation
for eventual running at $\sqrt{s} = 500$~GeV will be done in 2006.

\subsection{Polarimeters}
\label{section:polarimetry}

For RHIC to provide full-energy polarized beams, the polarization
must be measurable at various stages of acceleration in order to
identify and address possible origins of depolarization at each
step.  Only RHIC polarimetry will be discussed here and not the
various techniques used to measure the beam polarization further
upstream in the process of acceleration. The RHIC polarimetry
fulfills a three-fold purpose: (1) beam polarization measurements to
provide feedback for the accelerator physicists; (2) beam
polarization measurements to provide polarization values as input
for the various experiments; and (3) experimental study of polarized
elastic scattering.

There are two types of polarimeters installed in RHIC, designed to
measure the beam polarization in the vertical direction. The
proton-carbon ($p$C) polarimeter takes advantage of a known
analyzing power, $A_N^{p\textrm{C}} \approx 0.01$, in the elastic
scattering of polarized protons with carbon atoms ($p^{\uparrow } +
\textrm{C}\rightarrow p^{\uparrow } + \textrm{C}$) in order to
measure the beam polarization.  This analyzing power originates from
interference between electromagnetic and hadronic elastic scattering
amplitudes, which are finite due to the anomalous magnetic moment of
the proton. Thus the term "Coulomb-nuclear interference" (CNI) is
used to describe the process. Refer for example to
\cite{Bourrely:1977aw} for a discussion of CNI in hadronic reactions
at high energies. For the $p$C polarimeter at RHIC, the left-right
(azimuthal) asymmetry of the recoil carbon atoms is measured using
an array of silicon detectors, allowing calculation of the beam
polarization as in Eq.~\ref{eq:CNIPol},

\begin{equation}\label{eq:CNIPol}
P_{\textrm{beam}} =
\frac{1}{A_N^{p\textrm{C}}}\frac{\sqrt{N^{\uparrow}_{L}N^{\downarrow}_{R}}
-
\sqrt{N^{\downarrow}_{L}N^{\uparrow}_{R}}}{\sqrt{N^{\uparrow}_{L}N^{\downarrow}_{R}}
+ \sqrt{N^{\downarrow}_{L}N^{\uparrow}_{R}}}
\end{equation}
where $N^{\uparrow}_{L}$ ($N^{\downarrow}_{R}$) indicates the number
of recoil carbons scattering to the left (right) of the proton beam
from bunches polarized up (down). (Compare to
Eq.~\ref{eq:sqrtFormula} in Section~\ref{section:sqrtFormula}.)

The analyzing power in the process now utilized by the $p$C
polarimeter to measure the RHIC beam polarization was initially
measured by AGS experiment E950 \cite{Alekseev:2002ym}.  However,
the AGS measurement was only made to $\pm 30$\%, at a beam energy of
22~GeV, and further calibration is necessary to understand the RHIC
beam polarization at 100~GeV.

Calibration of the $p$C polarimeter to within an absolute beam
polarization of $\sim 5$\% can be provided by a polarized
hydrogen-jet-target polarimeter \cite{Zelenski:2005mz}.  Such a
device was commissioned at RHIC in 2004. The measurement is in some
ways similar to that of the $p$C polarimeter.  The left-right
transverse single-spin asymmetry of elastically scattered
\emph{protons} due to the CNI process is measured using silicon
strip detectors.  However, the analyzing power for this process of
elastic proton-proton scattering is not known. Instead, the
hydrogen-jet-target polarization, $>90\%$, is known to better than
2\% in absolute polarization.  The beam polarization can be obtained
by measuring first the asymmetry due to the polarized target
($\varepsilon_{tgt}$), averaging over the spin states of the beam,
and then similarly measuring the asymmetry due to the polarized beam
($\varepsilon_{beam}$), averaging over spin states of the target,
which are varied in time. The same (unknown) analyzing power, $A_N$,
applies to both cases, and the beam polarization, $P_{beam}$, can be
calculated as in Eq.~\ref{eq:jetPol}.

\begin{equation}\label{eq:jetPol}
A_N = \frac{\varepsilon_{tgt}}{P_{tgt}} =
\frac{\varepsilon_{beam}}{P_{beam}} \Rightarrow P_{beam} =
P_{tgt}\frac{\varepsilon_{beam}}{\varepsilon_{tgt}}
\end{equation}
Once the beam polarization is determined, Eq.~\ref{eq:CNIPol} can be
rearranged and used in turn to measure the analyzing power of the
process.  Results on the measurement of the analyzing power of the
jet polarimeter can be found in \cite{Okada:2005gu}.

Low rates for this process mean that measurements performed over a
long time are necessary. Because of this, the $p$C polarimeter,
which can make measurements in less than ten seconds and provide
immediate information on the stability or decay of the beam
polarization from a few data points taken over the several hours of
a fill, remains essential. Results obtained from the two different
polarimeters can be compared offline, and in this way the $p$C
polarimeter can be calibrated.

In order to check the \emph{absolute} sign of the spin direction for
each bunch, it is possible to measure bunch-by-bunch asymmetries,
making left-right asymmetry measurements from events coming from
only a single bunch number, and then take advantage of the fact that
the sign of the CNI asymmetry is known from theory.

Note that if bunch-by-bunch polarimeter information were not
available, as was the case in 2002, and a few individual bunches had
the incorrect spin direction assigned to them, this would make the
beam polarization appear to be less than its true value.  The
experiments in turn would utilize these same, incorrect spin
assignments, diluting any raw asymmetry in particle production they
may observe.  However, the larger beam polarization correction would
exactly compensate for the diluted raw asymmetry, and a true physics
asymmetry would still result.  If instead all bunches were assigned
the incorrect spin direction, the measured CNI asymmetry would have
the opposite sign with respect to theoretical predictions.

\subsection{Spin rotators}

The naturally stable spin direction through acceleration and storage
is transverse to the proton's momentum, in the vertical direction.
Spin rotator dipole magnets, commissioned in 2003, have been used to
achieve both radial and longitudinal spin \cite{MacKay:2003}.  The
spin vector is rotated away from vertical immediately before the
collision point and then back to vertical immediately afterwards.
The rotators are located outside the interaction regions of the
PHENIX and STAR experiments, giving both experiments the ability to
choose independently whether they want longitudinally or
transversely polarized collisions.  The BRAHMS experiment, having no
spin rotators available, focuses on transverse spin measurements.
The local nature of the spin rotator magnets means that the STAR and
PHENIX experiments must each have their own way of checking the
direction of the spin vector at their respective interaction
regions. In PHENIX this ability is provided by the local
polarimeter, described in Section~\ref{section:localPol}.

\subsection{Spin flippers}
\label{section:spinFlipper}

In addition, RHIC is equipped with a "spin flipper," an AC dipole
with a radial magnetic field installed in an area common to both
beams, which flips the spin $180^\circ$ bunch by bunch
\cite{Bai:2003jv}. Spin flipping involves detuning one of the two
snakes to alter the spin tune and then sweeping the spin flipper
frequency of the AC dipole through the resonance to effect a
controlled spin reversal. The ability to flip all the spin vectors
bunch by bunch is important in order to reduce potential systematic
errors from any correlations that may exist between a bunch and its
spin direction. An example of such possible correlations would be a
systematically larger bunch length for bunches of a particular spin
sign, which could lead to subtle differences in the effective
luminosity seen by the luminosity detectors and the
final-state-particle detectors (e.g. the PHENIX central arms, see
Section~\ref{section:centralArms}). Frequent or carefully timed use
of the spin flipper throughout a machine fill could also facilitate
balancing the relative luminosity at all interaction regions; any
bunch crossings later discarded for any reason from offline analysis
would represent the various spin configurations in an approximately
balanced way, and the relative luminosity would not be greatly
affected.

Preliminary commissioning of the spin flippers was performed in 2002
\cite{Bai:2003ce} but was not continued in subsequent years. Final
commissioning and the beginning of regular usage is planned for 2006
or 2007.

\section{The PHENIX experiment and detector}
\label{section:PHENIX}

The PHENIX collaboration, one of the two large collaborations at
RHIC, is comprised of more than 500 people from over 60 institutions
around the world. The PHENIX experiment was designed to specialize
in the measurement of photons, electrons, and muons as well as
high-$p_T$ probes in general over a limited acceptance, with good
mass resolution and particle identification capabilities. It has an
extremely high rate capability as well as sophisticated trigger
systems, allowing measurement of rare processes.  PHENIX produced a
wealth of physics results from the first years of RHIC running and
anticipates the ability to continue taking advantage of improvements
in machine luminosity to further develop its program to measure rare
probes.

The PHENIX detector consists of two mid-rapidity ($|\eta | < 0.35$,
$\eta = -\ln \tan \frac{\theta}{2}$) spectrometer arms, primarily
for identifying and tracking charged particles as well as measuring
electromagnetic probes, forward spectrometer arms for identifying
and tracking muons, and interaction detectors.  An overview of the
PHENIX detector is given in \cite{Adcox:2003zm}.

\subsection{Interaction and vertex detectors}

PHENIX has two detectors used to determine when a collision occurs
and to measure the position of the event vertex in the beam
direction. The beam-beam counters (BBC's) are quartz \v{C}erenkov
detectors, located at $\pm 1.44$~m from the nominal interaction
point, which measure charged particles over 2$\pi$ and cover a
pseudorapidity range of $3.0 < |\eta | < 3.9$. The zero-degree
calorimeters (ZDC's) are hadronic calorimeters sensitive primarily
to forward neutrons.  They also cover 2$\pi$ in azimuth, are located
at $\pm 18$~m, and extend over $4.7 < |\eta | < 5.6$.  The ZDC's
were developed for RHIC rather than specifically for PHENIX and are
used by all four RHIC experiments.

A minimum-bias (MB) trigger occurs when there is a minimum of one
photomultiplier tube (PMT) fired in each of the two BBC's.  In
200-GeV proton-proton collisions, the BBC's see approximately 50\%
of the total inelastic $p+p$ cross section and provide a vertex
resolution of $\sim 2$~cm. The ZDC's are sensitive to a much smaller
fraction of the interactions and are not used as part of the MB
trigger in $p+p$. The primary roles of the ZDC's in proton-proton
collisions are in local polarimetry (see
Section~\ref{section:localPol}) and for systematic studies of the
relative luminosity between bunch crossings with different spin
configurations (see Appendix~\ref{section:relLumi}). In gold-gold
collisions both the BBC's and ZDC's observe a significant fraction
of the inelastic cross section (the BBC's observe 92\%), and
correlations in the measured ZDC energy and BBC charge are used to
determine the centrality (degree of overlap) of the heavy-ion
collisions.

In the 2001-02 $p+p$ run at RHIC, when the data for the present work
were taken, there was an additional interaction detector installed,
the normalization trigger counter (NTC).  The NTC consisted of two
scintillation counters located at $\pm 40$~cm from the center of the
interaction diamond and covering the full azimuth.  This detector
was installed to improve cross section measurements in $p+p$
collisions;  it was sensitive to $\sim 85$\% of the total inelastic
cross section. For further discussion of the PHENIX BBC's and NTC,
see \cite{Allen:2003zt}; for a more detailed description of the RHIC
ZDC's, see \cite{Adler:2003sp}.

\subsection{PHENIX central arms}
\label{section:centralArms}

\begin{figure}
\centering
\includegraphics[height=0.6\textheight]{%
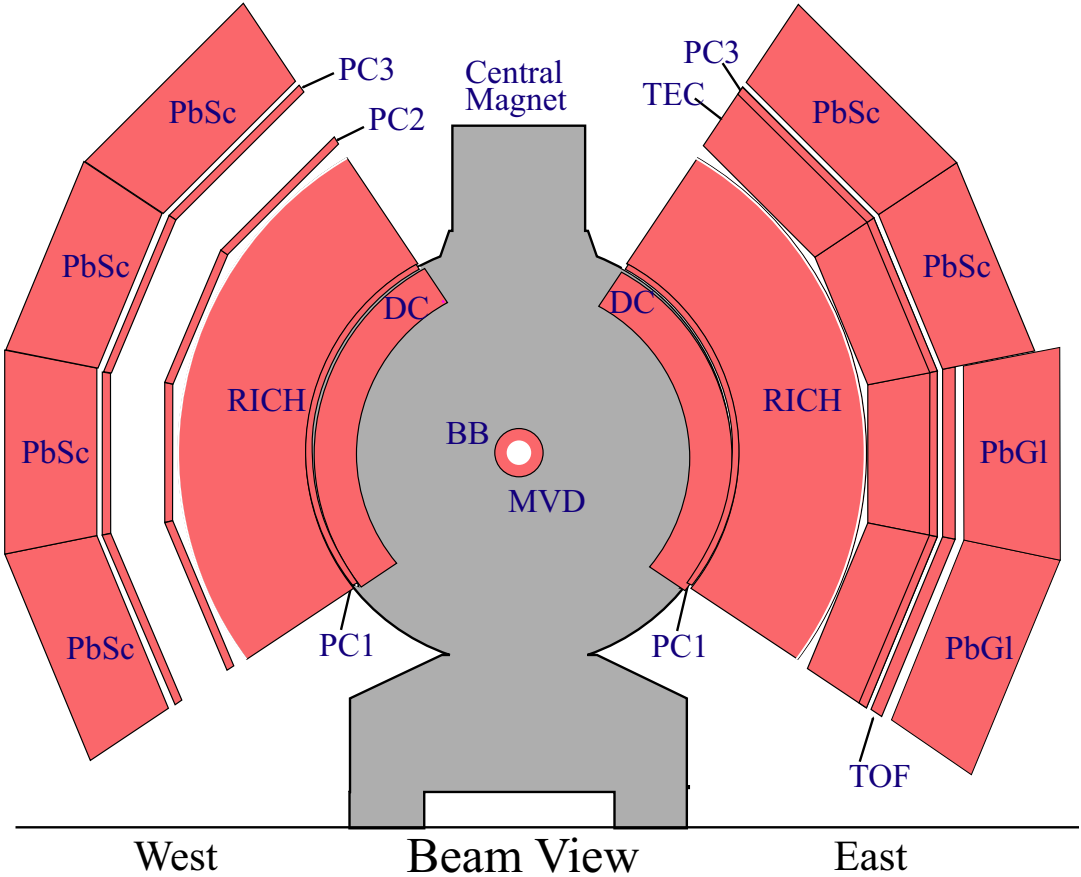} \caption[PHENIX central arms]{PHENIX central
arms in 2001-02.} \label{figure:centralArms}
\end{figure}

The PHENIX central arms cover a pseudorapidity range of $|\eta | <
0.35$ and two 90-degree intervals in azimuth, offset 33.75 degrees
from vertical. See Figure~\ref{figure:centralArms} for a schematic
diagram of the PHENIX central arms as instrumented in the 2001-02
RHIC run.

\subsubsection{Tracking detectors}

The PHENIX central tracking system is comprised of a drift chamber
(DC) and pad chambers (PC's).  The DC is located at $r = 2.0-2.4$~m
from the beam pipe and covers 2~m along the beam direction.  It
measures the trajectories of charged particles in the $r-\phi$
direction, providing the $p_T$ of the particle as well as $\phi$
information, necessary for invariant-mass reconstruction of particle
pairs.  The innermost plane of the PC is located immediately outside
the DC; the outermost plane is immediately in front of the
electromagnetic calorimeter (EMCal, see below).  There is a middle
PC plane located behind the ring-imaging \v{C}erenkov detector (see
below) in the West arm only.  The PC's provide three-dimensional
spatial point information for pattern recognition.  The outermost
plane of the PC is particularly effective in providing a charged
veto for particles that shower in the EMCal.

The tracking detectors are placed outside the magnetic field; there
is only a residual field of $< 0.6$~kG present in the DC.  With no
inner tracking available, tracks are assumed to point back to the
event vertex, provided by the BBC's. Thus charged particles that do
not originate at the vertex have incorrectly reconstructed momentum,
leading to low-momentum, long-lived particle decays and conversion
electrons as sources of background in high-$p_T$ charged particle
analysis.  The tracking detectors are described in more detail in
\cite{Adcox:2003zp}.

\subsubsection{Ring-imaging \v{C}erenkov detector}

A ring-imaging \v{C}erenkov detector (RICH) is placed between the
layers of tracking subsystems.  Its main purpose is electron-pion
discrimination.  It has been filled with \textrm{CO$_2$} gas in the
first five years of RHIC running.  The momentum threshold for
production of \v{C}erenkov radiation by charged pions in
\textrm{CO$_2$} is 4.7~GeV/$c$, while for electrons it is
0.017~GeV/$c$.  In the $p_T$ range between these two values, there
is excellent $e/\pi$ separation.  Further discussion of the RICH can
be found in \cite{Aizawa:2003zq}.

\subsubsection{Electromagnetic calorimeter}

The EMCal is designed to measure the energy and position of
electrons and photons.  It also contributes significantly to
particle identification through energy-position measurements as well
as timing information.  It is the outermost subsystem in the central
arms, located at $\sim5$~m in $r$.  The EMCal in the West arm is
divided into four sectors of a lead-scintillator (PbSc) sampling
calorimeter; in the East arm there are two sectors of PbSc and two
of a lead-glass (PbGl) \v{C}erenkov calorimeter.  The two kinds of
detectors rely on significantly different physics processes and as
such can be useful in making systematic comparisons.  The PbGl
excels in energy measurements and is less sensitive to hadrons,
while the PbSc has better timing resolution.  The nominal energy
resolution of the PbSc is 8.1\%/$\sqrt{E~\textrm{(GeV)}}\oplus
2.1\%$; the PbGl has a nominal energy resolution of
6\%/$\sqrt{E~\textrm{(GeV)}}$.  The intrinsic timing resolution is
$\sim 200$~ps for the PbSc and $\sim 300$~ps for the PbGl.  While
designed as an electromagnetic calorimeter, the EMCal is
approximately 1 nuclear interaction length and offers some
sensitivity to hadrons. Detailed description of the EMCal is
available in \cite{Aphecetche:2003zr}.

\subsection{Local polarimeter}
\label{section:localPol}

\begin{figure}
\centering
\includegraphics[height=0.4\textheight]{%
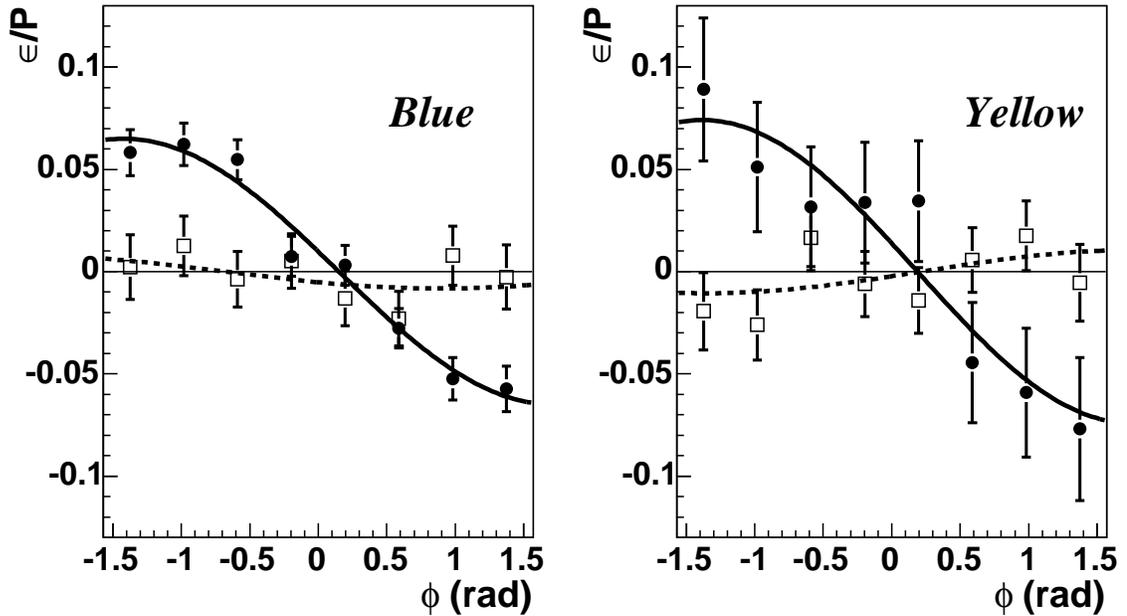} \caption[PHENIX local polarimeter measurement.]{PHENIX
local polarimeter measurement.  The raw asymmetry in neutron
production divided by the degree of beam polarization is shown vs.
azimuthal angle.
 The solid points are with the spin rotators off (vertical
polarization); the open points are with the spin rotators on
(longitudinal polarization).} \label{figure:localPol}
\end{figure}

In order to provide local polarimetry, a shower maximum detector
(SMD) was added to the PHENIX ZDC's.  The ZDC/SMD combination
measures both the energy and position of forward neutrons.  A large
($\sim -11\%$) transverse single-spin azimuthal asymmetry for
neutrons with $p_{T}<0.5$~GeV/$c$ has been observed
\cite{Bazilevsky:2003bm}.  This asymmetry can be exploited to
measure the degree to which the beam polarization is vertically
transverse, radially transverse, or longitudinal.  Maxima and minima
of particle production are expected at right angles in $\phi$ to the
spin vector. For longitudinal polarization, no azimuthal asymmetry
is expected; a non-zero longitudinal single-spin asymmetry would
violate parity and is forbidden in particle production via strong
and electromagnetic processes. Figure~\ref{figure:localPol} shows
the observed asymmetry as a function of $\phi$ for the cases of
vertical and longitudinal beam polarization.  The expected maxima
and minima at $\pm \frac{\pi}{2}$ can be seen for vertical
polarization, and the asymmetry is approximately consistent with
zero in the case of longitudinal polarization.

\subsection{Muon arms}

\begin{figure}
\centering
\includegraphics[height=0.5\textheight]{%
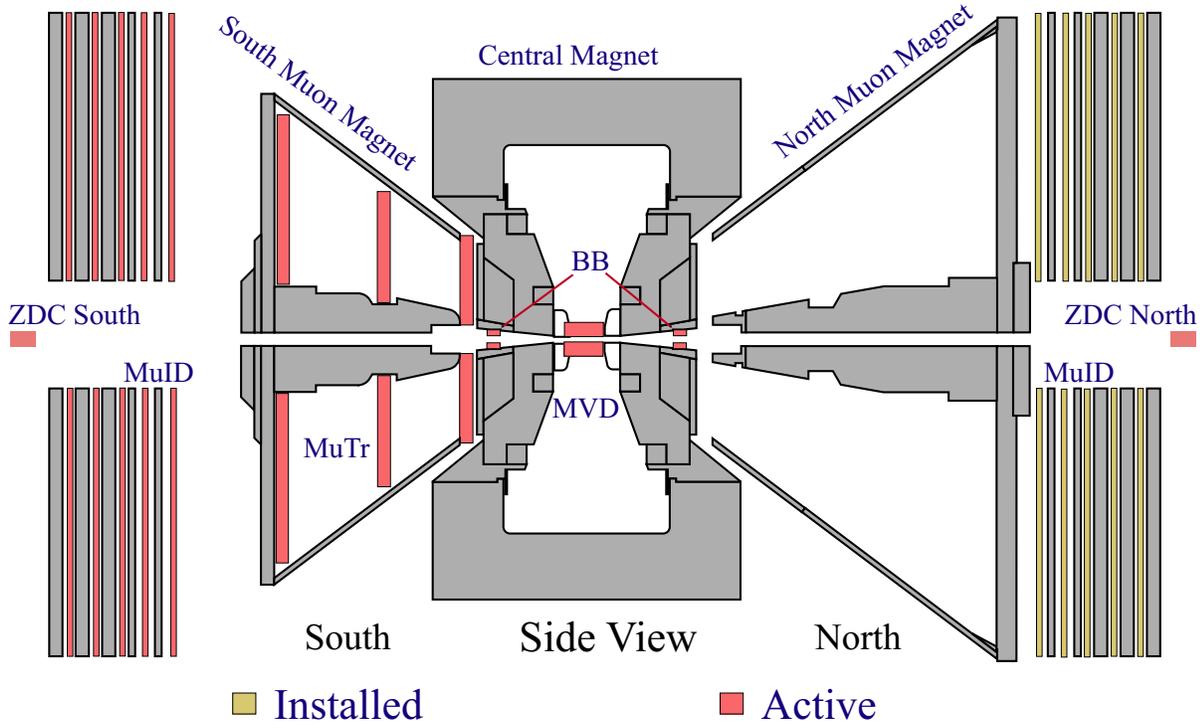} \caption[PHENIX muon arms]{PHENIX muon arms in
2001-02.} \label{figure:muonArms}
\end{figure}

The forward spectrometer arms in PHENIX are designed to identify and
measure prompt and decay muons.  They cover 2$\pi$ in azimuth and a
pseudorapidity range of $1.2 < |\eta| < 2.4$.  A high-resolution
tracker in a radial magnetic field is followed by alternating layers
of absorber and low-resolution tracking for muon identification. The
tracker magnet also serves as a hadron absorber.  The design is to
reject pions and kaons at a level of $10^{-3}$.  In
Figure~\ref{figure:muonArms}, the muon arms are shown as
instrumented for the 2001-02 RHIC run. Note that only the south arm
was available for data taking in this year; the north arm was
completed in 2003.  See \cite{Akikawa:2003zs} for further
information regarding the muon arms.

\subsection{Detectors and triggers used in the analysis for this thesis}

To measure the energy and position of photons from neutral pion
decays, the present analysis utilized both the PbSc and PbGl
electromagnetic calorimeters.  The central arm tracking detectors
were used to eliminate EMCal clusters associated with charged
tracks.  The BBC's were used to provide a MB trigger.  The
EMCal-RICH trigger (ERT, described below) was used to select events
with high-$p_T$ photons.

\subsubsection{EMCal-RICH trigger}

The hardware-level rare-event triggers implemented in PHENIX include
an EMCal-RICH trigger (ERT).  The ERT can be used as a high-energy
photon trigger or in conjunction with the RICH as an electron
trigger.  The EMCal towers are grouped into trigger tiles, over
which the deposited energy is summed, and an event is recorded if
the total energy in a tile is above a preset threshold.  The tiles
are configured to be non-overlapping sets of $2\times 2$-tower
regions and overlapping sets of $4\times 4$-tower regions. Up to
three different energy thresholds may be set for the $4\times
4$-tower tile size. The electron trigger utilizes the $2\times
2$-tower tile size.

\subsubsection{EMCal calibration}

There was an initial series of EMCal calibrations performed, both
before detector installation and \textit{in situ} in the PHENIX
interaction region, descriptions of which can be found in
\cite{Aphecetche:2003zr}.  During regular data taking, a pulsed UV
laser is utilized to provide EMCal calibration information. In
offline analysis, three major techniques used to calibrate the EMCal
energy are examining the location of the $\pi^0$ mass peak position,
the energy-momentum ratio for electrons, and the location of the
minimum-ionizing peak.

\chapter{Data and analysis}
\section{Overview}

The left-right transverse single-spin asymmetry, $A_N$, of
mid-rapidity neutral pions was measured from the data taken during
the first polarized proton run at RHIC, in late 2001 and early 2002.
To make a single-spin measurement with two polarized beams, the spin
states of only one beam at a time were taken into account, averaging
over the spin states of the other. Neutral pions were reconstructed
via their decay to two photons. The general procedure was to obtain
the spin-dependent $\pi^0$ yields for each machine fill, calculate
the raw (uncorrected) asymmetries for each fill as a function of
$p_T$, make fill-by-fill polarization corrections, and then average
the asymmetries over all fills. The results were corrected for
estimated contributions to the asymmetry from the background under
the \piz\ invariant-mass peak by measuring and subtracting the
asymmetry of the background immediately around the peak in mass.
Various studies were subsequently performed as checks. The final
results of this analysis have been published in \textit{Physical
Review Letters} \cite{Adler:2005in}.

\section{Data selection and quality}

Prior to this spin-asymmetry analysis, a polarization-averaged cross
section measurement was made for neutral pion production at PHENIX
and published in \cite{Adler:2003pb}.  The analysis completed for
this thesis started with the data sample utilized in the cross
section analysis, and then additional quality cuts relevant to a
spin-dependent analysis were made.

The data used were from ERT \trig-triggered events. The \trig\
energy threshold was $\sim0.8$ GeV in the 2001-02 run. The triggered
sample had much better statistics at higher transverse momenta than
the MB sample, but the MB sample was also analyzed for comparison
purposes. 18.7 million triggered events were analyzed, corresponding
to approximately 880 million sampled MB events. The detector
subsystems involved were the BBC and the EMCal, including both the
lead scintillator (PbSc) and lead glass (PbGl). Additionally, the
drift chamber and pad chambers were used to veto clusters in the
EMCal produced by charged particles. Basic checks on the quality of
output from all involved subsystems were performed.

\subsection{Fill and run selection}

In making $\pi^0 \rightarrow \gamma + \gamma$ spin-asymmetry or
cross section measurements, it is essential to ascertain that the
EMCal was working properly when the analyzed runs were taken. The
uppermost sector of PbSc in the West arm had been omitted from the
cross section analysis due to problems with electronics noise during
data taking. A number of runs had been eliminated from that analysis
because of EMCal towers with unusually high or low numbers of hits
that were not among the known and understood hot or dead EMCal
towers. In addition, a sequence of runs had been omitted because of
the ERT.  The energy threshold for the ERT \trig\ trigger had been
adjusted in the early part of the data-taking period, and only runs
taken after the threshold was stable at $\sim 0.8$~GeV were
included.  Runs from a total of twenty machine fills were included
in the final cross section analysis.

For the present analysis, six of the twenty fills used in the cross
section analysis were removed. Two were removed because no
polarization measurement was available. Three additional fills were
removed because a number of bad runs from these fills were found in
a study investigating the stability of the MB trigger. One fill was
removed because it had an unusual spin pattern.

\subsection{Event and crossing selection}

An offline BBC event vertex within $\pm $30~cm of the nominal
interaction point was required for all events.  The acceptance of
the central arms is approximately constant for collisions taking
place in this region. The BBC vertex resolution was $\sim 2$~cm. The
online vertex cut for the MB trigger in the 2001-02 run was $\pm
$75~cm.

Spin-sorting the events requires keeping track of the direction of
the spin vector for the polarized bunch in the bunch crossing that
produced the event (simply the term \emph{crossing} will be used
henceforth). Note that the same pairs of bunches in the two rings
collide at the same interaction points each time, but the same pairs
of bunches do not necessarily collide at the different interaction
points, i.e.~at different experiments. Data quality was checked on a
crossing-by-crossing basis.  For the entire 2001-02 $p+p$ data set,
four crossings out of a nominal 60 were consistently removed from
this analysis.  These four crossings had unusually low luminosities
because they were regularly affected by either injection or steering
activities in the ring. In addition, the ten crossings in which only
one beam had filled bunches while the other had a five-crossing
beam-abort gap were removed from all fills. Any events occurring in
these crossings were beam-background collisions rather than
beam-beam collisions.

\begin{figure}
\centering
\includegraphics[height=0.25\textheight]{%
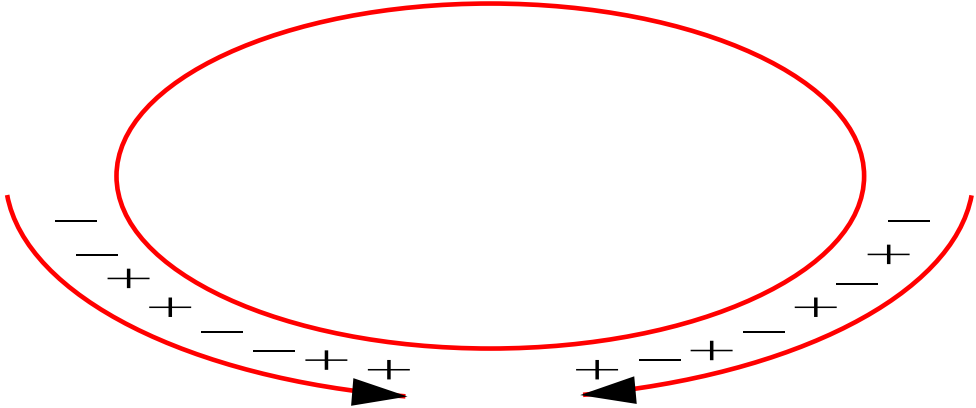} \caption[Schematic illustration of the spin
pattern in RHIC.]{Schematic illustration of the first spin pattern,
with no zero-polarization crossings.  In one beam, alternate bunches
have opposite spin directions; in the other beam, the spin direction
changes every two bunches.} \label{figure:spinPattern}
\end{figure}

There were two bunch-by-bunch spin patterns utilized in the 2001-02
run, one in the initial period of data taking and the other in the
latter period. The second pattern included bunches with zero
polarization so that they would be available for systematic checks,
in particular for the $p$C polarimeter.  For fills with the second
polarization pattern, one additional crossing was eliminated from
this analysis because it had zero polarization.  The other
zero-polarization crossings corresponded to two of the
low-luminosity bunches that had already been eliminated.  A
schematic illustration of the first spin pattern is given in
Figure~\ref{figure:spinPattern}.  Note that the spin patterns are
selected to provide approximately equal numbers of same-spin and
opposite-spin crossings for each experiment.  They are also chosen
so that the spin combinations are different every crossing.  This
rapid change in spin combinations, occurring on the order of every
hundred nanoseconds, greatly reduces potential time-dependent
systematic uncertainties.

Crossing-by-crossing luminosity measurements from the $p$C
polarimeter provided a fill-dependent bad crossing list.  A total of
four individual crossings from two fills were discarded from the
analysis based on this information.

\section{EMCal-RICH trigger}

\begin{figure}
\centering
\includegraphics[angle=0,height=0.5\textheight]{%
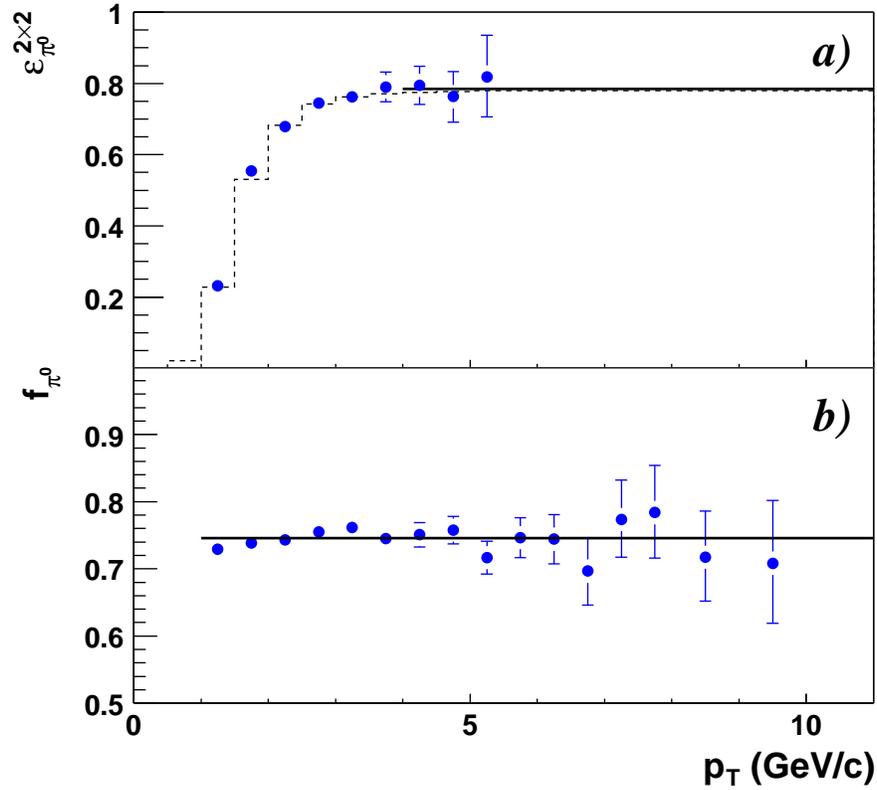} \caption[Trigger efficiency for neutral pions.]{a)
The \trig\ trigger efficiency for neutral pions, as a function of
pion transverse momentum.  The dashed line shows a Monte Carlo
simulation based on trigger tile efficiencies, and the solid line
indicates an upper limit on the $\pi^0$ efficiency based on the
number of active trigger tiles.  b) The fraction of the $\pi^0$
yield satisfying the MB trigger condition.  The solid line is the
fit of the data to a constant. The figure is taken from
\cite{Adler:2003pb}.} \label{figure:trigEfficiency}
\end{figure}

All data used for the final asymmetry results were from events
accepted by the \trig-tower tile trigger for high-$p_T$ (high energy
at mid-rapidity) photons. The \trig\ trigger had an average
rejection factor of 47, i.e.~only accepted on average one in 47 MB
events. It had a 78\% efficiency for neutral pions above $p_T
\approx 3.5$~GeV/$c$, as can be seen in
Fig.~\ref{figure:trigEfficiency}.

The effects of the trigger on the mean $p_T$ of the photon pairs
falling under the $\pi^0$ peak can be seen in
Table~\ref{table:meanPtMBTrig}.  The trigger raised the mean $p_T$
in the 1-2~GeV/$c$ bin significantly because it was still well below
its maximum efficiency in this transverse momentum range.  The
trigger had little effect on the mean $p_T$ of higher $p_T$ bins.

\begin{table}[tbp] \centering
\begin{tabular}{|c|c|c|}
  \hline
  $p_T$ bin & $<p_T>$ MB & $<p_T>$ \trig \\
  (GeV/$c$) & (GeV/$c$) & (GeV/$c$) \\
  \hline
  1-2 & 1.27 & 1.38 \\
  2-3 & 2.32 & 2.33 \\
  3-4 & 3.33 & 3.35 \\
  4-5 & 4.39 & 4.36 \\
  \hline
\end{tabular}
\caption[Trigger effects on mean $p_T$ of photon pairs.]{Mean
$p_{T}$ of photon pairs under the $\pi^{0}$ mass peak for MB and
\trig-triggered data.} \label{table:meanPtMBTrig}
\end{table}

\section{Reduction of background}
\begin{figure}
\centering
\includegraphics[angle=0,height=0.6\textheight]{%
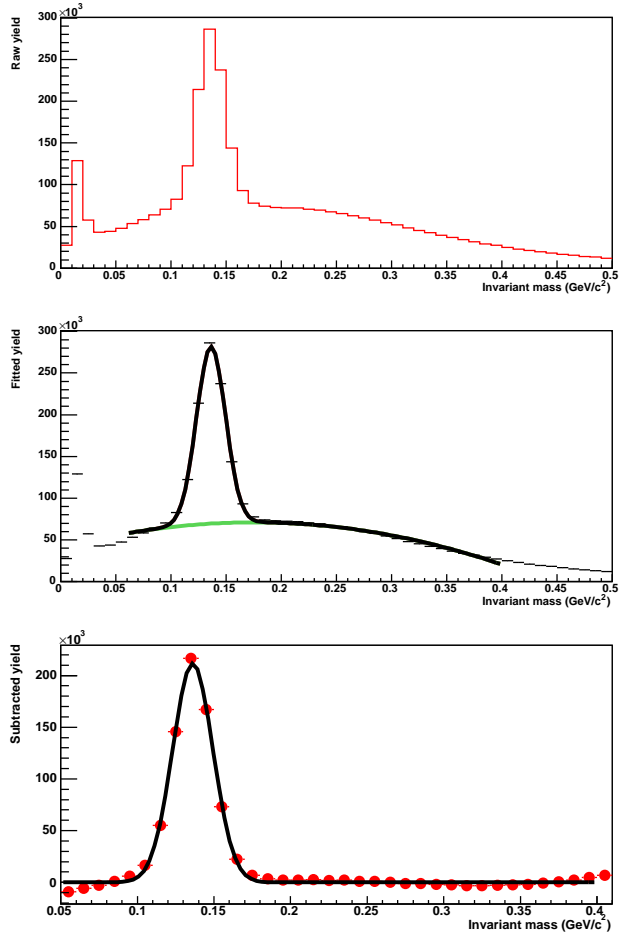} \caption[Fitted photon-pair invariant-mass
spectrum.]{Top panel: Invariant-mass spectrum for $1 < p_T <
2$~GeV/$c$ photon pairs. Middle panel: Fitted spectrum. Bottom
panel: Subtracted spectrum.} \label{figure:invMass}
\end{figure}

In order to understand the background contribution to the \piz\ mass
peak and obtain \piz\ yields, the invariant-mass spectrum for photon
pairs was fitted.  The mass peak was fitted to a Gaussian and the
combinatorial background to a second-degree polynomial. The \piz\
yields per $p_{T}$ bin, given in Table~\ref{table:yieldsWidths},
were obtained by subtracting the background from the total number of
pairs in the peak.  An example fitted and subtracted invariant-mass
spectrum for $1 < p_T < 2$~GeV/$c$ is shown in
Figure~\ref{figure:invMass}. Mass bins of 10 MeV/$c^2$ were used in
this analysis.

\begin{table}[tbp] \centering%
\begin{tabular}{|c|c|c|} \hline

$p_T$ (GeV/$c$) & $\pi^0$ yield & Peak width (MeV/$c^2$) \\
\hline

1-2 & 658k & 13.2 \\ \hline

2-3 & 143k & 11.2 \\ \hline

3-4 & 22k & 10.4 \\ \hline

4-5 & 4k  & 10.6 \\ \hline

\hline
\end{tabular}%
\caption[$\pi^0$ yields and peak widths.]{$\pi^0$ yields obtained
after background subtraction; 1$\sigma$ \piz\ peak widths from a
Gaussian fit.}%
\label{table:yieldsWidths}
\end{table}%

\begin{figure}
\centering
\includegraphics[angle=0,height=0.45\textheight]{%
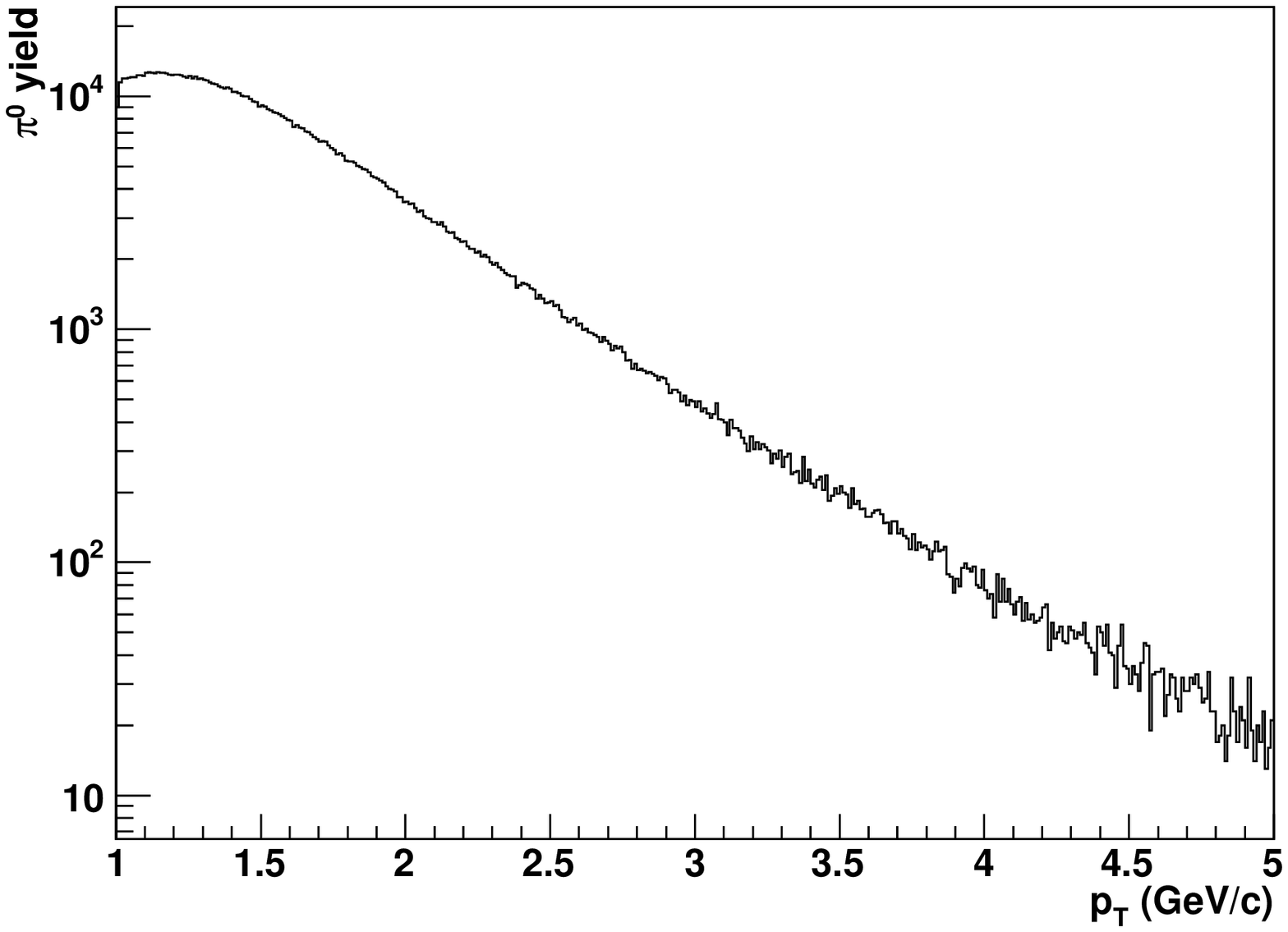} \caption[$p_T$ spectrum for photon pairs falling
under the \piz\ mass peak.]{Raw transverse momentum spectrum for
photon pairs falling under the \piz\ mass peak.}
\label{figure:momSpectrum}
\end{figure}

The 1$\sigma$ peak widths from a Gaussian fit to the 120-160
MeV/$c^{2}$ mass region are shown in Table \ref{table:yieldsWidths}.
The transverse momentum spectrum for the pairs falling under the
\piz\ mass peak can be seen in Figure~\ref{figure:momSpectrum}.

Several cuts were made in order to reduce the background in the
photon pair sample.  Mismatched true photons, coming from two
different particles, were the main source of background. Other
sources of background included electrons, hadrons that deposited
energy in the EMCal, or secondary particles not coming from the
event vertex, all of which could lead to false combinatorial pairs
whose mass fell under the \piz\ mass peak, either in combination
with each other or with true photons from neutral pions.  The cuts
were:

\begin{itemize}
\item Minimum energy cut of 0.1 GeV in the PbSc, 0.2 GeV in the PbGl.  This cut
effectively eliminated pairs in which one photon carried nearly all
the energy and the other very little.  The same minimum energy cut
was used for the $\pi ^{0}$ $A_{LL}$ analysis, the results of which
were published in \cite{Adler:2004ps}.

\item Charged veto cut.  All EMCal clusters within a 10-cm radius of a
projected charged-track position onto the EMCal were excluded.

\item Shower shape cut in the PbSc to select clusters displaying the expected
shape for energy deposits from photon hits. The analogous
information for the PbGl was not available.
\end{itemize}

While a time-of-flight (TOF) cut could have potentially offered
additional hadron-photon discrimination, the EMCal TOF was not well
calibrated in the 2001-02 data set, so no timing cut was performed.

The background fraction, $r$, was obtained by taking the ratio of
the fitted background to the total number of pairs falling within
the mass peak. This fraction before and after the extra cuts were
performed is shown in Table~\ref{table:backgroundReduction}.
Significant reduction was achieved in the lowest two $p_T$ bins.

\begin{table}[tbp] \centering%
\begin{tabular}{|c|c|c|} \hline
$p_T$ (GeV/$c$) & \multicolumn{2}{|c|}{Background fraction (\%)} \\
\hline
 & Before cuts & After cuts \\ \hline
1-2 & 58 & 34 \\ \hline

2-3 & 23 & 12 \\ \hline

3-4 & 12 & 6 \\ \hline
4-5 &  9 & 5 \\
 \hline

\end{tabular}%
\caption[Reduction of background contribution.]{Reduction of
background contribution to the \piz\ mass peak, taken as 120-160
MeV/$c^2$, before and after background-removal cuts.}%
\label {table:backgroundReduction}
\end{table}%

\section{Asymmetry calculation}
\subsection{Overview}
The two counter-circulating RHIC beams are frequently referred to as
"blue" and "yellow," named after the colored stripes painted on the
respective magnet systems. The blue beam orbits clockwise, the
yellow beam counter-clockwise. Both beams are typically polarized,
as they were in the 2001-02 data-taking period. In order to make a
single-spin asymmetry measurement, the spin direction of the bunches
in only one beam was considered at a time, averaging over the spin
direction of the bunches in the other. Results from the blue and
yellow beams were obtained separately and subsequently combined.

The formula used to calculate asymmetry values is given in
Eq.~\ref{eq:lumiFormula},

\begin{equation}
\label{eq:lumiFormula}
  A_N = \frac{1}{P_{\textrm{beam}}}\frac{1}{\langle |\cos \varphi| \rangle}\frac{N^{\uparrow
}-\mathcal{R}N^{\downarrow }}{N^{\uparrow }+\mathcal{R}N^{\downarrow }} \\
\end{equation}
in which $P_{\textrm{beam}}$ is the beam polarization,
$\frac{1}{\langle |\cos \varphi| \rangle}$ is an azimuthal
acceptance correction factor (see below), $N^\uparrow$
($N^\downarrow$) is the neutral pion yield from crossings with the
polarized bunch spin up (down), and $\mathcal{R} =
\mathcal{L}^\uparrow / \mathcal{L}^\downarrow$ is the relative
luminosity between crossings having the polarized bunch with spin up
versus down. As $A_N$ is a left-right asymmetry,
Eq.~\ref{eq:lumiFormula} must be used separately for the two
detector arms. As given, it applies to yields to the left of the
polarized beam; an overall minus sign is required for yields
observed to the right of the polarized beam.  Asymmetry results for
the left and right detector arms were obtained separately and then
combined.  The correction factors and relative luminosity are
discussed further below.

\begin{figure}
\centering
\includegraphics[angle=0,height=0.5\textheight]{%
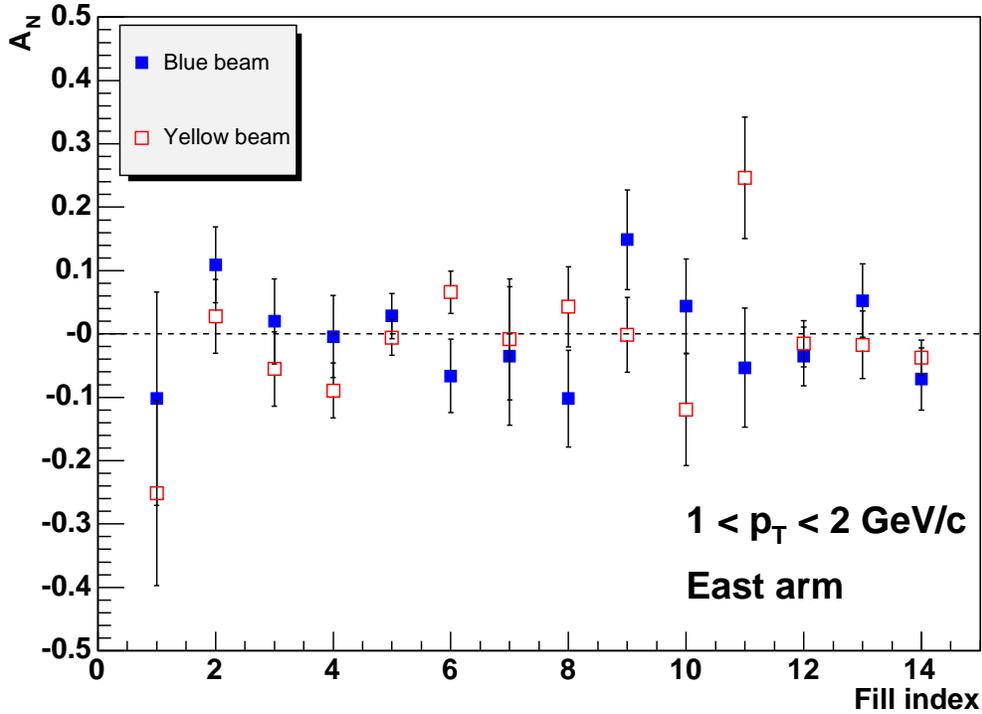} \caption[Fill-by-fill
asymmetries.]{Fill-by-fill asymmetry results for the two beams. }
\label{figure:asymVsFill}
\end{figure}

\begin{figure}
\centering
\includegraphics[angle=0,height=0.5\textheight]{%
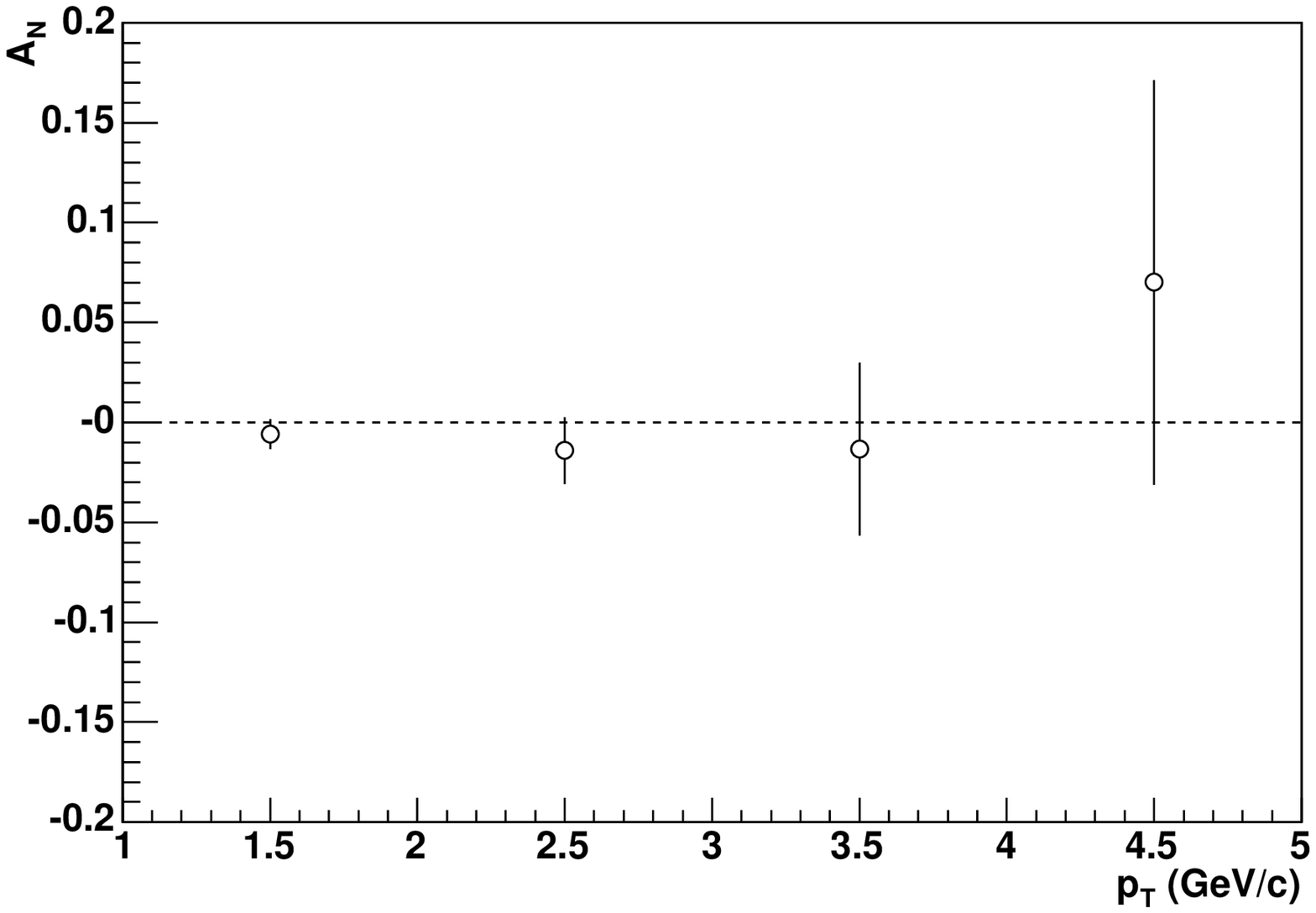} \caption[Combined results for both beams
and production to the left and right.]{Combined asymmetry results
for the two beams and particle production to the left and right. }
\label{figure:combinedLeftRight}
\end{figure}

The beam polarization varies fill by fill.  Thus the asymmetry is
determined for every fill, then averaged over all fills. An example
of fill-by-fill asymmetries is given in
Figure~\ref{figure:asymVsFill}; this figure also gives an indication
of the fill-by-fill stability of the measured asymmetry. Large
observed variation among fills could indicate systematic errors.
(Note that all uncertainties given in figures and tables are
statistical unless stated otherwise.)  Table \ref{table:lumi2x2}
shows results obtained from Eq.~\ref{eq:lumiFormula} for the two
detector arms for triggered events. Note that the uncertainties are
slightly smaller for the yellow beam due to higher average
polarization. Combined results for both detector arms are shown in
Figure~\ref{figure:combinedLeftRight}, and the values for both arms
and beams combined are given in Table~\ref{table:combinedLumi}. All
results given in this section are before correction to the
asymmetries for the asymmetry of the background, which is described
in Section~\ref{section:backgroundSubtraction}.

\begin{table}  [tbp] \centering
\begin{tabular}{|c|c|c|c|c|c|} \hline
 $p_T$ (GeV/$c$) & Beam & \multicolumn{2}{|c|}{Left} &
 \multicolumn{2}{|c|}{Right} \\ \hline
  & & $A_N$ & $\sigma_{A_N}$ & $A_N$ & $\sigma_{A_N}$  \\
\hline

1-2 & Blue & -0.008 & 0.015 & -0.012 & 0.019
 \\

 & Yellow & -0.005 & 0.015 & 0.003 & 0.011
 \\ \hline

2-3 & Blue & 0.028 & 0.035 & -0.066 & 0.039 \\

 & Yellow & -0.016 & 0.030 & -0.003 & 0.028 \\ \hline

3-4 & Blue & -0.101 & 0.094 & 0.106 & 0.099 \\

 & Yellow & 0.033 & 0.077 & -0.092 & 0.073 \\ \hline

4-5 & Blue & -0.02 & 0.23 & 0.16 & 0.22 \\

 & Yellow & 0.01 & 0.17 & 0.14 & 0.18 \\ \hline
\end{tabular}%
\caption[Asymmetry results.]{Asymmetry results for particles
observed in the left detector arm (west arm for blue, east arm for
yellow) and right detector arm (east for blue, west for yellow).}
\label{table:lumi2x2}
\end{table}%

\begin{table}  [tbp] \centering
\begin{tabular}{|c|c|c|} \hline
 $p_T$ (GeV/$c$) & $A_N$ & $\sigma_{A_N}$   \\
\hline

1-2 & -0.006 & 0.008
  \\ \hline

2-3 & -0.014 & 0.017
  \\ \hline

3-4 & -0.013 & 0.043
  \\ \hline

4-5 & 0.070 & 0.101
  \\ \hline

\end{tabular}
\caption[Combined results for both beams and production to the left
and right.]{Combined asymmetry results for the two beams and
particle production on the two sides of the polarized beam.}
\label{table:combinedLumi}
\end{table}

\subsection{Determination of relative luminosity}

The relative luminosity between crossings with spin-up bunches and
crossings with spin-down bunches for the polarized beam in
consideration was obtained from the number of MB events recorded by
the BBC. A typical value in this analysis was $\mathcal{R} =
\mathcal{L}^\uparrow / \mathcal{L}^\downarrow =$~1.09 for the yellow
beam, $\mathcal{R} =$~0.92 for the blue beam. Error in the relative
luminosity as measured in this case for azimuthal transverse
single-spin asymmetries could potentially come from an azimuthal
dependence of the BBC efficiencies.  The error would be proportional
to the physics asymmetry of the particles hitting the BBC (measured
to be $\lesssim 1$\%) times the difference in the efficiency of the
left and right halves of the detector. There is no current
measurement of this value, but it is expected to be small, also at
the level of a few times $10^{-2}$ or less. This would lead to a
potential systematic error on $\mathcal{R}$ on the order of a few
times $10^{-4}$, but this is only a rough estimate. Rather than
providing a quantitative error directly on the relative luminosity
measurement, an alternative method of asymmetry calculation that
does not rely upon measurement of the relative luminosity is used to
estimate the uncertainty on the asymmetry values calculated using
Eq.~\ref{eq:lumiFormula}. See Section~\ref{section:sqrtFormula} for
a description of this alternative method.  Refer to
Appendix~\ref{section:relLumi} for a discussion of potential sources
of error in the determination of the relative luminosity for
different single- and double-spin asymmetry measurements.

\subsection{Fill-by-fill polarization correction}

The average beam polarization in the 2001-02 run was $15\pm 5$\%,
with the 5\% representing a systematic scale uncertainty, discussed
below. Unpolarized protons in the beam act to dilute the physics
asymmetries being measured. It is necessary to correct for this
dilution, which is done by the factor of
$\frac{1}{P_{\textrm{beam}}}$ in Eq.~\ref{eq:lumiFormula}.

The beam polarization varies fill by fill and is typically different
for the two beams.  In the 2001-02 run, the yellow beam frequently
had slightly higher polarization than the blue beam.  The
polarization values for both beams for all fills included in this
analysis are given in Table~\ref{table:polarizationValues}.
Statistical uncertainties on the beam polarization were on the order
of $10^{-3}$ in absolute polarization.  These uncertainties were
negligible compared to the statistical uncertainties on the yields
as well as compared to the systematic uncertainty on the
polarization (see separate discussion below) and were not
incorporated into the final error on the asymmetry values.

\begin{table}\centering
\begin{tabular}{|c|c|c|} \hline
Fill & $P_{\textrm{blue}}$ & $P_{\textrm{yellow}}$ \\
\hline 2222 & 0.12 & 0.14 \\ \hline 2226 & 0.22 & 0.23
\\ \hline 2233 & 0.17 & 0.19 \\ \hline 2235 & 0.16 & 0.24 \\ \hline 2244 & 0.14 & 0.18 \\
\hline 2251 & 0.09 & 0.16 \\ \hline 2266 & 0.09 & 0.10
\\ \hline 2275 & 0.12 & 0.14 \\ \hline 2277 & 0.08 & 0.11 \\ \hline 2281 & 0.13 & 0.11
\\ \hline 2289 & 0.15 & 0.15 \\ \hline 2290 & 0.17 & 0.21 \\ \hline 2301 & 0.15 & 0.17 \\
\hline 2304 & 0.09 & 0.16 \\ \hline
\end{tabular}
\caption[Fill-by-fill beam polarization values.]{Fill-by-fill beam
polarization values for the 14 RHIC fills used.}
\label{table:polarizationValues}
\end{table}

\subsection{Acceptance correction}
\label{section:acceptanceCorrection}

The transverse single-spin asymmetry, $A_N$, is an azimuthal or
"left-right" asymmetry.  One can consider only particle production
to the left or right of the polarized beam and calculate the
asymmetry in production from spin-up versus spin-down bunches, as in
Eq.~\ref{eq:lumiFormula}.  Performing the calculation in this way,
detector acceptance effects cancel.  Alternatively, one can consider
only particle production from up- or down-polarized bunches and
calculate the asymmetry in production to the left versus the right,
as given in Eq.~\ref{eq:LRAsym}

\begin{equation}
\label{eq:LRAsym} A_N = \frac{1}{P_{\textrm{beam}}}\frac{1}{\langle
|\cos \varphi|
\rangle}\frac{N_{L}-\mathcal{R}_{\textrm{acc}}N_{R}}{N_{L}+\mathcal{R}_{\textrm{acc}}N_{R}}
\end{equation}
for bunch polarization in the upward direction. Here $N_{L}$ and
$N_{R}$ are the number of neutral pions produced in the left and
right detector arms with respect to the polarized beam direction,
and $\mathcal{R}_{\textrm{acc}} = \frac{\alpha_L}{\alpha_R}$ is the
relative acceptance of the left and right detector arms. In this
way, luminosity effects cancel. From Eq.~\ref{eq:LRAsym}, the
left-right nature of the asymmetry is clear.  However, maximal
effects, i.e.~the greatest and least particle production, are at
$90^\circ$ from the direction of the spin vector, which was vertical
in the entire 2001-02 run.  Thus integrating particle yields over
the entire azimuthal coverage of the central arm spectrometers would
lead to a dilution of the true physics asymmetry.  The factor of
$\frac{1}{\langle |\cos \varphi| \rangle}$ in both
Eq.~\ref{eq:lumiFormula} and \ref{eq:LRAsym} corrects for this
dilution.  Note that $\varphi = 0^\circ$ is in the horizontal plane,
implying no dilution at $\varphi = 0^\circ$ or $\varphi =
180^\circ$.

The value of $\langle |\cos \varphi| \rangle $ over the idealized
azimuthal coverage of the detector arms can be calculated
analytically by Eq.~\ref{eq:phiIntegral}.

\begin{equation}
\label{eq:phiIntegral}
 \langle |\cos \varphi| \rangle =\frac{\int
|\cos \varphi | d\varphi}{\int d\varphi}
\end{equation}
To account more carefully for dead areas in the EMCal and in the
ERT, which would require detailed attention to the distribution of
these areas in order to make an analytical calculation, the average
values of $|\cos \varphi |$ used in the analysis were calculated
directly from the data using Eq.~\ref{eq:phiSum}, where $j$ is a sum
over all photon pairs with an invariant mass which fell under the
\piz\ mass peak. A comparison of results for $\langle |\cos \varphi
|\rangle $ determined analytically and from the data is given in
Table~\ref{table:cosPhi}.

\begin{equation}
\label{eq:phiSum}
 \langle |\cos \varphi |\rangle =\frac{\sum
\limits_{j=1}^{N} |\cos \varphi _{j}|}{N}
\end{equation}

\begin{table}[tbp] \centering%
\begin{tabular}{|c|c|c|c|}
\hline $\langle |\cos \varphi |\rangle $ & Ideal & Actual: MB &
Actual: \trig
\\ \hline
West Arm & 0.943 & 0.955 & 0.949 \\
East Arm & 0.883 & 0.880 & 0.874 \\
Both Arms & 0.909 & 0.920 & 0.913 \\ \hline
\end{tabular}%
\caption[Azimuthal acceptance correction factors.]{Comparison of
results for the average value of $|\cos \varphi |$ for ideal and
actual detector acceptances. Results for the west and east arms are
significantly different because the uppermost sector was not
included in the west.  Differences between the MB and triggered data
are due to the distribution of dead or masked tiles in the ERT.}
\label{table:cosPhi}
\end{table}

\subsection{Subtraction of background asymmetry}
\label{section:backgroundSubtraction}

Subtraction of the asymmetry of the background is performed as given
in Eq.~\ref{eq:backgroundSubtraction}, taking into account the
fraction of background under the $\pi^{0}$ mass peak.
$A_N^{\textrm{peak}}$ indicates the asymmetry of all photon pairs
falling under the \piz\ peak; note that it has generally been
written simply as $A_N$ up until this point.
Equation~\ref{eq:backgroundSubtractionError} gives the prescription
for calculation of the final statistical uncertainty on the \piz\
asymmetry after subtraction of the background asymmetry.

\begin{equation}
A_{N}^{\pi^{0}}=\frac{A_{N}^{\textrm{peak}}-rA_{N}^{\textrm{bg}}}{1-r}
\label{eq:backgroundSubtraction}
\end{equation}

\begin{equation}
\sigma _{A_{N}^{\pi ^{0}}}=\frac{\sqrt{\sigma
_{A_{N}^{\textrm{peak}}}^{2}+r^{2}\sigma
_{A_{N}^{\textrm{bg}}}^{2}}}{1-r}
\label{eq:backgroundSubtractionError}
\end{equation}
The same technique to handle background in a \piz\ asymmetry
analysis was used for the longitudinal double-spin asymmetry
\cite{Adler:2004ps}. Refer back to
Table~\ref{table:backgroundReduction} for the fraction of background
in each $p_{T}$ bin.  After cuts to reduce the background, it ranged
from 34\% in the 1-2 GeV/$c$ $p_T$ bin to 5\% in the 4-5 GeV/$c$
bin.

It is not possible to measure the asymmetry of the background under
the peak directly.  Differentiation between true neutral pions and
combinatorial background is only possible statistically and not on
an event-by-event basis.  Therefore it must be estimated in order to
correct for it. The asymmetries of two different background
invariant-mass regions were studied as estimates of the asymmetry of
the background under the \piz\ peak: 50-MeV/$c^2$ regions
immediately around the $\pi^0$ mass peak (60-110 MeV/$c^2$ and
170-220 MeV/$c^2$), and the invariant-mass region above the $\pi^0$
but below the $\eta$ (250-450 MeV/$c^2$). The final results
published in \cite{Adler:2005in} subtracted the asymmetry of the two
50-MeV/$c^2$ regions.  It was felt that this invariant-mass region,
being closer to that directly under the peak, was likely to reflect
the asymmetry of background under the peak more accurately. However,
the background region used was found to have little effect on the
final results. The similarity in the background asymmetries for the
two different invariant-mass regions lent confidence to their
validity in estimating the asymmetry of the background under the
\piz\ peak. See Section~\ref{section:backgroundStudy} for further
comparison of the results obtained from the two different background
regions.

\begin{figure}
\centering
\includegraphics[angle=0,height=0.5\textheight]{%
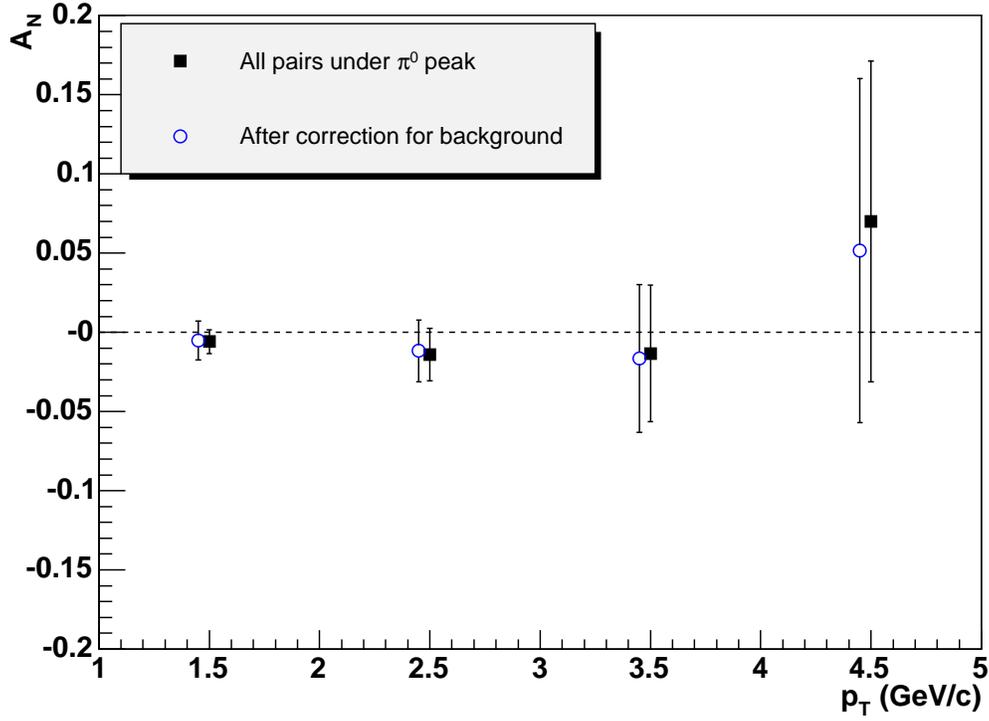} \caption[Asymmetry results before and
after correction for background.]{Asymmetry results before and after
correction for the asymmetry of the background in the invariant-mass
regions immediately surrounding the \piz\ peak. Background-corrected
points are shifted down by 50 MeV/$c$ from the center of the bin for
readability.} \label{figure:peakSideband}
\end{figure}

The asymmetries measured for photon pairs falling in the background
invariant-mass region immediately surrounding the peak are shown in
Table~\ref{table:sidebands}.  The asymmetries after
background-asymmetry subtraction for the two 50-MeV/$c^2$ regions
are given in Table~\ref{table:subtractedSidebands}.  The asymmetry
results before and after subtraction of the background asymmetry are
shown in Figure~\ref{figure:peakSideband}.  It can be seen that
correction for the background asymmetry made only a small difference
in the results.  The mean $p_{T}$ in each bin was adjusted to
account for possible differences between the mean $p_T$ of the true
neutral pions and the background pairs under the peak. The final
value is calculated by Eq.~\ref{eq:meanPt}
\begin{equation}
p_{T}^{\pi^{0}}=\frac{p_T^{\textrm{peak}}-rp_{T}^{\textrm{bg}}}{1-r}
\label{eq:meanPt}
\end{equation}
and given for each $p_T$ bin in Table~\ref{table:meanPtFinal}.

\begin{table}  [tbp] \centering
\begin{tabular}{|c|c|c|} \hline
 $p_T$ (GeV/$c$) & $A_N^{\textrm{bg}}$ & $\sigma_{A_N^{\textrm{bg}}}$ \\
\hline

1-2 & -0.007 & 0.009  \\ \hline

2-3 & -0.031 & 0.034
  \\ \hline

3-4 & 0.036 & 0.123
  \\ \hline

4-5 & 0.42 & 0.39
  \\ \hline

\end{tabular}
\caption[Background asymmetries.]{Asymmetry results of background
photon pairs falling within 50-MeV/$c^2$ regions around the
$\pi^{0}$ mass peak.} \label{table:sidebands}
\end{table}

\begin{table}  [tbp] \centering
\begin{tabular}{|c|c|c|} \hline
 $p_T$ (GeV/$c$) & $A_N^{\pi^0}$ & $\sigma_{A_N^{\pi^0}}$  \\
\hline

1-2 & -0.005 & 0.012
 \\ \hline

2-3 & -0.012 & 0.020
 \\ \hline

3-4 & -0.016 & 0.047
 \\ \hline

4-5 & 0.052 & 0.109
 \\ \hline

\end{tabular}
\caption[Background-subtracted \piz\
asymmetries.]{Background-subtracted \piz\ asymmetries, using
50-MeV/$c^2$ mass regions around the $\pi^{0}$ peak as the
background.}
\label{table:subtractedSidebands}%
\end{table}%

\begin{table}[tbp] \centering%
\begin{tabular}{|c|c|c|} \hline
\multicolumn{3}{|c|}{$\langle p_T \rangle$ (GeV/$c$)} \\ \hline

$\pi ^{0}$ peak & Background & Final \\ \hline

1.40 & 1.31 & 1.45 \\ \hline

2.34 & 2.28 & 2.34 \\ \hline

3.35 & 3.33 & 3.36 \\ \hline

4.38 & 4.37 & 4.38 \\ \hline

\end{tabular}%
\caption[Mean $p_T$ values of the background and after correction
for background.] {Mean $p_T$ values of the background and of neutral
pions after correction for background.} \label{table:meanPtFinal}
\end{table}%

\subsection{Calculation of statistical uncertainties}

\begin{figure}
\centering
\includegraphics[angle=0,height=0.4\textheight]{%
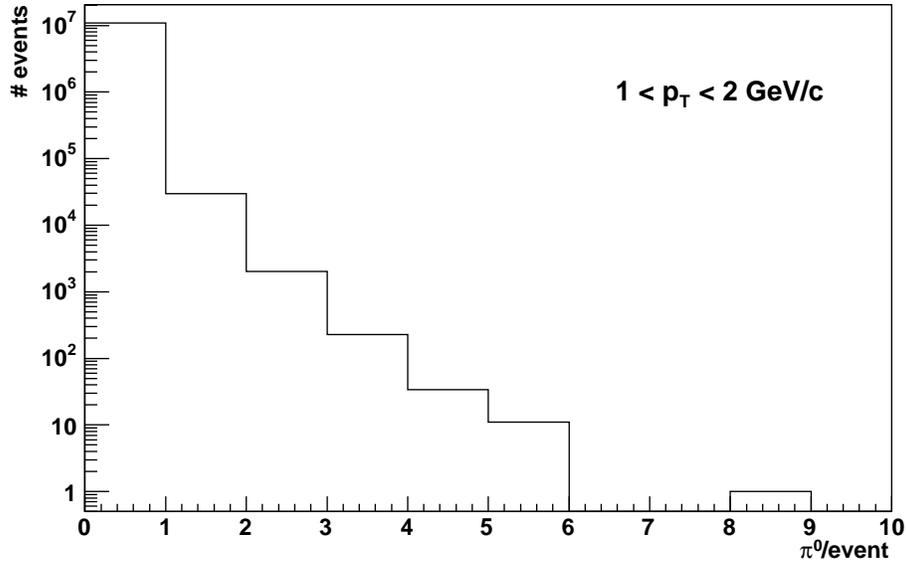} \caption[$\pi^0$ multiplicity per MB event,
$1 < p_T < 2$~GeV/$c$]{Number of photon pairs per MB event in
$\pi^0$ invariant-mass range, $1 < p_T < 2$~GeV/$c$, west arm. }
\label{figure:mult1To2GeV}
\end{figure}

\begin{figure}
\centering
\includegraphics[angle=0,height=0.4\textheight]{%
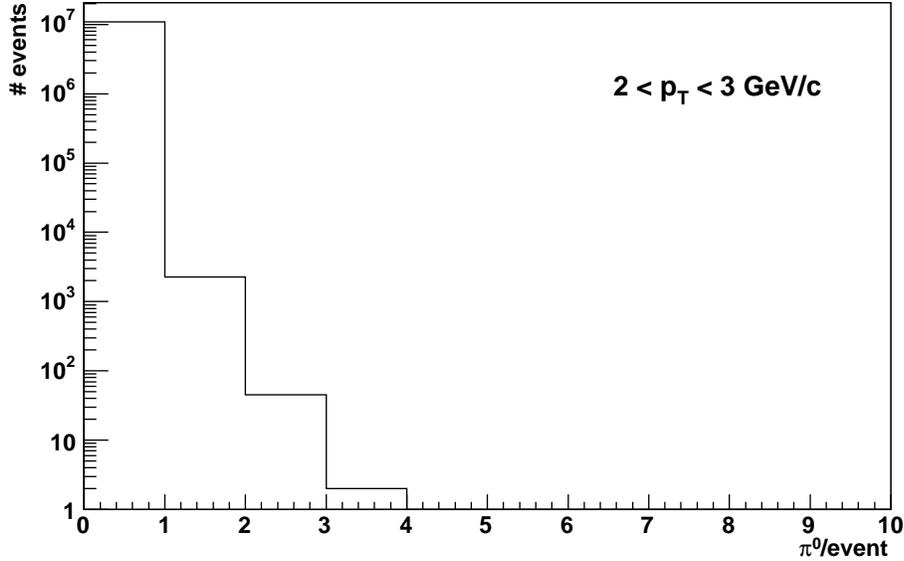} \caption[$\pi^0$ multiplicity per MB event,
$2 < p_T < 3$~GeV/$c$]{Number of photon pairs per event in $\pi^0$
invariant-mass range, $2 < p_T < 3$~GeV/$c$, west arm. }
\label{figure:mult2To3GeV}
\end{figure}

In order to calculate the statistical error, accounting for the fact
that multiple neutral pions could be produced per collision, the
multiplicity distributions of $\pi^{0}$'s per event for the
different detector arms were determined.  See Figures
\ref{figure:mult1To2GeV} and \ref{figure:mult2To3GeV} for sample
multiplicity distributions in the west arm. Note that these
distributions in fact indicate the multiplicity per event of photon
pairs with an invariant mass between 120 and 160 MeV/$c^{2}$, thus
including both real $\pi^{0}$'s as well as false combinatorial
pairs.  From these multiplicity distributions, the degree to which
the distribution is non-Poisson was calculated in a simplistic way
by taking the ratio of the distribution's RMS to the square root of
its mean. For a Poisson distribution, this ratio would be 1.

Table \ref{table:multFactors} shows the degree to which the photon
pair yield is non-Poisson, calculated as $\sigma _{k}/\sqrt{<k>}$,
where $<k>$ is the mean number of pairs per triggered event and
$\sigma _{k}$ is the RMS of this multiplicity distribution.  These
values were determined for the MB data, and then the uncertainty on
each individual yield, $N$, was taken to be $ \frac{\sigma
_{k}}{\sqrt{<k>}}\sqrt{N}$.  The 4-5 GeV/$c$ background bin suffers
from very low statistics. Rather than using the calculated value of
1.41 for the west arm, 1.04, the same as for the 3-4 GeV/$c$
background bin, was used.

\begin{table}[tbp] \centering%
\begin{tabular}{|c|c|c|c|c|} \hline

$p_T$ (GeV/$c$) & \multicolumn{2}{|c|}{$\pi^0$ peak} & \multicolumn{2}{|c|}{Background} \\
\hline
    & West & East & West & East \\ \hline

1-2 & 1.08 & 1.06 & 1.16 & 1.11 \\ \hline

2-3 & 1.02 & 1.02 & 1.08 & 1.05 \\ \hline

3-4 & 1.01 & 1.01 & 1.04 & 1.02 \\ \hline

4-5 & 1.02 & 1.00 & 1.41 & 1.00 \\ \hline

\end{tabular}%
\caption[Deviation of particle production from a Poisson
distribution.]{Degree to which the yield is non-Poisson for MB
photon pairs falling in the 120-160 MeV/$c^2$ and (60-110 or 170-220
MeV/$c^2$) invariant-mass regions.}%
\label{table:multFactors}
\end{table}%

\subsection{Asymmetry scale uncertainty}

In the 2001-02 run, only the $p$C polarimeter was available in RHIC.
As stated in Section~\ref{section:polarimetry}, the analyzing power
in the process utilized by the $p$C polarimeter to measure the beam
polarization was originally measured by AGS experiment E950 to $\pm
30\%$ \cite{Alekseev:2002ym}.  The total systematic error on the
measurement of the beam polarization was derived from a relative
systematic uncertainty on the RHIC beam measurement of 15\%, the
30\% relative uncertainty on the analyzing power of the process, and
a relative uncertainty of 10\% in the change in analyzing power from
a beam energy of 22 GeV at the AGS to 100 GeV at RHIC
\cite{Jinnouchi:2003cp}.  Adding these uncertainties in quadrature
gave a total relative systematic uncertainty on the beam
polarization of $\pm 35\%$.  This uncertainty is a scale
uncertainty; it affects asymmetry values and statistical errors,
generally proportional to
$\frac{1}{P_{\textrm{beam}}}\frac{1}{\sqrt{N}}$, in the same way,
preserving the significance of each point from zero.

\section{Studies and checks}
\subsection{Alternative asymmetry calculation}
\label{section:sqrtFormula}

As described above in Section~\ref{section:acceptanceCorrection},
azimuthal transverse single-spin asymmetries can be considered for a
single polarization direction comparing particle production to the
left and right of the beam, or for particle production on a single
side of the beam comparing different polarization directions.
Equation~\ref{eq:sqrtFormula} combines yields from up- and
down-polarized bunches and from the left and right halves of the
detector such that systematic errors are reduced.
\begin{equation}\label{eq:sqrtFormula}
A_{N} =\frac{1}{P_{\textrm{beam}}}\frac{1}{\langle |\cos \varphi|
\rangle}\frac{\sqrt{N_{L}^{\uparrow }N_{R}^{\downarrow }}-
\sqrt{N_{L}^{\downarrow }N_{R}^{\uparrow }}}{\sqrt{N_{L}^{\uparrow
}N_{R}^{\downarrow }}+\sqrt{N_{L}^{\downarrow }N_{R}^{\uparrow }}}
\end{equation}
In particular, the acceptance and luminosity asymmetries cancel out
to several orders. See \cite{Spinka:1999vv} for a detailed
discussion of this and other methods of calculation for transverse
single-spin asymmetries.  It should be noted that while
Eq.~\ref{eq:lumiFormula} is mathematically exact,
Eq.~\ref{eq:sqrtFormula} is an approximation, albeit an excellent
one for the purposes of this analysis. It should also be noted that
Eq.~\ref{eq:sqrtFormula} is only suitable for transverse single-spin
analysis, while Eq.~\ref{eq:lumiFormula} has an equivalent for
longitudinal double-spin analysis (see
Eq.~\ref{eq:lumiFormulaDoubleLongitudinal}).

The asymmetry is determined for every fill, then averaged over all
fills, as in the calculations using Eq.~\ref{eq:lumiFormula}.
Table~\ref{table:sqrtAsym} shows the results obtained from
Eq.~\ref{eq:sqrtFormula} for triggered events.

\begin{table}  [tbp] \centering
\begin{tabular}{|c|c|c|c|} \hline
 $p_T$ (GeV/$c$) & Beam & $A_N$ & $\sigma_{A_N}$   \\
\hline

1-2 & Blue & -0.010 & 0.012 \\

 & Yellow & -0.001 & 0.009  \\ \hline

2-3 & Blue & -0.017 & 0.026  \\

 & Yellow & -0.009 & 0.020  \\ \hline

3-4 & Blue & -0.002 & 0.068  \\

 & Yellow & -0.032 & 0.053  \\ \hline

4-5 & Blue & 0.06 & 0.16  \\

 & Yellow & 0.08 & 0.12  \\ \hline
\end{tabular}
\caption[Asymmetry results, alternative calculation.]{Asymmetry
results as determined by Eq.~\ref{eq:sqrtFormula}.}
\label{table:sqrtAsym}
\end{table}

\begin{figure}
\centering
\includegraphics[angle=0,height=0.5\textheight]{%
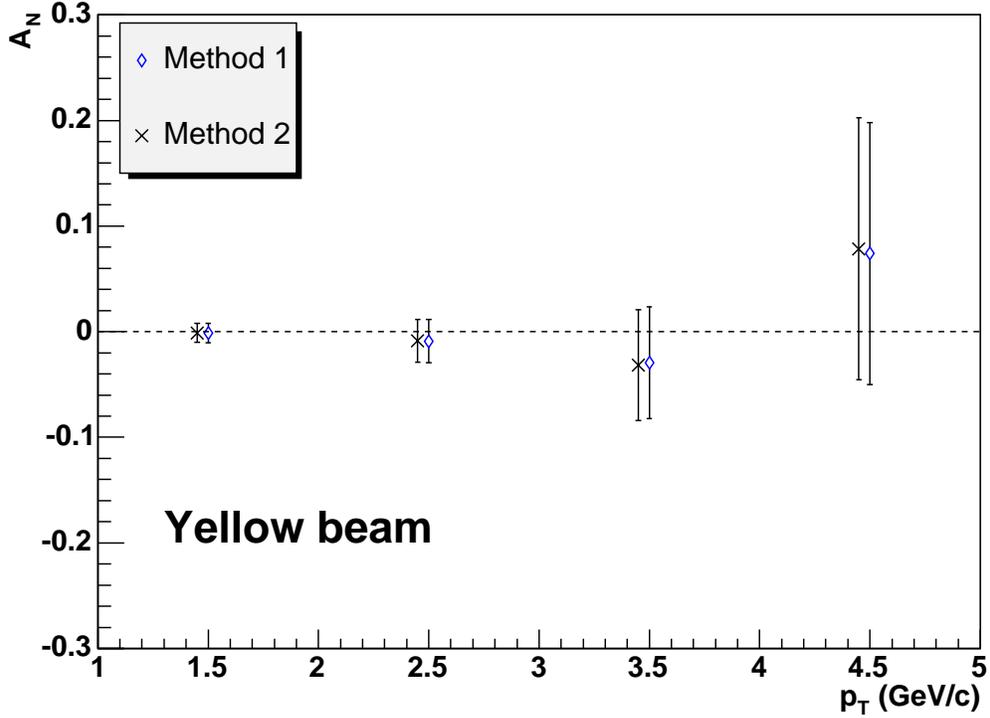} \caption[Comparison of two methods of asymmetry
calculation.]{Comparison of asymmetry results obtained using
Eq.~\ref{eq:lumiFormula} (Method 1) and Eq.~\ref{eq:sqrtFormula}
(Method 2), shown here for the yellow beam. Points for Method 2 are
shifted down by 50 MeV/$c$ from the center of the bin for
readability.} \label{figure:sqrtLumi}
\end{figure}

The results of the asymmetry calculations obtained using
Eq.~\ref{eq:lumiFormula} and Eq.~\ref{eq:sqrtFormula} can be seen
together in Figure~\ref{figure:sqrtLumi}.  The figure is for the
yellow beam; results for the blue beam are similar.  The two methods
agree extremely well. The dominant systematic uncertainty in the
results from Eq.~\ref{eq:lumiFormula} is expected to be from the
determination of the relative luminosity; therefore, systematic
uncertainties are calculated from a direct quantitative comparison
of the asymmetry results obtained from these two methods of
calculation.

A systematic uncertainty, $\sigma_{sys}$, was calculated for each
bin from $A_N^{1} - A_N^{2} \equiv \Delta$, $\sigma_{A_N^{1}}$, and
$\sigma_{A_N^{2}}$.  The index '1' refers to results from
Eq.~\ref{eq:lumiFormula}; '2' to results from
Eq.~\ref{eq:sqrtFormula}.  The uncertainty on the difference was
calculated following a prescription for results obtained by applying
two different methods to data sets that are 100\% correlated
(exactly the same data) \cite{Barlow:2002yb}. In this case,
$\sigma_{\Delta} = \sqrt{| \sigma_{A_N^{1}}^2 + \sigma_{A_N^{2}}^2
-2\rho \sigma_{A_N^{1}} \sigma_{A_N^{2}}|}$, with the correlation
$\rho = 1$.  $\sigma_{sys} = |\Delta| - \sigma_{\Delta}$ was taken
in cases where $|\Delta|
> \sigma_{\Delta}$, and $\sigma_{sys} = 0$ was taken in cases where
$|\Delta| < \sigma_{\Delta}$.  See
Tables~\ref{table:lumiSqrtComparisonBlue} and
\ref{table:lumiSqrtComparisonYellow} for the values used in the
calculation and the results. The systematic error was also
calculated as the square root of the difference of the squares,
$\sigma_{sys} = \sqrt{|\Delta|^{2} - \sigma_{\Delta}^{2}}$, to see
if it affected the results.  At the level of $10^{-3}$, which is the
precision quoted in \cite{Adler:2005in}, it did not.  Final
systematic uncertainties calculated using this method, averaged over
the two beams, are shown in Table~\ref{table:sysError}.

\begin{table}
\begin{tabular}{|c|c|c|c|c|c|c|c|} \hline
$p_T$\  & $A_N^{1}$  & $A_N^{2}$  & $\Delta$ & $\sigma_{A_N^{1}}$  & $\sigma_{A_N^{2}}$ & $\sigma_{\Delta}$ & $\sigma_{sys}$\\
(GeV/$c$) & ($\times 10^2$) & ($\times 10^2$) & ($\times 10^2$) &
($\times 10^2$) & ($\times 10^2$) & ($\times 10^2$) & ($\times
10^2$)  \\ \hline  1-2       & -0.291 & -0.292 & 0.001 & 0.816 &
0.816 & 0.000 & 0.001 \\ \hline 2-3       & -1.79  & -1.79  & 0.00 &
2.07  & 2.07  & 0.00  & 0.00 \\ \hline 3-4       &  1.61  &  1.61 &
0.00  & 5.51  & 5.51  & 0.00  & 0.00 \\ \hline 4-5       &  4.23 &
4.17  & 0.06  & 13.0  &13.0   & 0.00  & 0.06 \\ \hline
\end{tabular}
\caption[Agreement of asym. results from the two methods, blue
beam.]{Agreement of asymmetry results from the two methods of
calculation, blue beam. Note that these values were obtained before
final cuts were performed.  See text for further explanation.}
\label{table:lumiSqrtComparisonBlue}
\end{table}

\begin{table}
\begin{tabular}{|c|c|c|c|c|c|c|c|} \hline
$p_T$  & $A_N^{1}$  & $A_N^{2}$  & $\Delta$ & $\sigma_{A_N^{1}}$  & $\sigma_{A_N^{2}}$ & $\sigma_{\Delta}$ & $\sigma_{sys}$\\
(GeV/$c$) & ($\times 10^2$) & ($\times 10^2$) & ($\times 10^2$) &
($\times 10^2$) & ($\times 10^2$) & ($\times 10^2$) & ($\times
10^2$)  \\ \hline

1-2       &  -0.974 & -0.975 &  0.001 & 0.636 &  0.635 & 0.001 &
0.000 \\ \hline

2-3       &  -2.50  & -2.50 &  0.00  &  1.61  & 1.61  & 0.00  & 0.00
\\ \hline

3-4       & -3.29  & -3.30  &  0.01 &  4.27  &  4.28  & 0.01  & 0.00
\\ \hline

4-5       &   6.81  & 6.99   & -0.18  & 10.0 & 10.1   & 0.1   & 0.1
\\ \hline
\end{tabular}
\caption[Agreement of asym. results from the two methods, yellow
beam.]{Agreement of asymmetry results from the two methods of
calculation, yellow beam.  The data samples are 100\% correlated, so
agreement at better than the statistical level is expected. Note
that these values were obtained before final cuts were performed.
See text for further explanation.}
\label{table:lumiSqrtComparisonYellow}
\end{table}

\begin{table} \centering
\begin{tabular}{|c|c|} \hline
$p_T$   & Avg $\sigma_{sys}$ \\
(GeV/$c$) & ($\times 10^2$) \\ \hline

1-2       & 0.000 \\ \hline

2-3       & 0.00  \\ \hline

3-4       & 0.00  \\ \hline

4-5 & 0.1 \\ \hline
\end{tabular}
\caption[Systematic uncertainty on the asymmetry.]{Systematic
uncertainty on the neutral pion asymmetry, calculated for each bin
and averaged over both beams.} \label{table:sysError}
\end{table}

While the comparison of the asymmetry results from the two different
methods of calculation provided a strong check on the relative
luminosity, the main identified potential source of systematic error
present in the asymmetries as calculated by
Eq.~\ref{eq:lumiFormula}, a number of other checks on the results
were performed.  These other checks are described in the following
sections.

\subsection{Left and right detector arms}

\begin{figure}
\centering
\includegraphics[angle=0,height=0.5\textheight]{%
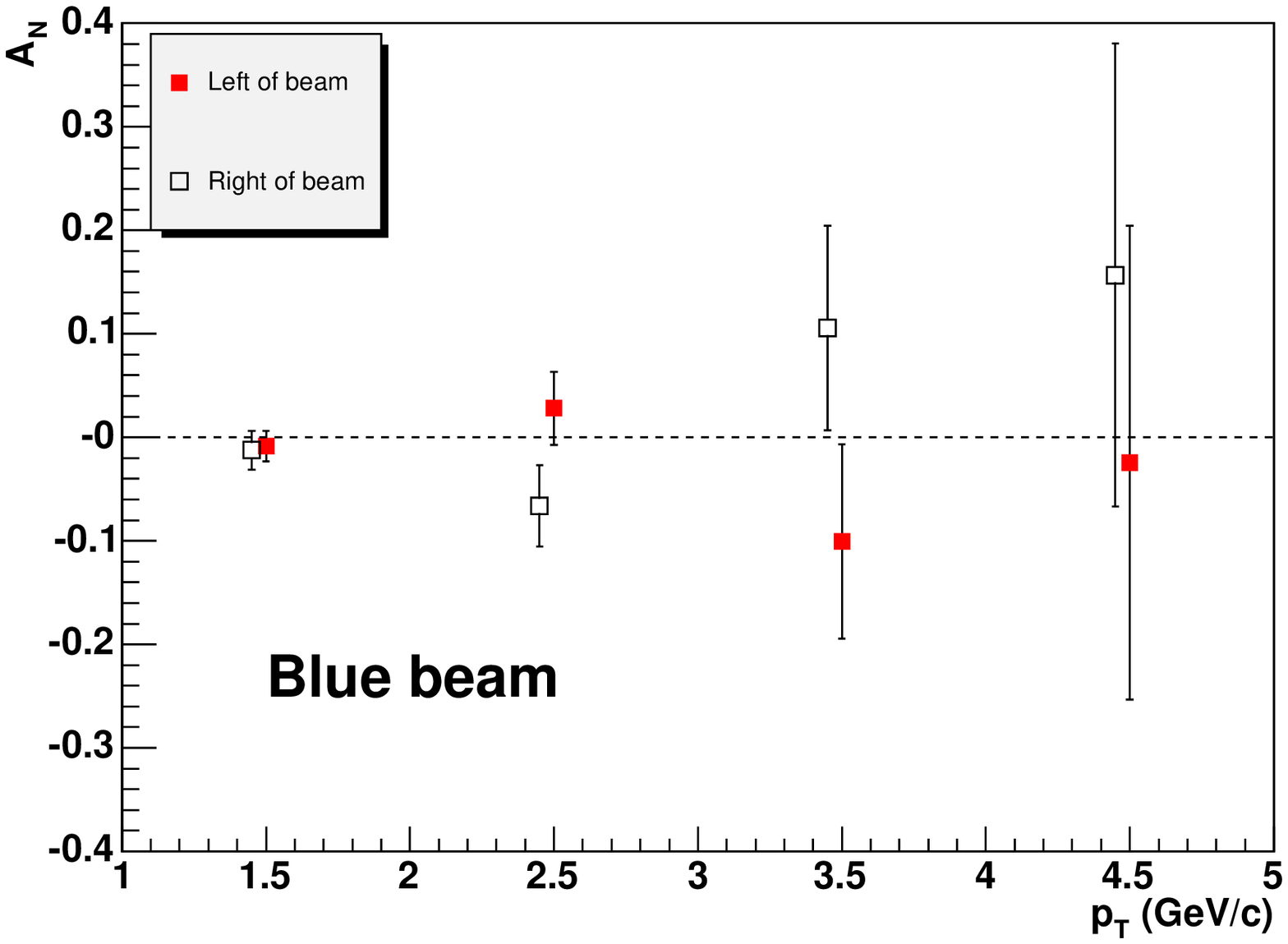} \caption[Asym. comparison to the left and the
right, blue beam polarized.]{Comparison of results for particle
production to the left and right side, blue beam polarized. Points
for the right of the beam are shifted down by 50 MeV/$c$ from the
center of the bin for readability. Statistical agreement is
expected.} \label{figure:leftRightAgreementBlue}
\end{figure}

\begin{figure}
\centering
\includegraphics[angle=0,height=0.5\textheight]{%
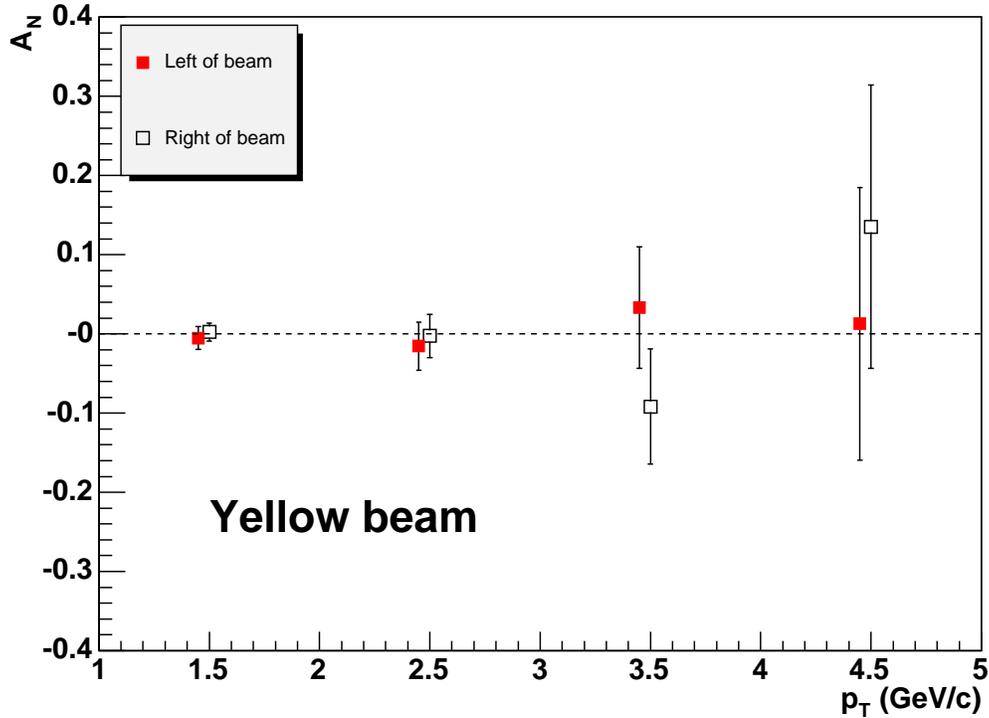} \caption[Asym. comparison to the left and the
right, yellow beam polarized.]{Comparison of results for particle
production to the left and right side, yellow beam polarized.
Points for the right of the beam are shifted down by 50 MeV/$c$ from
the center of the bin for readability. Statistical agreement is
expected.} \label{figure:leftRightAgreementYellow}
\end{figure}

Calculating the asymmetry using Eq.~\ref{eq:lumiFormula} necessarily
gives separate results for the left and right sides of the polarized
beam.  There is no overlap in the particle yields from the two
detector arms; therefore, the expected agreement between the two
results is that for uncorrelated samples. The uncertainty on the
difference in results is taken to be $\sigma_{\Delta} = \sqrt{|
\sigma_{A_N^{\textrm{left}}}^2 + \sigma_{A_N^{\textrm{right}}}^2
-2\rho \sigma_{A_N^{\textrm{left}}}
\sigma_{A_N^{\textrm{right}}}|}$, with the correlation $\rho = 0$.
As can be seen from Table~\ref{table:leftRightAgreement}, the
difference in the results from the two detector arms was within the
uncertainty on the difference, and the results were in agreement.
Results for both detector arms can be seen together for the blue and
yellow beams in Figures~\ref{figure:leftRightAgreementBlue} and
\ref{figure:leftRightAgreementYellow}, respectively.

\begin{table} \centering
\begin{tabular}{|c|c|c|c|c|c|c|}
  \hline
  $p_T$ (GeV/$c$) & $A_N^{\textrm{left}}$ & $A_N^{\textrm{right}}$ & $\Delta$ & $\sigma^{\textrm{left}}$ & $\sigma^{\textrm{right}}$ & $\sigma_\Delta$ \\
  \hline

  1-2 & -0.014 & -0.007 & -0.007 & 0.011 & 0.009 & 0.0142 \\
  \hline

  2-3 & -0.034  & -0.021 & -0.013 & 0.026 & 0.024  & 0.0354 \\
  \hline

  3-4 &  0.002  & -0.072 & 0.074 & 0.068 & 0.065 & 0.0941 \\
  \hline

  4-5 &  0.01  &  0.13 & -0.12 & 0.15 & 0.16  &  0.219 \\
  \hline
\end{tabular}
\caption[Agreement of results for production to the left and
right.]{Evaluation of agreement of asymmetry results from particle
production to the left and right of the polarized beam, shown for
the yellow beam. $\Delta$ is the difference between the asymmetries
for the two arms. $\sigma_\Delta$ is the uncertainty on the
difference.} \label{table:leftRightAgreement}
\end{table}

\subsection{Two independent beams}

\begin{figure}
\centering
\includegraphics[angle=0,height=0.5\textheight]{%
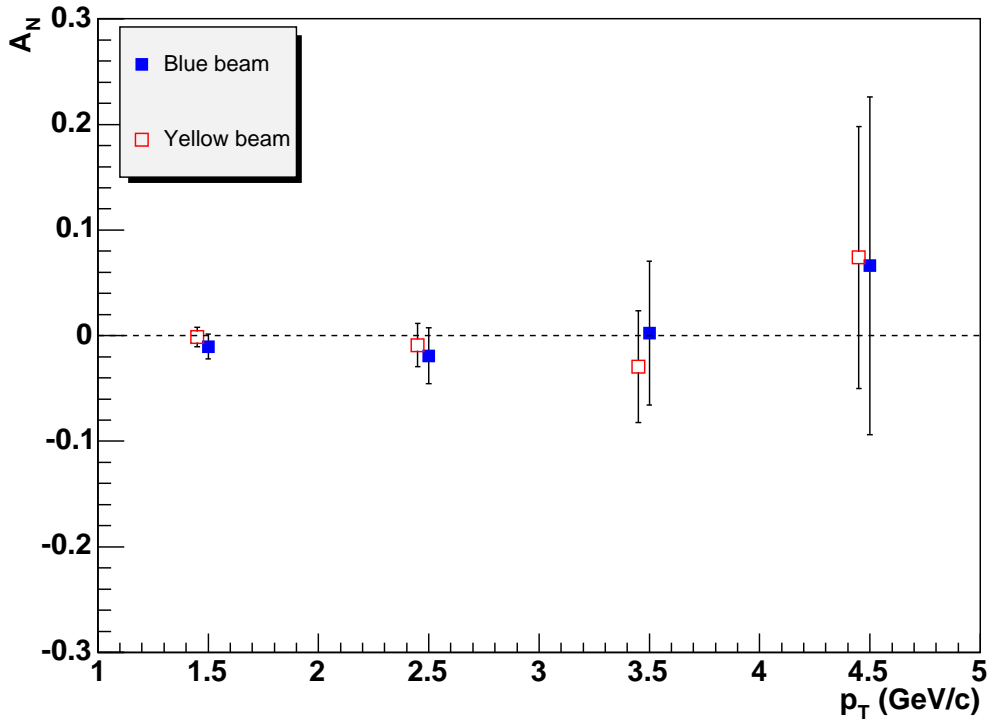} \caption[Asym. comparison for the two polarized
beams.]{Comparison of results obtained for the two polarized beams.
Points for the yellow beam are shifted down by 50 MeV/$c$ from the
center of the bin for readability. Statistical agreement or better
is expected.} \label{figure:blueYellowAgreement}
\end{figure}

While the results for the blue and yellow beams use the same events
and yields, they are combined in a different way, taking into
account the spin direction of either one beam or the other, leading
to (nearly) statistically independent measurements.  As the
correlation between results from the two beams is believed to be
small but is unknown, it is assumed to be zero in evaluating the
agreement of the results. Thus as for the case of comparing results
from the two detector arms, the uncertainty on the difference of the
results from the two beams is taken to be $\sigma_{\Delta} = \sqrt{|
\sigma_{A_N^{\textrm{blue}}}^2 + \sigma_{A_N^{\textrm{yellow}}}^2
-2\rho \sigma_{A_N^{\textrm{blue}}}
\sigma_{A_N^{\textrm{yellow}}}|}$, with $\rho = 0$.  It can be seen
in Table~\ref{table:blueYellowAgreement} that the results from the
two beams agree as expected.  Results for both beams can be seen
together and evaluated by eye in
Figure~\ref{figure:blueYellowAgreement}.

\begin{table} \centering
\begin{tabular}{|c|c|c|c|c|c|c|}
  \hline
  $p_T$ (GeV/$c$) & $A_N^{\textrm{blue}}$ & $A_N^{\textrm{yellow}}$ & $\Delta$ & $\sigma^{\textrm{blue}}$ & $\sigma^{\textrm{yellow}}$ & $\sigma_\Delta$ \\
  \hline
  1-2 & -0.01035 & -0.00129 & -0.00906 & 0.01184 & 0.00922 & 0.01501 \\
  \hline
  2-3 & -0.0190 &  -0.0090 &  -0.0100 &  0.0264 &  0.0205 &  0.03342 \\
  \hline
  3-4 &  0.0026 &  -0.0292 &   0.0318 &  0.0681 &  0.0529 &  0.08623 \\
  \hline
  4-5 &  0.0662 &   0.0741 &  -0.0079 &  0.160  &  0.124  &  0.20243 \\
  \hline
\end{tabular}
\caption[Agreement of results for individual beams.]{Evaluation of
agreement of asymmetry results from blue and yellow beams. $\Delta$
is the difference between the asymmetries for the two beams.
$\sigma_\Delta$ is the uncertainty on the difference.}
\label{table:blueYellowAgreement}
\end{table}

\subsection{Triggered and minimum-bias data}

\begin{figure}
\centering
\includegraphics[angle=0,height=0.5\textheight]{%
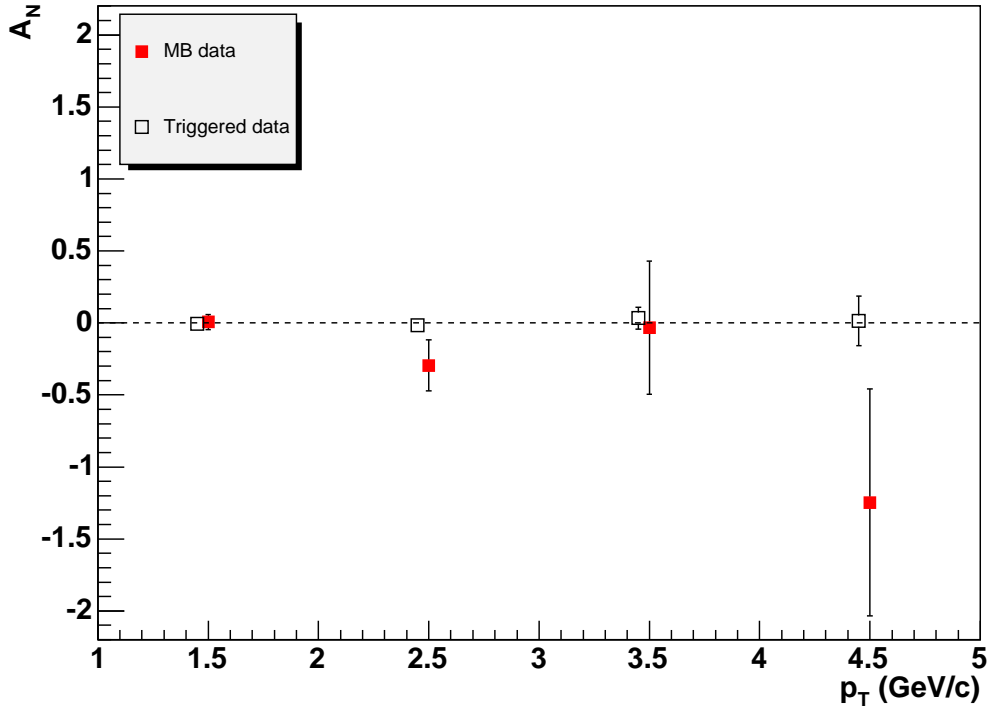} \caption[Asym. comparison for minimum-bias and triggered
events.]{Comparison of results for minimum-bias and triggered
events, shown here for the yellow beam and east detector arm. Points
for the triggered data are shifted down by 50 MeV/$c$ from the
center of the bin for readability. The correlation between the
samples is unknown.} \label{figure:bbc2x2Agreement}
\end{figure}

Results for the physics asymmetry, $A_N$, can also be compared for
the triggered and minimum-bias data samples.  In this case the
correlation between the samples is poorly understood.  In the higher
$p_T$ bins, where the trigger was more efficient, nearly all pions
in the MB sample should have fired the trigger and been present in
the triggered sample as well, making the MB sample nearly a direct
subset of the triggered sample. In the lower $p_T$ bins, this
correlation should be lower but still significant.   Because of the
unknown correlation and the fact that the MB sample is severely
statistically inferior to the triggered sample, no direct evaluation
of their agreement was performed.  However, it can be seen in
Figure~\ref{figure:bbc2x2Agreement} that the MB and triggered
results do not exhibit notable disagreement.

\subsection{Different background regions in invariant mass}
\label{section:backgroundStudy}

As described above, the asymmetry of two different background mass
regions in invariant mass was investigated:  50-MeV/$c^2$ regions
around the $\pi^{0}$ mass peak (60-110 MeV/$c^2$ and 170-220
MeV/$c^2$) and the mass region between the $\pi^{0}$ and $\eta $
(250-450 MeV/$c^2$). Table~\ref{table:backgroundComparison} shows
the background asymmetries as calculated by Eq.~\ref{eq:lumiFormula}
for the two background regions.  The asymmetries are similar.  The
background asymmetries are consistent with zero for both background
regions for $2 < p_T < 5$. In the 1-2~GeV/$c$ $p_T$ bin, both
regions suggest a slightly negative asymmetry.

\begin{table}[tbp] \centering
\begin{tabular}{|c|c|c|c|c|}
  \hline
  $p_T$ (GeV/$c$) & \multicolumn{2}{|c|}{$A_N$} &
  \multicolumn{2}{|c|}{$\sigma_{A_N}$} \\ \hline

   & bg 1 & bg 2 & bg 1 & bg 2 \\
   \hline

  1-2  & -0.008 & -0.014 & 0.005  & 0.004 \\ \hline
  2-3  & -0.006 & 0.008 & 0.020  & 0.013 \\ \hline
  3-4  & -0.012 & 0.015 & 0.079  & 0.055 \\ \hline
  4-5  &  0.00  & 0.03  & 0.21   & 0.14 \\
  \hline
\end{tabular}
\caption[Comparison of background asymmetries.]{Asymmetry results of
photon pairs falling within the two 50-MeV/$c^2$ regions around the
mass peak (bg 1) and within 250-450 MeV/$c^2$ (bg 2). Note that this
check was performed before the final cuts on the data sample; thus,
the results from bg 1 shown here differ from the final background
results.} \label{table:backgroundComparison}
\end{table}

\begin{figure}
\centering
\includegraphics[height=0.5\textheight]{%
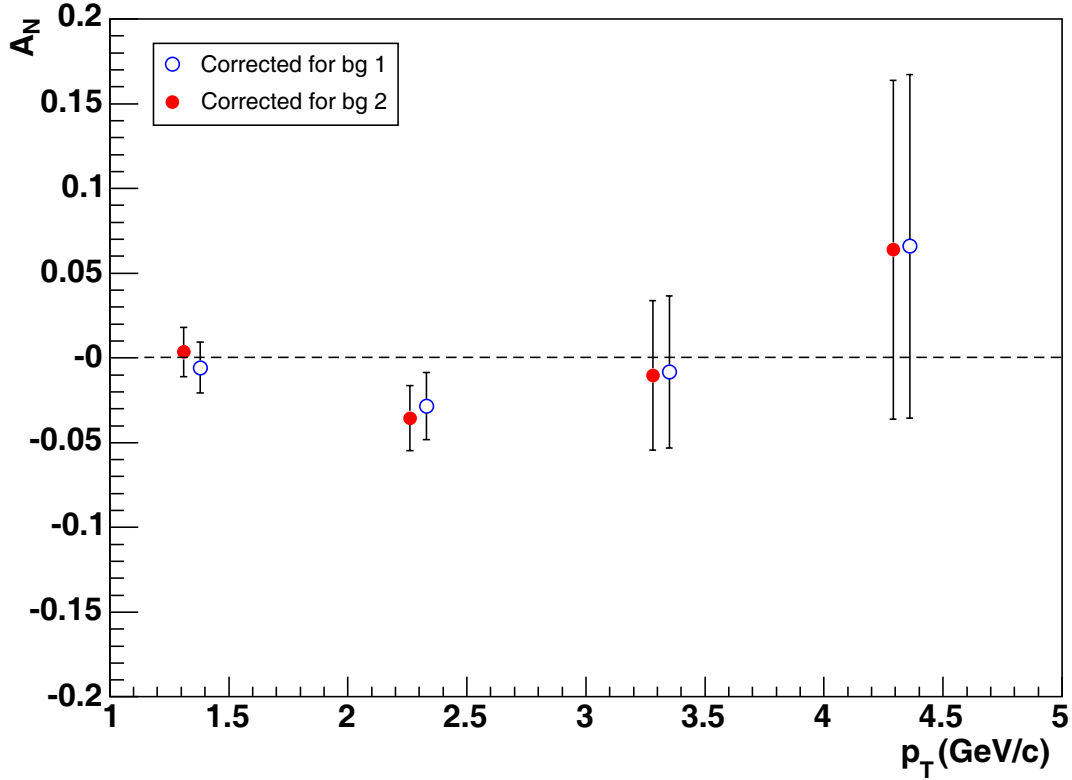} \caption[Asym. comparison after
subtracting two different background regions.]{Comparison of
asymmetries obtained after subtracting the asymmetry of two
different background invariant-mass regions. "bg 1" indicates the
asymmetry after subtraction of the asymmetry around the peak; "bg 2"
is after subtraction of the asymmetry of the 250-450 MeV/$c^2$
background region.  Points for bg 2 are shifted down by 50 MeV/$c$
from the center of the bin for readability. Note that this check was
performed before the final cuts on the data sample; thus the results
differ from the final ones.} \label{figure:subtractedBackgrounds}
\end{figure}

In Figure \ref{figure:subtractedBackgrounds} a direct comparison of
the asymmetry results after subtraction of the two background
asymmetries is shown.  The background region used has little effect
on the final asymmetry.

\subsection{Different neutral pion invariant-mass integration regions}

\begin{figure}
\centering
\includegraphics[angle=0,height=0.5\textheight]{%
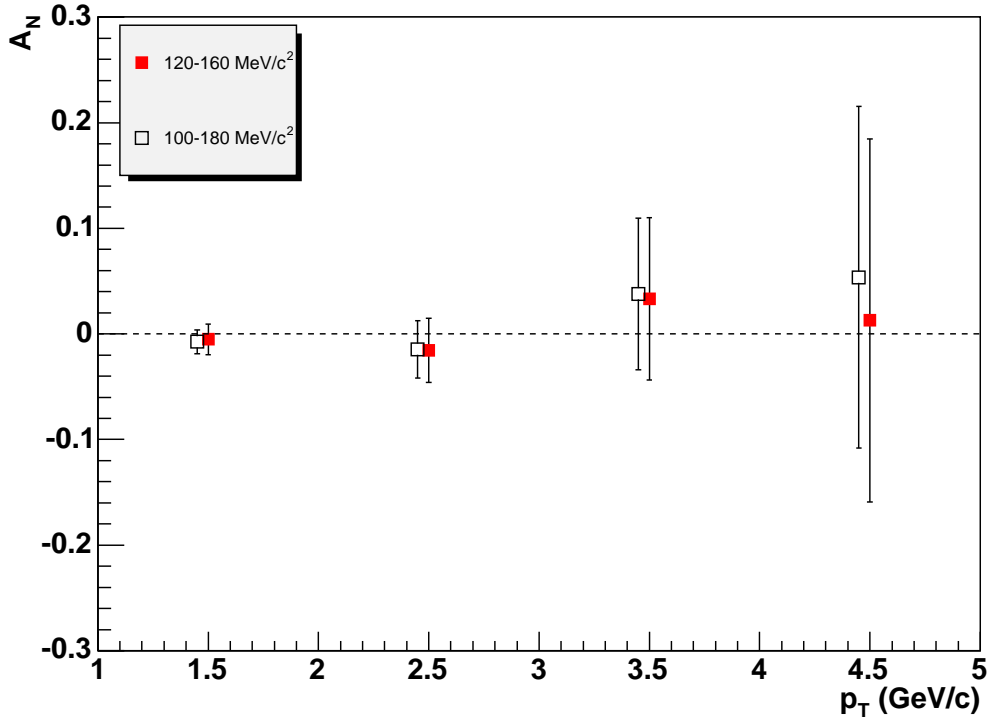} \caption[Asym. comparison for different \piz\ peak
integration regions.]{Comparison of results for different \piz\ peak
integration regions.  Points for the wider integration region are
shifted down by 50 MeV/$c$ from the center of the bin for
readability.} \label{figure:massWindowAgreement}
\end{figure}

As an additional check on the sensitivity of $A_N$ to the background
under the \piz\ peak, three different integration regions for the
\piz\ mass were examined:  120-160 MeV/$c^2$, 110-170 MeV/$c^2$, and
100-180 MeV/$c^2$.  Figure~\ref{figure:massWindowAgreement} shows a
comparison of asymmetries obtained for peak integration from
120-160~MeV/$c^2$ and 100-180~MeV/$c^2$.  There is little effect on
the result from the amount of background included, providing
additional evidence that the background under the peak does not
affect the asymmetry greatly.

\subsection{Bunch shuffling}
A technique called "bunch shuffling" can be utilized to check for
uncorrelated bunch-to-bunch and fill-to-fill systematic errors.  For
each bunch crossing the spin direction is reassigned randomly, and
then the new asymmetry with false spin dependence is recalculated.
This procedure is repeated many times. With random reassignment of
the spin direction to each crossing, a Gaussian asymmetry
distribution centered around zero is expected. One must take care in
choosing the exact procedure used for this study.  If done
correctly, the root-mean-square (RMS) width of the "shuffled"
(non-physics) asymmetry distribution should correspond to the
statistical uncertainty on the physics asymmetry, since fluctuations
in the calculated asymmetry should not be due to any spin dependence
but rather to statistical fluctuations in event-by-event particle
production. If the RMS width of the shuffled asymmetry distribution
is larger than the statistical uncertainty on the physics asymmetry,
it should reflect the presence of elements that broaden the
distribution beyond statistical fluctuations in particle production,
e.g.~some source of bunch-to-bunch systematic error in the physics
asymmetry.  It should be noted, however, that the expected
quantitative agreement between the statistical uncertainty and RMS
width of the shuffled distribution is not entirely understood.  As
is discussed in Section~\ref{section:shufflingMC}, a study done for
this thesis found that it is also possible to obtain widths smaller
than the statistical uncertainties.

It is easier to consider the validity of various procedures assuming
large asymmetries, for example 1 (100\%) .  Taking one bunch at a
time and randomly assigning its spin direction to be up or down,
with no further constraints, it is theoretically possible to make
the spin assignments exactly as the true, original spin directions
($\textrm{up}_{\textrm{phys}} \rightarrow
\textrm{up}_{\textrm{shuf}}$, $\textrm{down}_{\textrm{phys}}
\rightarrow \textrm{down}_{\textrm{shuf}}$), or exactly opposite
($\textrm{up}_{\textrm{phys}} \rightarrow
\textrm{down}_{\textrm{shuf}}$, $\textrm{down}_{\textrm{phys}}
\rightarrow \textrm{up}_{\textrm{shuf}}$), yielding a shuffled
asymmetry distribution that can range from -1 to +1, even with a
wealth of statistics. This method would generally give distribution
widths wider than the statistical uncertainties on the physics
asymmetry values. Selecting any half of the bunches to be assigned
up and the other half to be assigned down creates a similar
situation for true relative luminosity values close to 1, i.e.~a
nearly equal number of events coming from bunches with spin up and
spin down: it is possible to assign the shuffled spins to be nearly
the same as the original physics spin directions.

The procedure utilized in this analysis was to assign half of the
original up spins to down and keep the others as up, and similarly
assign half of the original down spins to up while keeping the
others down. Thus the particle production gets redistributed evenly,
at the \emph{crossing} (not event) level, between up and down spins,
and even for large physics asymmetries, shuffled asymmetry will
always be (nearly) zero. Repeating this procedure many times should
yield a distribution around zero that is due to
\emph{event-by-event} statistical fluctuations in particle
production (i.e. 0, 1, 2, etc. neutral pions produced in a given
event, with a certain probability distribution).

\begin{figure}
\centering
\includegraphics[angle=0,height=0.5\textheight]{%
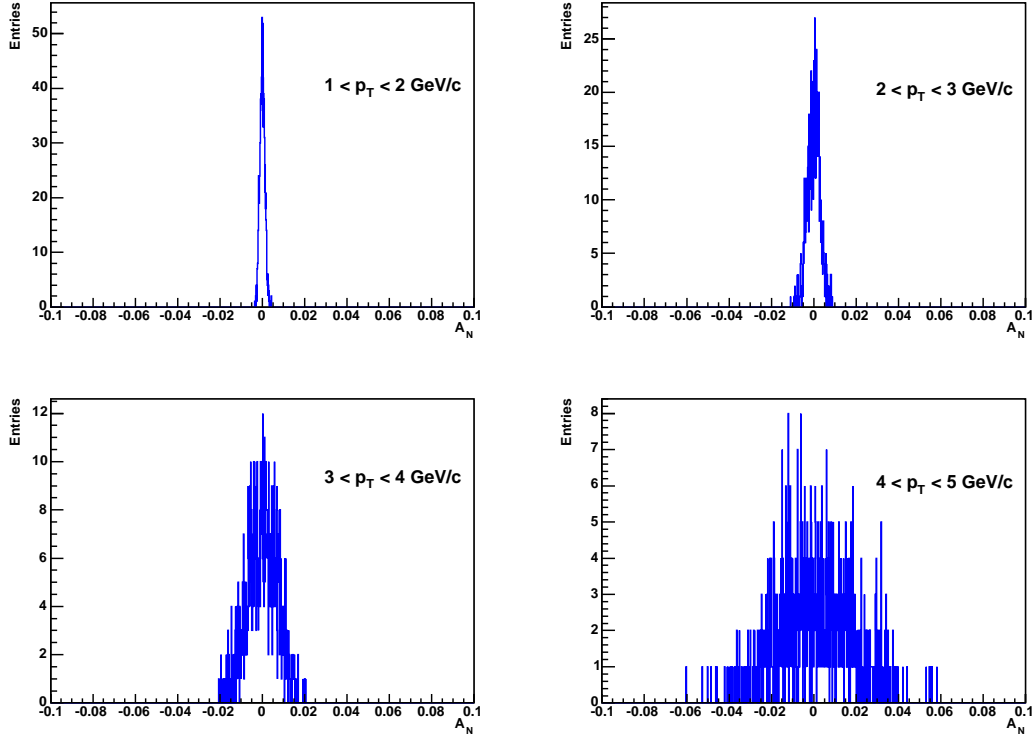} \caption[Bunch-shuffled asymmetry
distributions.]{Bunch-shuffled asymmetry distributions for triggered
data, shown here for the blue beam. } \label{figure:shuffledAsym}
\end{figure}

\begin{figure}
\centering
\includegraphics[angle=0,height=0.5\textheight]{%
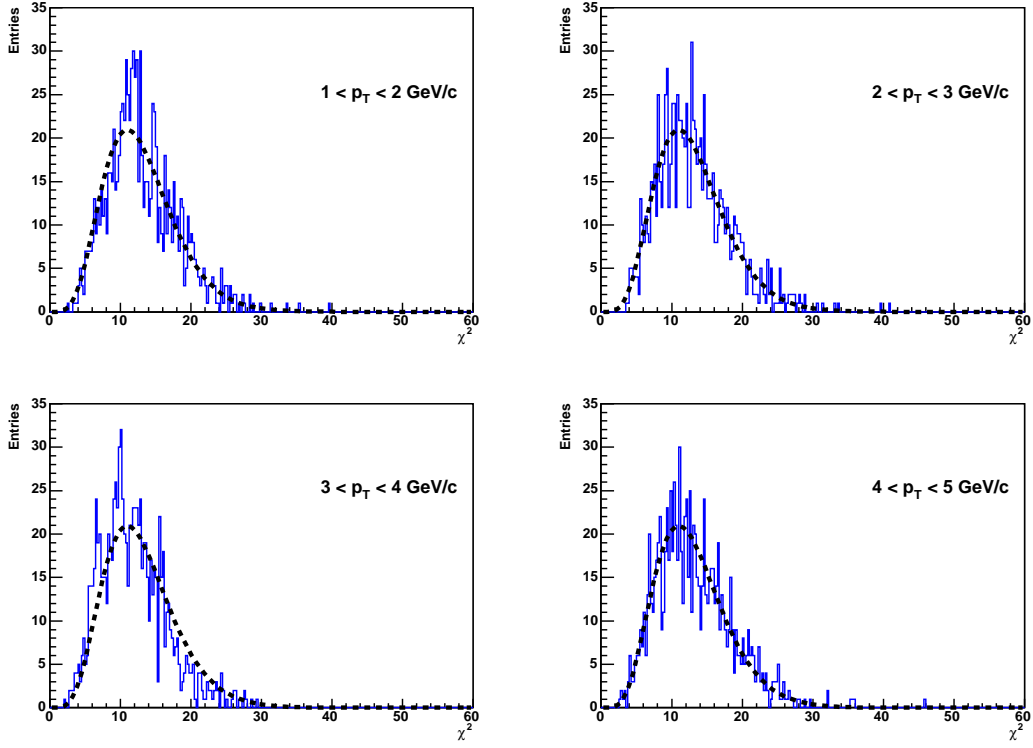} \caption[Bunch-shuffled $\chi^2$
distributions.]{Bunch-shuffled $\chi^2$ distributions from triggered
data, shown here for the blue beam. The dashed lines indicate the
expected distributions.} \label{figure:shuffledChiSq}
\end{figure}

For this analysis, Eq.~\ref{eq:sqrtFormula} was used to calculate
the shuffled asymmetries, as it avoided the complication of
recalculating the relative luminosity for each iteration. The bunch
shuffling procedure was performed 1000 times. Refer to Figure
\ref{figure:shuffledAsym} for examples of shuffled asymmetry
distributions from the data set used for this analysis; the expected
increase of the distribution width for the smaller-statistics
(higher-$p_T$) bins can be clearly seen.  In Table
\ref{table:shuffledMeans} it can be seen that the mean of the
shuffled asymmetry distributions was zero at the level of $10^{-4}$
or better. The $\chi^2$ distributions for a fit to a constant
asymmetry across all shuffled fills can be seen in Figure
\ref{figure:shuffledChiSq}. The dotted line shows the expected
$\chi^2$ distribution for 13 degrees of freedom (asymmetry values
for 14 fills, fit to a constant), which are in agreement with the
data.

\begin{table}  [tbp] \centering
\begin{tabular}{|c|c|c|} \hline
 $p_T$ (GeV/$c$) & Beam & $\overline{A}_N^{\textrm{shuf}}$  \\ \hline

1-2 & Blue & $-9.3 \times 10^{-6}$ \\

 & Yellow &  $-2.2 \times 10^{-5}$ \\ \hline

2-3 & Blue & $-7.8 \times 10^{-7}$ \\

 & Yellow & $-2.2 \times 10^{-5}$ \\ \hline

3-4 & Blue & $3.8 \times 10^{-5}$ \\

 & Yellow & $1.0 \times 10^{-4}$ \\ \hline

4-5 & Blue & $3.0 \times 10^{-4}$ \\

 & Yellow & $-4.3 \times 10^{-4}$ \\ \hline
\end{tabular}
\caption[Shuffled asymmetry distribution means.]{Mean values of
shuffled asymmetry distributions.} \label{table:shuffledMeans}
\end{table}

Table \ref{table:shuffledAgreement} shows the statistical
uncertainties on the physics asymmetries compared to the RMS widths
of the shuffled asymmetry distributions for the blue and yellow
beams. In order for the uncertainties on the physics asymmetries to
reflect simply statistical distributions in particle production, the
beam polarization was not factored in.  For this reason, the
statistical uncertainties are exactly the same for the blue and
yellow beams, the results for each using the same total set of
events.  The \% difference is calculated as $100 \times
(\sigma_{A_N^{\textrm{shuf}}} -
\sigma_{A_N^{\textrm{phys}}})/\sigma_{A_N^{\textrm{phys}}}$.

\begin{table}[tbp] \centering%
\begin{tabular}{|c|c|c|c|c|}
\hline $p_T$ (GeV/$c$) & Beam & $\sigma_{A_N^{\textrm{phys}}}$ &
$\sigma_{A_N^{\textrm{shuf}}}$ & \% difference \\ \hline
1-2 & Blue & 0.0014 & 0.0012 & -14 \\
 & Yellow & 0.0014 & 0.0012 & -14 \\ \hline
2-3 & Blue & 0.0032 & 0.0030 & -6.3 \\
 & Yellow & 0.0032 & 0.0028 & -13 \\ \hline
3-4 & Blue & 0.0082 & 0.0070 & -15 \\
 & Yellow & 0.0082 & 0.0069 & -16 \\ \hline
4-5 & Blue & 0.0191 & 0.0179 & -6.3 \\
 & Yellow & 0.0191 & 0.0179 & -6.3 \\ \hline
\end{tabular}%
\caption[Comparison of stat. uncertainties and widths of shuffled
distributions.] {Comparison of statistical uncertainties on the
physics asymmetries ($\sigma_{A_N^{\textrm{phys}}}$) and RMS widths
of shuffled asymmetry distributions
($\sigma_{A_N^{\textrm{shuf}}}$). } \label{table:shuffledAgreement}
\end{table}%

The shuffled widths are systematically \emph{smaller} than the
statistical uncertainties calculated for the physics asymmetry. This
outcome was at first a surprise, as the expectation was that the RMS
width should have been the same as or greater than the physics
statistical uncertainty. However, a subsequent Monte-Carlo study of
the bunch shuffling technique, discussed below, corroborated the
tendency for the width of shuffled distributions for a zero physics
asymmetry to be narrower than the statistical uncertainty.

\subsubsection{Bunch shuffling Monte Carlo}
\label{section:shufflingMC}

A Monte Carlo study of the bunch shuffling technique was performed
to better quantitatively understand the expected agreement between
the RMS widths of shuffled asymmetry distributions and the
statistical uncertainties of the physics asymmetry values, for
different simulated physics asymmetry values.

In the Monte Carlo study, 14 fills were assumed, the same number as
in the actual data set. In order to simulate the variation in the
total particle yield per fill, the actual total \piz\ yields per
fill for the 1-2 GeV/$c$ $p_T$ bin were used as the simulated yields
per fill.  The number of bunches per beam was taken to be 48,
selected because it was very close to the typical value of 46 in the
actual data set, and because it was divisible by four, so that it
was possible to assume half of the bunches were spin up and half
were spin down, then easily reassign exactly half in each spin group
to the wrong spin direction.  The desired physics asymmetry to
simulate was selected, which determined the average yield per bunch
with spins up and down in each fill. Simulated yields for each bunch
crossing were produced by sampling from Poisson distributions around
these asymmetry-dependent averages.  In this way, the uncertainty on
the yields themselves was purely Poisson and well understood.  The
uncertainty on the simulated asymmetry was purely statistical and
calculated by performing straightforward error propagation on the
asymmetry formula assuming only uncertainties on the yields; there
were no effects due to detectors, triggers, polarization
measurements, or any other factors incorporated. The simulated
yields were then input into the same software program written to
handle the bunch shuffling of the data. The RMS width of the
resulting shuffled asymmetry distribution was then compared to the
statistical error on the simulated "physics" asymmetry.

The effects of varying the simulated physics asymmetry, the total
statistics, and the number of bunches per beam were studied.  In
Table~\ref{table:shufflingMCMain} the results for two different
statistical sample sizes as well as three different simulated
asymmetry values are shown.  There are several items to note here.
Despite incorrectly assigning the spin direction for exactly one
half of the bunches, the width of the shuffled distribution
increases noticeably as a function of the simulated physics
asymmetry. The statistical uncertainty on the physics asymmetry is
itself weakly but \emph{inversely} dependent on the physics
asymmetry, leading to a shuffled width that is more than double the
simulated uncertainty for a 90\% simulated asymmetry. For the case
of a zero simulated asymmetry, corresponding to the results of the
present analysis (see Section~\ref{section:results}), a shuffled
width smaller than the simulated uncertainty is obtained for a
statistical sample equivalent to the \piz\ yield in the 1-2 GeV/$c$
$p_T$ bin in the present analysis. For ten times this statistical
sample, the study suggests that the shuffled widths tend to increase
with respect to the smaller sample, for at least the zero and 30\%
simulated asymmetry values. In order to investigate the possibility
that the smaller width for zero simulated asymmetry was due to
correlations in obtaining shuffled asymmetries many times for a
limited number of bunch crossings, a study with only 10 bunches per
beam was performed, the results of which are given in
Table~\ref{table:shufflingMCNumCrossings}. Performing 10,000
reassignments of the spin direction of five out of ten bunches
should have led to many duplicate reassignments and thus duplicate
shuffled asymmetry values in the distribution. (Note that the
difference in the number of iterations in the data versus the
simulation was due simply to technical reasons.)  For the smaller
statistical sample size, the shuffled distribution width was only
slightly smaller for 10 bunches per beam than 48 bunches; for the
larger sample size the reduction was more significant. While the
degree of correlation should not depend on the value of the physics
asymmetry, it may be that the reduction of the shuffled distribution
width due to correlations only makes it less than the statistical
uncertainty in the case of zero asymmetry because for larger
asymmetries, the dependence of the width on the asymmetry is a
larger effect.

As an additional means of investigating correlations as the cause of
shuffled distribution widths being smaller than the statistical
uncertainties, an attempt was made to remove all correlations.  To
achieve this, the yield for each crossing was resampled for each
iteration of the reassignment of spin directions.  This procedure
effectively incorrectly assigned the spin direction for the yields
from the bunches in many different fills ($\sharp$ effective fills =
$\sharp$ shuffling iterations), all having the same physics
asymmetry, rather than repeatedly reassigning the spin direction for
the yields in a single fill or small number of fills. The results of
this exercise are shown in
Table~\ref{table:shufflingMCNoCorrelations}; the shuffled width is
within 1\% of the statistical uncertainty on the simulated physics
asymmetry, suggesting that the smaller shuffled distribution widths
may be due to correlations.  Note that this procedure is not
applicable to real data, due to the fact that there is always a
limited number of fills.

\begin{table} \centering
\begin{tabular}{|c|c|c|c|c|}
  \hline
  Statistics & $A_N^{\textrm{sim}}$ & $\sigma_{A_N^{\textrm{sim}}} \times 10^4$ & $\sigma_{A_N^{\textrm{shuf}}} \times 10^4$ & \% difference \\
  \hline
  1 & 0 &  9.70 & 9.16 & -5.5 \\
  \hline
  1 & 0.3 & 9.20 & 9.18 & -0.2 \\
  \hline
  1 & 0.9 & 4.20 & 9.37 & 123 \\
  \hline
  10 & 0 & 3.05 & 3.09 & 1.3 \\
  \hline
  10 & 0.3 & 2.91 &  3.09 & 6.2 \\
  \hline
  10 & 0.9 & 1.33 & 2.99 & 125 \\
  \hline
\end{tabular}
\caption[Shuffling results for different sample sizes and simulated
asym. values.]{Bunch shuffling results for two different statistical
sample sizes and three different simulated asymmetry values,
assuming 48 bunches per beam. The "Statistics" column indicates the
factor times the 1-2 GeV/$c$ $p_T$ bin \piz\ yields in the actual
data sample.  10,000 shuffles were performed.}
\label{table:shufflingMCMain}
\end{table}

\begin{table} \centering
\begin{tabular}{|c|c|c|c|c|c|}
  \hline
  \# bunches & Statistics & $A_N^{\textrm{sim}}$ & $\sigma_{A_N^{\textrm{sim}}} \times 10^4$ & $\sigma_{A_N^{\textrm{shuf}}} \times 10^4$ & \% difference \\
  \hline
  10 & 1 &  0  & 9.65 &  9.01 &  -6.6 \\
  \hline
  48 & 1 & 0 &  9.70 & 9.16 & -5.5 \\
\hline
  10 & 10 & 0  & 3.05 &  2.89 &  -5.2 \\
  \hline
  48 & 10 & 0 & 3.05 & 3.09 & 1.3 \\
  \hline
\end{tabular}
\caption[Comparison of simulated shuffling results for 10 and 48
bunches.]{Comparison of simulated shuffling results for 10 and 48
bunches per beam. 10,000 shuffles were performed.}
\label{table:shufflingMCNumCrossings}
\end{table}

\begin{table} \centering
\begin{tabular}{|c|c|c|c|c|}
  \hline
  Statistics & $A_N^{\textrm{sim}}$ & $\sigma_{A_N^{\textrm{sim}}} \times 10^4$ & $\sigma_{A_N^{\textrm{shuf}}} \times 10^4$ & \% difference \\
  \hline
  1 &  0 &  9.66 &  9.74 &  0.9 \\
  \hline
\end{tabular}
\caption[Modified bunch shuffling simulation with correlations
removed.]{Modified bunch shuffling simulation with correlations
removed (see text).} \label{table:shufflingMCNoCorrelations}
\end{table}

While the Monte Carlo investigation confirmed the pattern seen in
applying the bunch shuffling technique to the actual data, the
merits and limitations of the technique are still not completely
understood.  Improved procedures for randomly reassigning the spin
direction of the bunches in order to obtain better agreement between
statistical uncertainties on the physics asymmetries and shuffled
asymmetry distribution widths may exist. A more thorough future
study would be valuable. This method of checking for uncorrelated
bunch-to-bunch and fill-to-fill systematic errors is particularly
important for double-spin asymmetry measurements, in which many of
the checks available to a single-spin analysis are not available.

\section{Results}
\label{section:results}

\begin{figure}
\centering
\includegraphics[height=0.5\textheight]{%
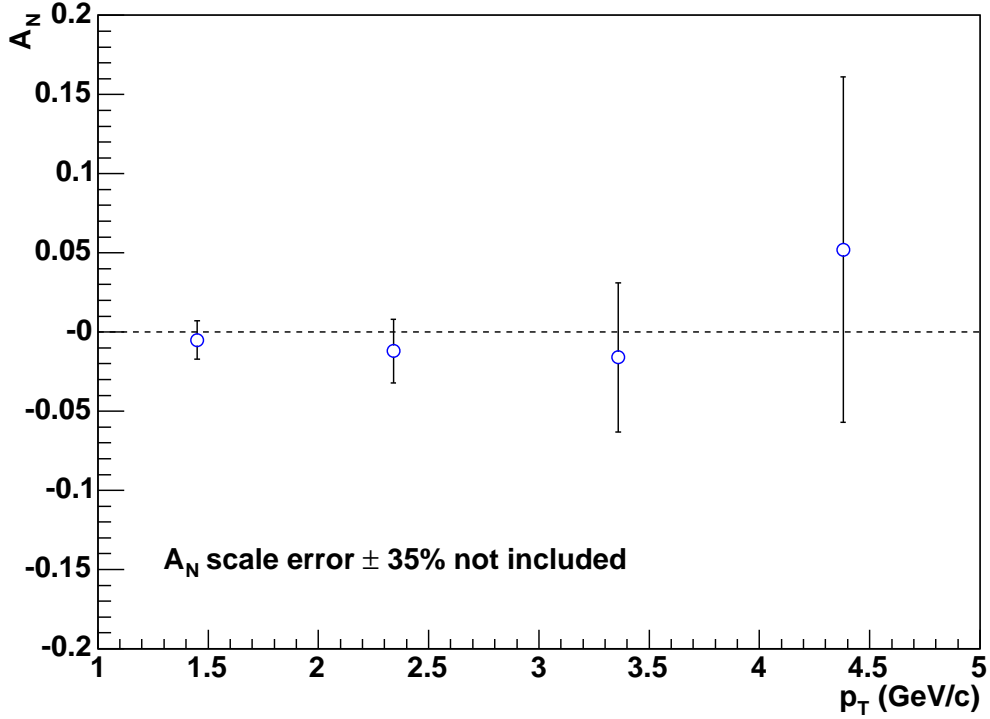} \caption[Final \piz\ asymmetry results.]{ Final
mid-rapidity neutral pion transverse single-spin asymmetry. The
error bars represent statistical uncertainties.}
\label{figure:finalAsym}
\end{figure}

Final asymmetry results for mid-rapidity neutral pions from 200-GeV
polarized $p+p$ collisions for $1 < p_T < 5$~GeV/$c$, as published
in \cite{Adler:2005in}, are given in Figure~\ref{figure:finalAsym}
and Table~\ref{table:finalPi0Results}. They utilized the triggered
data sample and the 50-MeV/$c^2$ regions in invariant mass around
the \piz\ peak for the background correction.  The asymmetries are
consistent with zero within a few percent for all $p_T$ bins. For
further discussion of these results and their implications, see
Chapter~\ref{section:conclusions}.

\begin{table} \centering
\begin{tabular}{|c|c|c||c|c|c|} \hline
$p_T$   & $\langle p_T \rangle$  & $r$  & $A_N^{\textrm{peak}}$  & $A_N^{\textrm{bg}}$ & $A_N^{\pi^0}$  \\
(GeV/$c$) & (GeV/$c$)                 & (\%) & (\%)          & (\%)
& (\%)          \\ \hline

1-2       & 1.45 & 34   & -0.6$\pm$0.8  & -0.7$\pm$ 0.9 & -0.5$\pm$
1.2 \\ \hline

2-3 & 2.34 & 12   & -1.4$\pm$1.7  & -3.1$\pm$ 3.4 & -1.2$\pm$ 2.0 \\
\hline

3-4 & 3.36 & 6 &  1.3$\pm$4.2  & 3.6$\pm$12.2 & -1.6$\pm$ 4.7
\\ \hline

4-5 & 4.38 &  5   & 7.0$\pm$10.1 & 42  $\pm$39   & 5.2$\pm$10.9 \\
\hline
\end{tabular}
\caption[Asymmetry results.]{Neutral pion transverse single-spin
asymmetry values and statistical uncertainties for all photon pairs
falling within the $\pi^0$ mass peak, for the background
($\textrm{bg}$), and for the $\pi^0$ background-corrected.  The
third column ($r$) indicates the background contribution under the
$\pi^0$ peak.  An $A_N$ scale uncertainty of $\pm$35\% is not
included.} \label{table:finalPi0Results}
\end{table}

\section{Comparison to charged hadron asymmetry results}
\label{section:charged}

\begin{figure}
\centering
\includegraphics[height=0.5\textheight]{%
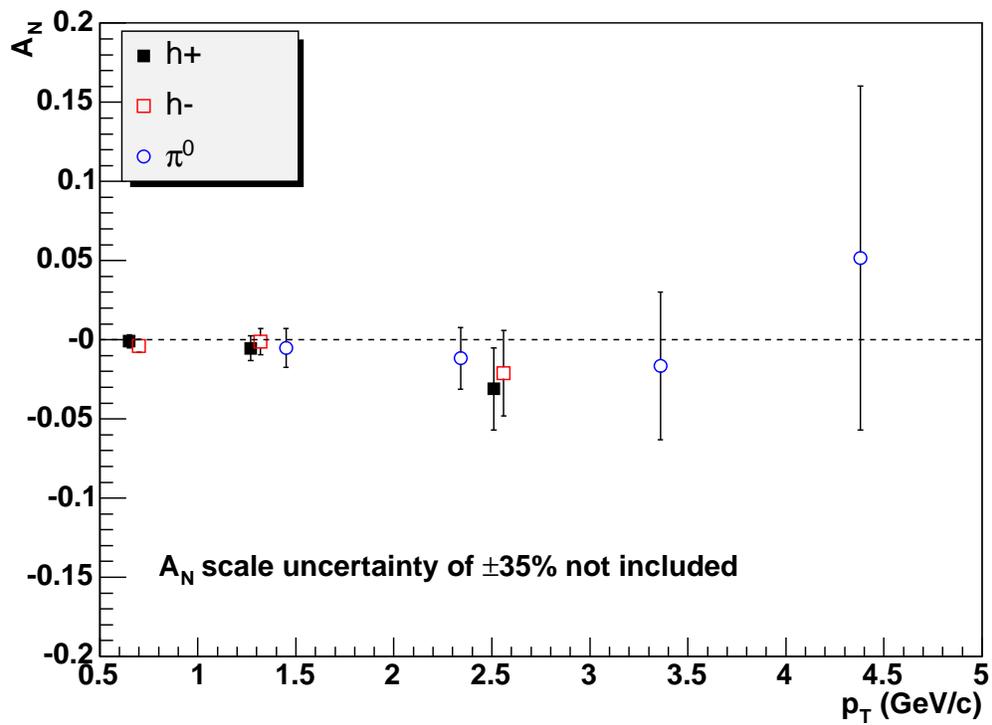} \caption[Comparison of $A_N$ for
mid-rapidity neutral pions and charged hadrons.]{ Mid-rapidity
neutral pion and charged hadron transverse single-spin asymmetries
versus mean $p_T$. Points for positive hadrons have been shifted
down by 50 MeV/$c$ to improve readability. The error bars represent
statistical uncertainties.} \label{figure:finalAsymChargedNeutral}
\end{figure}

A similar analysis of the transverse single-spin asymmetry, $A_N$,
of inclusive charged hadrons at mid-rapidity for $0.5 < p_T <
5.0$~GeV/$c$ was performed by F. Bauer for the 2001-02 data.  The
results of the neutral pion and charged hadron measurements have
been published together in \cite{Adler:2005in}.  For more details on
the charged hadron analysis, refer to the publication. Charged
hadron asymmetry results were obtained separately for positively and
negatively charged particles; they are presented in
Table~\ref{table:chargedAsymmetryResults}. While a clear charge
dependence has been observed in transverse single-spin asymmetries
for forward production (refer back to
Section~\ref{section:transverseStructure}), the results for
mid-rapidity neutral pions and both charges of hadron were all found
to be similarly consistent with zero.
Figure~\ref{figure:finalAsymChargedNeutral} shows the observed
asymmetries for both neutral pions and charged hadrons together.

\begin{table} \centering
\begin{tabular}{|c|c||c|c|} \hline
$p_T$  & $\langle p_T \rangle$ & $A_N^{h^-}$  & $A_N^{h^+}$      \\
(GeV/$c$) & (GeV/$c$)                & (\%)           & (\%)
\\ \hline

0.5-1     & 0.70  & -0.38$\pm$0.42 & -0.09$\pm$0.41
\\ \hline

1-2       & 1.32 & -0.12$\pm$0.82 & -0.54$\pm$0.78 \\ \hline

2-5 & 2.56 & -2.1$ \pm$2.7  & -3.1 $\pm$2.6  \\ \hline
\end{tabular}
\caption[Charged hadron asymmetry results.]{Charged hadron
transverse single-spin asymmetry values and statistical
uncertainties.  An $A_N$ scale uncertainty of $\pm$35\% is not
included.} \label{table:chargedAsymmetryResults}
\end{table}

\chapter{Future prospects for transverse spin physics}
\label{section:future}

The experiments at RHIC have the potential to perform a wide variety
of studies related to the transverse spin structure of the proton.
Several measurements beyond those completed have already been
planned, and others are being proposed and considered.  Because
there are currently a number of possible mechanisms under discussion
that may contribute to the large transverse SSA's observed, seeking
measurements sensitive to or dominated by a single mechanism is
especially valuable.  In this chapter, potential future measurements
at RHIC are discussed, with particular focus on possible
measurements by the PHENIX experiment.

\section{BRAHMS}

The BRAHMS experiment does not have spin rotator magnets around its
IP and therefore does not have access to longitudinally polarized
collisions.  Consequently, its polarized proton program is
completely focused on transverse spin physics.  BRAHMS was designed
to specialize in the measurement of identified charged particles at
forward and mid-rapidity.  It is in a unique position at RHIC to
make a precision measurement of $A_N$ for identified charged
particles ($\pi^+ / \pi^-$, $K^+ / K^-$, $p / \bar{p}$) in the
forward and backward directions with respect to the polarized beam.
Preliminary results from pions and protons are already available
from the 2004 data \cite{Videbaek:2005fm}; the first kaon
asymmetries as well as more accurate pion and proton asymmetries are
expected from the data obtained in 2005.  The kaon asymmetries and
their charge dependence will provide additional information directly
in support of or against the hypothesis that the large asymmetries
observed are due to valence quarks.  Assuming they are, positive
kaons, with a quark content of $u\bar{s}$, would be expected to have
a positive asymmetry, and negative kaons, comprised of $\bar{u}s$,
would be expected to have an asymmetry of approximately zero or only
slightly negative. More precise $A_N$ measurements will also help to
distinguish among various model calculations that are consistent
with the present results and to put an upper bound on the Sivers
distribution function. Now that the BELLE experiment has made a
first measurement of the Collins FF \cite{Abe:2005zx}, transverse
SSA's will also be able to constrain transversity. The 2005 BRAHMS
results for pions are expected to cover $0.15 < |x_F| < 0.35$; lower
beam energies in future running would allow even higher $x_F$ values
to be reached. Current results from STAR suggest that the $\pi^0$
asymmetry may have a maximum at $x_F \approx 0.5$ or 0.6 and
decrease quickly beyond that; see below. Lower-energy running would
permit exploration of the behavior of the charged-particle
asymmetries as a function of $x_F$ and provide information on
transversity at different $x$ values. RHIC is capable of colliding
polarized proton beams at center-of-mass energies as low as 50~GeV.

\section{STAR}

The STAR detector has the largest acceptance among the RHIC
experiments.  Its main barrel, a time-projection chamber for
tracking charged particles surrounded by an electromagnetic
calorimeter, has full azimuthal coverage for $|\eta| < 1$, making it
particularly suitable for multi-particle correlation measurements
and jet reconstruction. In 2003 Boer and Vogelsang proposed a single
transverse-spin di-jet measurement that could isolate the Sivers
effect and probe the gluon Sivers function \cite{Boer:2003tx}. A
non-zero Sivers function implies a spin dependence in the $k_T$
distributions of the partons within the proton, which would lead to
an observable spin-dependent asymmetry in $\Delta \varphi$ of
(nearly) back-to-back jets. A discussion of how such an analysis
would be performed by STAR is given in \cite{Fatemi:2004dk}; there
will likely be a significant result after the 2006 run at RHIC.

As a follow-up to earlier neutral pion $A_N$ measurements with a
forward $\pi^0$ detector covering $3.3 < \eta < 4.2$
\cite{Adams:2003fx,Ogawa:2004vm}, STAR intends to make a precision
measurement of the $\pi^0$ transverse SSA over a large $x_F$ range.
The current results suggest that the asymmetry may decrease in
magnitude for $x_F \gtrsim 0.6$; additional data should make this
trend clearer, if it exists.  As is the case for BRAHMS $A_N$
measurements, further $A_N$ results from STAR will be able to
constrain the Sivers and transversity distributions.

Through recent upgrades, STAR now has electromagnetic calorimetry
covering $-1 < \eta < 2$, allowing observation of neutral pions and
direct photons within this acceptance.  Measurement of $A_N$ for
direct photons has been proposed as a probe of the Sivers function
for gluons \cite{Schmidt:2005gv}.  As mentioned previously, direct
photon production is dominated by quark-gluon Compton scattering ($q
+ g \rightarrow \gamma + X$) over a wide range in photon transverse
momentum at RHIC.  The fact that there is no fragmentation involved
in direct photon production eliminates contributions to the
asymmetry from the Collins effect. Alternatively or additionally,
STAR could measure correlated photon-jet pairs, which would offer
better understanding of the kinematics of the direct photon
production, i.e.~the $x$ values of the two partons. Jet production,
similar to direct photon production, does not consider fragmentation
to particular hadrons and is insensitive to the Collins effect.
SSA's of photons and jets in events with correlated photon-jet pairs
would thus access the gluon and quark Sivers functions,
respectively, with some ability to identify the $x$ values at which
these functions are probed.

\section{PHENIX}

As discussed in Section~\ref{section:PHENIX}, PHENIX specializes in
the measurement of leptonic and photonic probes and high-$p_T$
particles over a limited acceptance. From the modest transverse-spin
data sample taken in 2005 (0.16 $\textrm{pb}^{-1}$, $\sim 48\%$
average polarization), PHENIX has already begun analysis to obtain
improved mid-rapidity $A_N$ results for neutral pions and charged
hadrons, which are expected to provide tighter constraints on the
gluon Sivers function. Future higher-statistics data samples for
these particles at mid-rapidity will provide greater sensitivity to
transversity and the Collins effect.  See
Chapter~\ref{section:conclusions} for further relevant discussion.

There is also analysis underway to obtain first results for $A_N$ of
single muons, largely from open charm decay but with significant
contributions from light-hadron decays. The current $x_F$ reach for
this measurement is $\sim 0.15$, lower than the $x_F$ values at
which significant asymmetries have been observed for other particle
species; higher $x_F$ values would become accessible with
lower-energy running.  A forward hadron $A_N$ measurement using the
PHENIX muon spectrometers ($1.2 < |\eta| < 2.4$) may also be
possible using muons from hadron decays and the charged hadrons
themselves that punch through the steel absorber in front of the
muon tracker; however, careful studies will need to be done in order
to understand the particle ratios in this sample.

In 2006, PHENIX intends to perform a transverse single -spin
di-hadron measurement similar to the STAR di-jet measurement
isolating the Sivers effect, following the proposal in
\cite{Boer:2003tx}.  This analysis would study the spin dependence
of the azimuthal angle between nearly back-to-back \piz-hadron
pairs, triggering on a decay photon from the $\pi^0$ in order to
obtain a higher-statistics sample. Although dilution of the effect
is anticipated for hadron rather than jet pairs, studies have shown
that the effect should still be measurable. In examining
back-to-back hadrons rather than jets, fragmentation to the
final-state hadrons must be considered, and some sensitivity to the
Collins mechanism is introduced. However, as described in
Chapter~\ref{section:conclusions}, there is a large contribution
from gluon fragmentation to $\pi^0$ production for $p_T \lesssim
5$~GeV/$c$, to which the Collins mechanism does not apply. The
effect is expected to be maximized when the initial spin of the
proton is in the same direction as the final-state jet axis; thus,
to match the PHENIX central arm acceptance, data will be taken with
radially polarized collisions.

Also similar to possible measurements at STAR, PHENIX could measure
$A_N$ of direct photons, sensitive to the Sivers function.  Future
upgrades extending the azimuthal coverage for tracking to $2\pi$ in
the inner region and adding forward electromagnetic calorimetry
($0.9 < |\eta| < 3.0$) are expected to expand the coverage for this
measurement as well as make $\gamma$-jet and jet-jet measurements
feasible.

It has been proposed to study the transverse SSA of $D$ meson
production at RHIC as a measurement of the gluon Sivers function
\cite{Anselmino:2004ux,Anselmino:2004nk}. This measurement would be
significant at any point over a wide range of $x_F$ values ($-0.2 <
x_F < 0.6$).  $D$ mesons are produced principally via the reaction
$g + g \rightarrow c + \bar{c}$, with contributions from $q +
\bar{q} \rightarrow c + \bar{c}$ becoming important only at very
large $x_F$ ($x_F \gtrsim 0.6$).  In neither process is the final
$c$ or $\bar{c}$ polarized, excluding the Collins mechanism from
contributing to any asymmetry that may be observed. Any transverse
SSA seen for mid- to moderate rapidity $D$ production would thus be
a direct indication of a non-zero gluon Sivers function.  PHENIX is
currently capable of measuring open charm decays statistically via
single electrons in the central arms and single muons in the muon
spectrometers. In the future, a silicon vertex detector upgrade will
make it possible to identify $D$ mesons event-by-event. A
silicon-pixel and silicon-strip barrel detector, covering the
central arm pseudorapidity region and $2\pi$ in azimuth, is expected
to be installed in 2009. A silicon-strip endcap detector, covering
the muon arm acceptance, has been proposed for 2011.

The flavor separation of the Sivers function for $u$, $d$,
$\bar{u}$, and $\bar{d}$ quarks via $A_N$ of forward or backward $W$
boson production has been suggested \cite{Schmidt:2005}.  PHENIX
already has a $W$ physics program planned for the future, once
500-GeV polarized collisions are achieved by RHIC, making $W$
measurements possible. The processes of interest will be:
\begin{eqnarray}
\nonumber
u + \bar{d} &\rightarrow& W^+ \rightarrow \mu^+ + \nu_\mu \\
\nonumber
d + \bar{u} &\rightarrow& W^- \rightarrow \mu^- +
\bar{\nu}_\mu
\end{eqnarray}
PHENIX will observe the final-state muons; an upgrade to trigger on
the highest-$p_T$ muons, which will come principally from $W$
decays, is expected to be installed in 2009.  The muon trigger
upgrade will also improve the pattern recognition in the muon arms,
reducing background and making measurements such as single muons
from $D$ decays, or muon pairs from resonance decays or the
Drell-Yan process ($q + \bar{q} \rightarrow \ell^+ + \ell^-$),
cleaner.

The double transverse-spin asymmetry, $A_{TT} =
\frac{\sigma^{\uparrow \uparrow} - \sigma^{\uparrow
\downarrow}}{\sigma^{\uparrow \uparrow} + \sigma^{\uparrow
\downarrow}}$, is another observable sensitive to transverse spin
quantities. $A_{TT}$ for the Drell-Yan process provides direct
access to the transversity distribution. The transversity
distributions for the quark and the antiquark provide the necessary
convolution of two chiral-odd functions to be an allowed process in
QCD, with no fragmentation or final-state interactions involved.
Although this asymmetry is expected to be at the sub-percent level
for $\sqrt{s} = 200$~GeV, it could reach several percent for
$\sqrt{s} < 100$~GeV. PHENIX already has an effective di-muon
trigger that would be suitable for measuring Drell-Yan pairs;
however, studies would need to be done to understand the current
backgrounds.  The muon trigger upgrade mentioned above would provide
a cleaner sample. Investigation would be necessary to optimize the
beam energy in order to obtain the best compromise between
luminosity and the size of the predicted asymmetry.  A first direct
measurement of transversity would be an exciting milestone in the
field of transverse nucleon spin structure.

\chapter{Conclusions}
\label{section:conclusions}

The transverse single-spin asymmetry, $A_N$, for neutral pion
production at $x_F\approx$0.0 for $1 < p_T < 5$~GeV/$c$ from
polarized proton-proton interactions at $\sqrt{s} = 200$~GeV has
been presented.  It is consistent with zero within statistical
errors of a few percent. The measurement, together with a similar
measurement for charged hadrons, represents the first mid-rapidity
$A_N$ result at high $p_T$ and collider energies; both have been
published in \textit{Physical Review Letters} \cite{Adler:2005in}.
The measurement is in a kinematic regime where the theoretical
framework of perturbative QCD has been demonstrated to describe the
polarization-averaged cross sections for neutral pions and charged
hadrons well for $p_T \gtrsim 1.5$~GeV/$c$. A pQCD framework is
expected to be applicable in the interpretation of the polarized
results down to similar transverse momentum values.

The present result is consistent with the mid-rapidity results for
neutral pions from $p+p$ collisions at $\sqrt{s} = 19.4$~GeV
\cite{Adams:1994yu}. The measurement is complementary to the
measurement of $A_N$ for forward neutral pion production in $p+p$
collisions at $\sqrt{s} = 200$~GeV \cite{Adams:2003fx}, which
observed asymmetries reaching $\sim 30\%$.

\begin{figure}
\centering
\includegraphics[height=0.6\textheight]{%
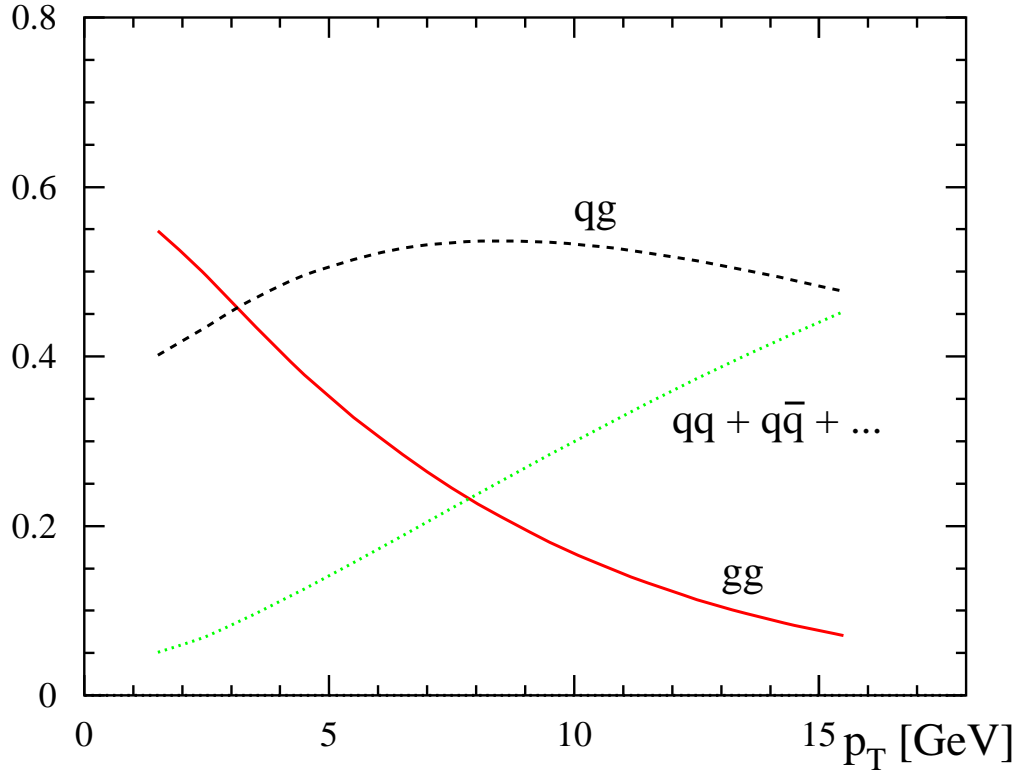} \caption[Relative contributions of
partonic processes to \piz\ production.]{Relative fractional
contributions of partonic processes to mid-rapidity \piz\ production
at $\sqrt{s} = 200$~GeV, calculated by W. Vogelsang.}
\label{figure:partonicProcesses}
\end{figure}

Neutral pion production at forward rapidity is expected to originate
from partonic processes involving valence quarks ($x \gtrsim 0.1$),
whereas the particle production at mid-rapidity presented here is
dominated by gluon-gluon and quark-gluon processes ($x \lesssim
0.1$). Figure~\ref{figure:partonicProcesses} shows the fractional
contribution of different partonic scattering processes to the
production of neutral pions at mid-rapidity at 200 GeV, calculated
by W. Vogelsang. As evident from the figure, \piz\ production in the
transverse momentum range covered by the measurement presented here
is approximately half from gluon-gluon scattering and half from
gluon-quark scattering. As such, the asymmetry is not very sensitive
to mechanisms involving quarks, e.g.~the Collins effect.

Independently of the suppression or dilution of the Collins effect
in the present results due to a dominance of gluon scattering, it
has been stated by Anselmino \textit{et al.} that the Collins effect
is suppressed more generally
\cite{Anselmino:2004ky,Anselmino:2004nm}. They argue that when all
partonic intrinsic motion is taken into account in the formalism
without simplifications or approximations, the complicated azimuthal
angle dependencies and numerous different phases involved in the
Collins mechanism lead to strong suppression of the final asymmetry.
They demonstrate that with saturated bounds on the non-perturbative
components (the transversity distribution function and Collins FF),
the Collins mechanism alone is insufficient to explain the large
asymmetries observed in forward charged pion production at $\sqrt{s}
= 19.4$~GeV \cite{Adams:1991cs}. There has been some criticism of
the formalism on which their argument is based (see for example the
alternative treatment presented in \cite{Bacchetta:2005rm}), but a
consensus in support of its validity or repudiating it has yet to be
reached within the theoretical community. In the framework of
Anselmino \textit{et al.}, the PHENIX mid-rapidity $A_N$ results,
i.e.~the \piz\ measurement presented here as well as the charged
hadron measurement discussed in Section~\ref{section:charged},
provide strong constraints on the Sivers distribution function for
gluons and indicate that it is small. A publication providing a
quantitative constraint on the gluon Sivers distribution from the
PHENIX data is forthcoming \cite{DAlesio:2005}.

Regardless of more detailed interpretations that may become
available, the results are consistent with the naive pQCD
expectation that transverse single-spin asymmetries are suppressed
at high $p_T$ and mid-rapidity \cite{Kane:1978nd,Qiu:1998ia}.

The transverse SSA results from the E704 experiment at Fermilab
\cite{Adams:1991cs,Adams:1994yu} and from the RHIC experiments
\cite{Adams:2003fx,Videbaek:2005fm,Adler:2005in} are strikingly
similar despite an order of magnitude difference in center-of-mass
energy. In the forward direction ($x_F > 0$), significant
asymmetries were observed at both energies for the production of
charged and neutral pions. In both cases a similar, clear dependence
of the asymmetry on the particle charge was observed: positive for
positive and neutral pions and negative for negative pions, with
$|A_N^{\pi^+}| \approx |A_N^{\pi^-}|$. Mid-rapidity ($x_F \approx
0$) neutral pion results are consistent with zero at both energies.
Asymmetries consistent with zero have been measured not only at
mid-rapidity but also in the backward direction ($x_F < 0$) at RHIC.
The $A_N$ results for forward, mid-rapidity, and backward particle
production considered all together suggest that the large SSA's
observed in the forward region are due to valence quarks.

Interest in transverse nucleon-spin structure and transverse SSA's
rose sharply when the first large asymmetries were observed in the
late 1970's, contradicting naive expectations from pQCD.  Despite
similar effects observed by more than one subsequent experiment in
the 1980's, it was not until higher-energy data became available
starting in the 1990's that a pQCD framework became potentially
applicable in interpreting the effects, and the theoretical ideas
most widely accepted today began to develop.

The study of transverse nucleon-spin structure has progressed
rapidly both theoretically and experimentally over the course of the
last several years. Notable experimental contributions have come
from the HERMES and COMPASS polarized deep-inelastic scattering
experiments, the BELLE experiment studying $e^+ + e^-$ annihilation,
and the STAR, PHENIX, and BRAHMS collaborations studying polarized
$p+p$ collisions at RHIC.

The recent measurement of the Collins FF for pions by BELLE
\cite{Abe:2005zx} represents a turning point in the study of the
transverse spin structure of the proton.  The non-zero result means
that the Collins mechanism remains as a possible origin of the large
transverse single-spin asymmetries observed.  Moreover, it allows
access to the transversity distribution function through single-spin
asymmetries for the first time.  Non-zero asymmetries already
measured by DIS and hadronic-collision experiments can now be
revisited and reinterpreted to obtain first constraints on
transversity, assuming the asymmetries contain contributions from
the Collins effect.

The startup of RHIC as the world's first polarized-proton collider
in late 2001 ushered in a new era in the study of nucleon spin
structure.  RHIC holds great potential for in-depth exploration of
both the transverse and longitudinal spin structure of the proton.
To date, two transverse spin publications
\cite{Adams:2003fx,Adler:2005in} and one longitudinal-spin
publication \cite{Adler:2004ps} have come out of the major
experiments at RHIC.  Additional preliminary results are also
available, as summarized in Section~\ref{section:polStructure}.

The transverse single-spin asymmetry for neutral pions presented
here represents an early measurement in a rigorous program to study
transverse proton spin structure at hard scales using a pQCD
framework at RHIC.  Conclusively explaining the large transverse
single-spin asymmetries, which have been observed over an extensive
range of energies, would help to link proton structure at soft and
hard scales. A number of planned and proposed future transverse spin
measurements have been described in Chapter~\ref{section:future}.
Some of the measurements proposed would be able to isolate
particular mechanisms. These measurements will be essential in order
to disentangle the several possible contributions to the large
observed asymmetries currently under discussion and to begin to
understand the transverse spin structure of the proton.  It should
be possible to measure transversity, the last remaining
leading-twist, $k_T$-integrated distribution function that is
completely unknown, and to start to measure the various
transverse-momentum-dependent (TMD) distribution functions such as
the Sivers function.

As discussed in Section~\ref{section:transverseStructure}, TMD
distribution functions are related to the orbital angular momentum
of partons within the proton. Thus measurement of the Sivers
function could shed light on parton OAM, which remains to date
nearly inscrutable. No proposal for a clear and direct experimental
measurement of OAM has yet been set forth. Therefore, measurement of
the Sivers function could provide a starting point for elucidating
this still-opaque aspect of nucleon angular-momentum structure.

Despite the fact that the proton is one of the most commonplace and
stable particles in existence, a fundamental component of ordinary
matter, the path to unraveling proton structure has been a long one,
and the journey is not yet finished. The final picture will not be
complete without full description of its momentum, helicity, and
transverse-spin structure. A comprehensive understanding of the
proton, in many ways the embodiment of QCD, implies an understanding
of the strong force, one of the four fundamental forces in nature
and the foundation for all of nuclear physics.

\begin{appendix}
\chapter{Relative luminosity considerations}
\label{section:relLumi}

In a collider environment, different considerations must be made in
determining the relative luminosity of crossings with different spin
configurations in the case of double- or single-spin asymmetry
measurements, and for transverse or longitudinal spin in the case of
single-spin asymmetry measurements.  Care must be taken in designing
a MB trigger that has no spin-dependent bias and can make an
accurate measurement of the relative luminosity. To make double-spin
asymmetry measurements, the principal concern is that there may be a
non-zero double-spin asymmetry in the production of particles that
fire the MB trigger.  For PHENIX, the MB trigger fires when at least
one charged particle produces a hit in each BBC ($3.0 < \eta <
3.9$).  A physics asymmetry in the production of such charged
particles in the kinematic range to fire the BBC's would thus lead
to a spin-dependent bias in the "minimum-bias" trigger.  While it is
extremely difficult to \emph{prove} that there is no such bias,
checks can be performed against alternative "minimum-bias" triggers,
sensitive to different physics processes.  The relative luminosity
measured by the different processes can be compared.  This technique
has been used in the measurement of the double-longitudinal
asymmetry in neutral pion production at PHENIX \cite{Adler:2004ps}.

A MB detector used simultaneously as a relative luminosity detector
will bias the measurement if there is a double-longitudinal
asymmetry in the particles to which it is sensitive.  If the
detector is capable of measuring the multiplicity of the produced
particles, it may be possible to understand this bias.  This will
not be the case if it has only a simple binary hit/no-hit response.

The physics asymmetry in the MB/relative luminosity detector could
also potentially be measured by looking for dependence of the
relative luminosity on beam polarization. If for example the
same-helicity cross section is higher than the opposite-helicity
cross section in the case of double-longitudinal observables, then
the measured relative luminosity should be enhanced when a higher
fraction of the beam is polarized.

An additional idea to measure the physics asymmetry in the
MB/relative luminosity detector would be to examine the MB rate
change upon changing the spin combinations in a single fill. Start
for example with all bunches in both beams having positive helicity
(designated by "+"), creating only $++$ collisions, and measure the
MB event rate.  Use the spin flipper described in
Section~\ref{section:spinFlipper} to flip all spins in a single beam
in order to achieve $+-$ collisions, and measure the MB event rate
again. Assuming the spin flipper does not affect other beam
conditions, a different MB rate implies a different physics cross
section for $++$ and $+-$ helicity combinations.  This procedure
could be repeated several times throughout a fill.  In order to
check for other beam effects, it would be possible to compare $++$
event rates after zero and two beam flips, or to compare the
original $++$ event rates to $-~-$ rates, with both beams flipped,
which should be same as $++$ rates if parity is conserved.

If the luminosity is high enough such that there is a non-negligible
probability of multiple collisions per bunch crossing occurring, it
may present a problem for both double- and single-spin asymmetry
measurements.  If the occurrence of multiple collisions per crossing
is not spin-dependent, the relative luminosity measurement is not
affected. If the probability of multiple collisions per crossing is
spin-dependent due to spin-dependent beam conditions, it can likely
be studied using the spin flipper.  If it is spin-dependent due to a
physics asymmetry, the problem is similar to the one discussed above
in the case of no more than a single interaction per crossing.

It should be noted that the analogous formula to
Eq.~\ref{eq:lumiFormula} for longitudinal double-spin asymmetries is
Eq.~\ref{eq:lumiFormulaDoubleLongitudinal},

\begin{equation}
A_{LL} = \frac{1}{P_1 P_2}\frac{N^{++} - \mathcal{R}N^{+-}}{N^{++} +
\mathcal{R}N^{+-}} \label{eq:lumiFormulaDoubleLongitudinal}
\end{equation}
where $P_1$ and $P_2$ are the polarizations of the two beams,
$N^{++}$ represents the particle yield from same-helicity crossings
($++$ and $--$), $N^{+-}$ represents the particle yield from
opposite-helicity crossings ($+-$ and $-+$), and $\mathcal{R}$ is
the relative luminosity between same- and opposite-helicity
crossings.  For longitudinal double-spin asymmetries, because there
is no formula analogous to Eq.~\ref{eq:sqrtFormula}, in which
luminosity differences cancel to several orders, it is essential to
determine the relative luminosity accurately.

The transverse double-spin asymmetry, $A_{TT}$, is expected to have
the form $A + B\cos 2\varphi$, where $A$ and $B$ are constants.
Therefore there is a potentially non-zero constant term as well as
an azimuthal dependence.  The same techniques for understanding the
relative luminosity as in the double-longitudinal case should be
applicable.

In the case of a transverse single-spin asymmetry, $A_N$, as in the
present analysis, the form is expected to be purely azimuthal,
expressible as $A\cos \varphi$.  In this case there is no concern
regarding physics asymmetries present in the detectors used as the
MB trigger and to measure the relative luminosity, as long as the
detectors cover the full $2\pi$ in azimuth.  A new concern arises,
however.  An azimuthal dependence of the detector efficiency could
produce bias in the "minimum-bias" trigger and the relative
luminosity measurement.  Comparing the asymmetry results obtained
using Eq.~\ref{eq:sqrtFormula}, which is largely insensitive to the
relative luminosity, provides a handle on this issue.  In addition,
many of the cross checks and systematic studies discussed above for
the double-spin asymmetries would also be relevant.

\end{appendix}

\end{document}